\documentclass[11pt,a4paper]{article}

\usepackage[margin=2.5cm]{geometry}

\usepackage[utf8]{inputenc}
\usepackage{microtype}
\usepackage{lmodern}
\usepackage{exscale}

\usepackage{amsmath}
\usepackage{amssymb}
\usepackage{mathtools}

\numberwithin{equation}{section}

\allowdisplaybreaks

\usepackage{hyperref}
\usepackage[dvipsnames]{xcolor}
\hypersetup{
    colorlinks,
    linkcolor={WildStrawberry},
    citecolor={WildStrawberry},
    urlcolor={WildStrawberry}
}

\usepackage{cite}

\usepackage{mathrsfs}

\usepackage{relsize,exscale}
\usepackage{tensor}

\usepackage[low-sup]{subdepth}

\newcommand{\scr}{\scriptscriptstyle}

\newcommand{\longsim}{\scalebox{1.8}[1]{$\sim$}}

\makeatletter
\newcommand{\dalembertian}{\mathop{\mathpalette\dalembertian@\relax}}
\newcommand{\dalembertian@}[2]{%
  \begingroup
  \sbox\z@{$\m@th#1\square$}%
  \dimen0=\fontdimen8
    \ifx#1\displaystyle\textfont\else
    \ifx#1\textstyle\textfont\else
    \ifx#1\scriptstyle\scriptfont\else
    \scriptscriptfont\fi\fi\fi3
  \makebox[\wd\z@]{%
    \hbox to \ht\z@{%
      \vrule width \dimen0
      \kern-\dimen0
      \vbox to \ht\z@{
        \hrule height \dimen0 width \ht\z@
        \vss
        \hrule height 2\dimen0
      }%
      \kern-2.5\dimen0
      \vrule width 2.5\dimen0
    }%
  }%
  \endgroup
}
\makeatother

\begin{document}

\begin{center}
{\bf \Large Two simple photon gauges in inflation}

\

\renewcommand{\thefootnote}{\fnsymbol{footnote}}

{Dra\v{z}en Glavan}\,\footnote{email: \href{mailto:glavan@fzu.cz}{\tt glavan@fzu.cz}}

\setcounter{footnote}{0} 

\medskip

{\it CEICO, FZU --- Institute of Physics of the Czech Academy of Sciences}
\\
{\it Na Slovance 1999/2, 182 21 Prague 8, Czech Republic}

\

\parbox{0.86\linewidth}{
Photon propagators for power-law inflation are constructed 
in two one-parameter families of noncovariant gauges, in an
arbitrary number of spacetime dimensions.
In both gauges photon propagators take relatively simple forms
expressed in terms of scalar propagators and their derivatives. 
These are considerably simpler compared to their general covariant gauge counterpart. 
This makes feasible performing dimensionally regulated loop computations 
involving massless vector fields in inflation.}

\end{center}

\

\section{Introduction}
\label{sec: Introduction}

Massless vector fields (photons) do not sense directly the cosmological expansion
due to their conformal coupling to gravity. However, they couple to other fields,
namely scalars and gravitons, that are not conformally coupled to gravity and experience
gravitational particle production already at tree 
level~\cite{Parker:1968mv,Parker:1969au,Parker:1971pt}. That coupling
breaks conformality of photons and communicate to them the effects of the 
expansion. This can be particularly important in primordial inflation where
gravitons~\cite{Starobinsky:1979ty} and scalars~\cite{Mukhanov:1981xt} 
are copiously produced, and the effects mediated to photons might be 
sizable (see~\cite{Turner:1987bw,Calzetta:1997ku,Kandus:1999st,Giovannini:2000dj,Dimopoulos:2001wx,Davis:2000zp,Maroto:2000zu,Dolgov:2001ce,Calzetta:2001cf,Prokopec:2002jn,Prokopec:2002uw,Prokopec:2003bx,Prokopec:2003iu,Kahya:2005kj,Kahya:2006ui,Prokopec:2006ue,Prokopec:2007ak,Prokopec:2008gw,Leonard:2012si,Leonard:2012ex,Chen:2016nrs,Chen:2016uwp,Chen:2016hrz,Kaya:2018qbj,Popov:2017xut,Glavan:2019uni} for interactions with inflationary scalars,
and~\cite{Leonard:2013xsa,Glavan:2013jca,Wang:2014tza,Glavan:2015ura,Wang:2015eaa,Glavan:2016bvp,Miao:2018bol,Glavan:2023tet} for interactions with 
inflationary gravitons).
Quantifying these effects is done by computing quantum loop corrections
(and possibly resumming them), which requires
photon propagators (two-point functions) as the main bulding blocks of perturbations
theory. Constructing propagators in simple gauges in power-law inflation is the 
subject matter of this work.

The de Sitter space is often a good approximation for inflationary spacetime. 
Photon propagators in de Sitter in a general covariant gauge have a long 
history, starting from the seminal paper~\cite{Allen:1985wd} where the Feynman 
gauge photon propagator was reported, followed by 
works~\cite{Tsamis:2006gj,Garidi:2006ey,Youssef:2010dw,Frob:2013qsa}
that extended the construction to an arbitrary gauge-fixing parameter. However, 
it was realized only relatively recently that none of the reported dde Sitter invariant 
propagators satisfy 
the Ward-Takahashi identity~\cite{Glavan:2022pmk}, except for the transverse propagator 
in the exact gauge~\cite{Tsamis:2006gj}. In fact, the Ward-Takahashi identity forces 
photon propagator to break the full de Sitter invariance in all average generally covariant gauges, 
and such propagator was constructed in~\cite{Glavan:2022dwb,Glavan:2022nrd}.

Despite breaking de Sitter invariance, the covariant gauge photon propagator still
maintains a very tractable form. However, its extension
to somewhat more realistic power-law inflation~\cite{Lucchin:1984yf,La:1989za}~\footnote{Even
though power-law inflation is excluded by observations~\cite{Planck:2018jri}, it is
still a better approximation for slow roll inflation that the de Sitter space, on account of 
its non-vanishing principal slow-roll parameter.}
has proven to be much more difficult, and has yielded
a rather complicated propagator~\cite{Domazet:2024dil}. While this propagator in principle
permits one to perform dimensionally regulated loop computations involving massless 
vectors in power-law inflation, it is much more desirable to construct simpler propagators 
in different noncovariant gauges, akin to the simple gauge propagator in de Sitter 
space~\cite{Woodard:2004ut}. This is the main motivation behind this work.

There is no fundamental reason why gauge fixing should respect general
covariance. Choice of gauge is a matter of convenience, and bears no influence on 
the gauge-independent observables.
Identifying simple gauge choices is of importance for practical 
matters of performing loop computations, as it can mean a considerable difference 
in the amount of time and effort required to obtain the final result
(e.g.~compare the length of the computation~\cite{Leonard:2013xsa} 
and~\cite{Glavan:2015ura}).
This work is devoted to 
constructing the photon propagator in non-covariant linear gauges characterized
by a two-parameter gauge-fixing term,
\begin{equation}
S_{\rm gf}[A_\mu] = \int\! d^{D\!}x \, \sqrt{-g} \, 
	\biggl[
	- \frac{1}{2\xi} \Bigl( \nabla^\mu A_\mu - 2 \zeta n^\mu A_\mu \Bigr)^{\!2} \,
	\biggr]
	\, ,
\label{gauges}
\end{equation}
where~$n_\mu \!=\! \delta_\mu^0 \mathcal{H}$, with~$\xi$ and~$\zeta$ the two 
gauge-fixing parameters. The quantization in such gauges has been worked out 
in~\cite{Glavan:2022pmk} for arbitrary~$\xi$ and~$\zeta$. However, 
constructing the propagator for arbitrary values of~$\zeta$ is of dubious
utility, as is already evident from the complexity of the propagator in the general covariant 
gauge~\cite{Domazet:2024dil}, for which~$\zeta\!=\!0$. This is why
two particular choices for~$\zeta$ that lead to simple propagators are identified here.
The first choice is~$\zeta\!=\!1$, dubbed the {\it conformal gauge} because in four
spacetime dimensions it maintains conformal coupling of the photon to the expanding
background~\cite{Huguet:2013dia}.\footnote{This gauge should not be confused with the
higher derivative Eastwood-Singer gauge~\cite{Eastwood:1985eh,Esposito:2008dw,Faci:2009un,Faci:2011rq,Huguet:2013tv}.}~The 
second choice is~$\zeta\!=\!\epsilon$, dubbed the {\it deceleration gauge}
since the value of the gauge-fixing parameter equals the principal
slow-roll parameter~$\epsilon$ that is related to the deceleration 
parameter~$q\!=\!\epsilon \!-\! 1$ often used in cosmology.

The construction of the propagators is based on the canonical quantization formalism
from~\cite{Glavan:2022pmk},\footnote{This approach can be seen as a reformulation
of the Gupta-Bleuler quantization procedure~\cite{Gupta:1949rh,Bleuler:1950cy} in 
the canonical language, adapted for covariant and noncovariant gauges alike, and 
divorced from symmetry requirements. For mathematically oriented Gupta-Bleuler
quantization in covariant gauges on globally hyperbolic spaces see 
e.g.~\cite{Finster:2013fva,Wrochna:2014gia}.}
that is summarized in Sec.~\ref{sec: Electromagnetism in FLRW}.
This way the problem is reduced to solving for photon mode functions, which is
done in Sec.~\ref{sec: Dynamics in power-law inflation} relying on results for scalar mode 
functions collected prior in the first part of
Sec.~\ref{sec: Scalar mode functions and two-point functions}.
Integrals over mode functions representing photon two-point functions
are then evaluated in Sec.~\ref{sec: Two-point functions}, 
relying on results for scalar two-point functions collected in the second part of 
Sec.~\ref{sec: Scalar mode functions and two-point functions}.
In the de Sitter limit~$\epsilon\!\to\!0$ the conformal gauge propagator
reduces to a one-parameter generalization of the simple photon gauge
propagator~\cite{Woodard:2004ut}, while the deceleration gauge propagator
reduces to the general covariant gauge 
propagator~\cite{Glavan:2022dwb,Glavan:2022nrd}, 
correctly reproducing the de Sitter breaking term. This limit is given in 
Sec.~\ref{sec: Various limits}, alongside other special limits that serve to establish 
connection with the literature. These include the flat space limit in which both gauges
reduces to the Lorentz invariant~$R_\xi$ gauge where the~$\zeta$ dependence
drops out.
Further checks of propagators are performed by computing 
the field strength tensor correlator and the energy-momentum tensor in
Sec.~\ref{sec: Simple observables}. The paper concludes with the discussion in 
Sec.~\ref{sec: Discussion} summarizing the results and outlining the possible 
applications of the worked out propagators. Further details and checks are relegated to
two appendices.

\section{Massless vector field in FLRW}
\label{sec: Electromagnetism in FLRW}

The photon propagator in power-law inflation is computed here as a mode sum.
In order to derive this mode sum it is necessary to have the canonical quantization rules,
and to solve for the field operators. 
Since canonical quantization in multiplier gauges such as~(\ref{gauges}) 
requires indefinite metric space of states, it is generally not directly possible to identify 
creation/annihilation operators as one would do for e.g.~scalar fields. This is due to the
presence of first-class constraints that need to be quantized as conditions on the space
of states. This section is devoted to recapping this canonical quantization procedure 
that laid out in detail in~\cite{Glavan:2022pmk}. It relies on the canonical formulation
of the classical theory that is given here after first introducing the basic definitions
of cosmological spaces and power-law inflation in particular.

\subsection{FLRW and power-law inflation}
\label{subsec: Power-law inflation}

Spatially flat cosmological backgrounds are characterized by the
Friedmann-Lema\^{i}tre-Robertson-Walker (FLRW) line element,
\begin{equation}
ds^2 = - dt^2 + a^2(t) d\vec{x}^{\,2} 
	= a^2(\eta) \bigl( - d\eta^2 + d\vec{x}^{\,2} \bigr) \, ,
\end{equation}
where~$t$ is the physical time, and~$\eta$ the conformal time, the two being related 
by~$dt \!=\! a d\eta$, and where equal time slices are~$(D\!-\!1)$-dimensional Euclidean 
spaces. The scale factor~$a$ encodes the evolution of the expansion.
The expansion rate is captured either by the physical Hubble rate, 
or by the conformal Hubble rate, defined respectively by
\begin{equation}
H = \frac{1}{a} \frac{da}{dt} \, ,
\qquad \quad
\mathcal{H} = \frac{1}{a} \frac{da}{d\eta} \, .
\end{equation}
The two Hubble rates are related as~$\mathcal{H}\!=\!Ha$.
Acceleration is captured by the principal slow-roll parameter,
\begin{equation}
\epsilon = - \frac{1}{H^2} \frac{dH}{dt}
	=
	1 - \frac{1}{\mathcal{H}^2} \frac{d\mathcal{H}}{d\eta}
	\, ,
\end{equation}
where~$\epsilon \!<\! 1$ corresponds to accelerated expansion, and~$\epsilon\!>\!1$
to decelerated expansion. Sometimes the deceleration 
parameter~$q\!=\!1 \!-\! \epsilon$ is used instead of the slow-roll parameter.
Power-law inflation is charaterized by a constant principal slow-roll 
parameter,~$0 \!<\! \epsilon \!=\! {\tt const.} \!<\! 1$. In this case the scale factor and the Hubble rate
take simple closed form expressions,
\begin{equation}
\mathcal{H} = \frac{H_0}{1 - (1\!-\!\epsilon) H_0 (\eta \!-\! \eta_0)} \, ,
\qquad \quad
a = \Bigl( \frac{\mathcal{H}}{H_0} \Bigr)^{\! \frac{1}{1-\epsilon}}
\, .
\end{equation}
where~$\eta_0$ is some initial time for which~$a(\eta_0)\!=\!1$ 
and~$\mathcal{H}(\eta_0) \!=\! H_0 \!=\! H(\eta_0)$. The de Sitter space limit is
obtained by~$\epsilon\!\to\!0$. 

When considering the two-point functions in later sections it will be convenient to express
them in terms of bi-local scalar variables. The first of these is the distance function,
\begin{equation}
y(x;x') = (1\!-\!\epsilon)^2 \mathcal{H} \mathcal{H}'
	\Bigl[ \| \vec{x} \!-\! \vec{x}^{\,\prime} \|^2 - (\eta\!-\!\eta')^2 \Bigr]
	\, ,
\label{ydef}
\end{equation}
that is related to the geodesic distance. In the de Sitter limit this quantity is invariant
under all de Sitter symmetries. Cosmological backgrounds, including power-law inflation, 
admits only the cosmological symmetries, and two more variables bi-local variables are necessary,
\begin{equation}
u(x;x') = (1\!-\!\epsilon) \ln(aa') \, ,
\qquad \quad
v(x;x') = (1\!-\!\epsilon) \ln(a/a') \, .
\label{uvDef}
\end{equation}
Here and henceforth primes on quantities denote them to be 
related to the primed coordinate in bi-local objects (e.g.~$a'\!=\!a(\eta')$),
while unprimed quantities are related to the unprimed coordinate.

\subsection{Classical gauge-invariant canonical formulation}
\label{subsec: Gauge-invariant canonical formulation}

The first step towards canonical quantization of the photon field~$A_\mu$, 
defined by the covariant Maxwell action,
\begin{equation}
S[A_\mu] = \int\! d^{D\!}x\, \sqrt{-g} \,
	\biggl[ - \frac{1}{4} g^{\mu\rho} g^{\nu\sigma} F_{\mu\nu} F_{\rho\sigma} \biggr]
	\, ,
\label{CovariantAction}
\end{equation}
where~$F_{\mu\nu} \!=\! \partial_\mu A_\nu \!-\! \partial_\nu A_\mu$ is the vector field strength tensor,
is to derive the classical canonical formulation. For that it is useful to first specialize the
action above to FLRW backgrounds and to decompose the indices into temporal and spatial ones,
\begin{equation}
S[A_\mu] = \int\! d^{D\!}x \, a^{D-4} \, \biggl[ 
	\frac{1}{2} F_{0i} F_{0i} - \frac{1}{4} F_{ij} F_{ij} \biggr] \, .
\end{equation}
Henceforth all decomposed indices are written as subscripts, and repeated spatial indices
are assumed to be summed over.
The intermediate step in deriving the canonical action is constructing the 
extended action~\cite{Gitman},\footnote{This step is equivalent to the standard 
Legendre transform, but is much better adapted to off-shell applications, and to theories 
with constraints.}
obtained by promoting time derivatives to independent velocity fields,
\begin{equation}
\partial_0 A_0 \longrightarrow V_0 \, ,
\qquad \quad
F_{0i} \longrightarrow V_i \, ,
\end{equation}
and introducing accompanying Lagrange multiplier fields,~$\pi_0$ and~$\pi_i$,
to ensure on-shell equivalence,
\begin{align}
&
\mathcal{S}\bigl[ A_0, V_0, \Pi_0, A_i, V_i, \Pi_i \bigr]
	= \int\! d^{D\!}x \, \biggl\{
	a^{D-4} \biggl[ \frac{1}{2} V_i V_i - \frac{1}{4} F_{ij} F_{ij} \biggr]
\label{ExtendedAction}
\\
&	\hspace{5cm}
	+ \Pi_0 \Bigl(\partial_0 A_0 - V_0 \Bigr)
	+ \Pi_i \Bigl( \partial_0 A_i - \partial_i A_0 - V_i \Bigr)
	\biggr\} \, .
\nonumber 
\end{align}
Then solving on-shell for as many velocity fields as possible, which in this case is only~$V_i$,
\begin{equation}
\frac{\delta \mathcal{S}}{ \delta V_i }
	\approx 0
\qquad \Longrightarrow \qquad
V_i \approx \overline{V}_i = a^{4-D} \Pi_i \, ,
\end{equation}
produces the desired canonical action upon plugging the solutions back into
the extended action,
\begin{equation}
\mathscr{S} \bigl[ A_0, \Pi_0, A_i, \Pi_i, \ell \bigr]
	\equiv \mathcal{S} \bigl[ A_0, V_0 \to \ell, \Pi_0, A_i, \overline{V}_i, \Pi_i \bigr]
	=\!\!
	\int\! d^{D\!}x \, \Bigl[ \Pi_0 \partial_0 A_0 + \Pi_i \partial_0 A_i 
	-
	\mathscr{H} - \ell \Psi_1 \Bigr] \, ,
\label{CanonicalAction}
\end{equation}
where the canonical Hamiltonian is
\begin{equation}
\mathscr{H} = \frac{1}{2} a^{4-D} \Pi_i \Pi_i
	+ \Pi_i \partial_i A_0
	+ \frac{1}{4} a^{D-4} F_{ij} F_{ij} \, .
\end{equation}
Note that the field~$V_0$ was relabeled to~$\ell$ in order to emphasize its
interpretation as the Lagrange multiplier 
generating the primary constraint~$\Psi_1 \!=\! \Pi_0$. Also note that
Dirac's notation for on-shell (weak) equalities~``$\approx$'' is used,
to distinguish them from the off-shell (strong) equalities denoted by~``$=$'',
that apply at the level of the action.

Canonical equations of motion follow from varying the canonical action~(\ref{CanonicalAction}) 
with respect to dynamical fields,
\begin{equation}
\partial_0 A_0 \approx \ell \, ,
\qquad
\partial_0 \Pi_0 \approx \partial_i \Pi_i \, ,
\qquad
\partial_0 A_i \approx a^{4-D} \Pi_0 + \partial_i A_0 \, ,
\qquad
\partial_0 \Pi_i \approx a^{D-4} \partial_j F_{ji} \, ,
\end{equation}
They can be put in the form of Hamilton's equations using
the Poisson brackets that follow from the symplectic part of the canonical 
action,
\begin{equation}
\bigl\{ A_0(\eta,\vec{x}) , \Pi_0(\eta,\vec{x}^{\,\prime}) \bigr\}
	= \delta^{D-1}(\vec{x} \!-\! \vec{x}^{\,\prime}) \, ,
\qquad
\bigl\{ A_i(\eta,\vec{x}) , \Pi_j(\eta,\vec{x}^{\,\prime}) \bigr\}
	= 
	\delta_{ij} \delta^{D-1}(\vec{x} \!-\! \vec{x}^{\,\prime}) \, ,
\label{PoissonBrackets}
\end{equation}
with the remaining brackets vanishing. Varying the canonical action with respect to the 
Lagrange multiplier~$\ell$ yields the primary constraint,
\begin{equation}
\Psi_1 = \Pi_0 \approx 0 \, ,
\label{PrimaryConstraint}
\end{equation}
while its conservation generates a secondary constraint,
\begin{equation}
\partial_0 \Psi_1 \approx \partial_i \Pi_i 
	\equiv \Psi_2 \approx 0 \, .
\label{SecondaryConstraint}
\end{equation}
The two constraints form a complete set of first-class constraints,
meaning that their Poisson brackets vanish on-shell,
\begin{equation}
\bigl\{ \Psi_I(\eta,\vec{x}) , \Psi_J(\eta,\vec{x}^{\,\prime}) \bigr\}
	\approx 0 \, ,
\qquad
I,J=1,2 \, ,
\end{equation}
and hence no further constraints are generated by the conservation of the 
secondary constraint.

\subsection{Classical gauge-fixed canonical formulation}
\label{subsec: Gauge-fixed canonical formulation}

Imposing a multiplier gauge~\cite{Henneaux:1992ig} (also called average gauge)
amounts to fixing~$\ell$ off-shell in the canonical action~(\ref{CanonicalAction}) 
to be a function of canonical variables.
This procedure is shortcutted by adding the gauge-fixing term~$S_{\rm gf}[A_\mu]$ 
to the gauge-invariant action~(\ref{CovariantAction}), thus defining the gauge-fixed action,
\begin{equation}
S_\star[A_\mu] = S[A_\mu] + S_{\rm gf}[A_\mu] \, .
\label{GaugeFixedAction}
\end{equation}
The gauge-fixing term in~(\ref{gauges}) considered here preserves cosmological 
symmetries, and evaluated on the FLRW background reads
\begin{equation}
S_{\rm gf}[A_\mu] = \int\! d^{D\!}x \, a^{D-4} \, \biggl[
	- \frac{1}{2\xi} \Bigl( \partial_0A_0 + (D\!-\!2\!-\! 2\zeta) \mathcal{H} A_0 - \partial_i A_i \Bigr)^{\!2} \,
	\biggr] \, .
\end{equation}
The canonical formulation of this action is derived 
following the same procedure as in Sec.~\ref{subsec: Gauge-invariant canonical formulation}.
The difference is that the gauge-fixed action is not longer gauge-invariant, so that
no constraints are generated. The resulting canonical gauge-fixed action takes the form
\begin{align}
\mathscr{S}_\star [ A_0, \Pi_0, A_i , \Pi_i ]
	= \int\! d^{D\!}x \, \Bigl[
	\Pi_0 \partial_0 A_0 + \Pi_i \partial_0 A_i
	- \mathscr{H}_\star
	\Bigr] \, ,
\label{CanonicalGaugeFixedAction}
\end{align}
where the gauge-fixed Hamiltonian is~\cite{Glavan:2022pmk}
\begin{equation}
\mathscr{H}_\star =
	\frac{a^{4-D}}{2} \Bigl( \Pi_i \Pi_i - \xi \Pi_0 \Pi_0 \Bigr)
	+ \Pi_i \partial_i A_0
	+ \Pi_0 \partial_i A_i
	- (D\!-\!2\!-\!2\zeta) \mathcal{H} \Pi_0 A_0
	+ \frac{a^{D-4}}{4} F_{ij} F_{ij} \, .
\end{equation}
It can now be seen that this corresponds to the following choice for the Lagrange multiplier in~(\ref{CanonicalAction}),
\begin{equation}
\ell \longrightarrow - \frac{\xi a^{4-D}}{2} \Pi_0 + \partial_i A_i
	- (D\!-\!2\!-\!2\zeta) \mathcal{H} A_0 \, .
\end{equation}
The gauge-fixed equations of motion now follow from varying the gauge-fixed canonical action
with respect to the dynamical fields
\begin{subequations}
\begin{align}
\partial_0 A_0 \approx{}&
	- \xi a^{4-D} \Pi_0 + \partial_i A_i
	- (D\!-\!2 \!-\! 2\zeta) \mathcal{H} A_0
	\, ,
\\
\partial_0 \Pi_0 \approx{}&
	\partial_i \Pi_i
	+ (D\!-\!2\!-\!2\zeta) \mathcal{H} \Pi_0
	\, ,
\\
\partial_0 A_i \approx{}&
	a^{4-D} \Pi_i
	+ \partial_i A_0
	\, ,
\\
\partial_0 \Pi_i \approx{}&
	\partial_i \Pi_0 + a^{D-4} \partial_j F_{ji}
	\, ,
\end{align}
\label{GaugeFixedEOM}%
\end{subequations}
and they can be written as Hamiton's equations utilizing
the same Poisson brackets~(\ref{PoissonBrackets}).

Note that the gauge-fixed action encodes for the gauge-fixed dynamics,
but does not encode the first-class constraints in~(\ref{PrimaryConstraint}) and~(\ref{SecondaryConstraint}).
These have to be demanded separately, in addition to the gauge-fixed action, 
at the level of initial conditions given at some initial value surface,
\begin{equation}
\Psi_1(\eta_0,\vec{x}) = \Pi_0 (\eta_0, \vec{x}) \approx 0 \, ,
\qquad
\Psi_2(\eta_0,\vec{x}) = \partial_i \Pi_i(\eta_0,\vec{x}) \approx 0 \, .
\label{InitialConstraints}
\end{equation}
Then the gauge-fixed dynamics ensures that these are conserved in time.

\subsection{Quantizing photon in FLRW}
\label{subsec: Quantizing electromagnetism in FLRW}

Quantizing the photon field in multiplier gauges requires two ingredients:
quantizing the dynamics, and quantizing the constraints. The former is straightforward 
and follows the rules of canonical quantization for theories without constraints. This 
entails promoting the dynamical fields to field operators,
\begin{equation}
A_\mu(x) \longrightarrow \hat{A}_\mu(x) \, ,
\qquad \quad
\Pi_\mu(x) \longrightarrow \hat{\Pi}_\mu(x) \, ,
\end{equation}
and their Poisson brackets~(\ref{PoissonBrackets}) to commutators,
\begin{equation}
\bigl[ \hat{A}_0(\eta,\vec{x}) , \hat{\Pi}_0(\eta,\vec{x}^{\,\prime}) \bigr]
	=
	i \delta^{D-1}(\vec{x} \!-\! \vec{x}^{\,\prime}) \, ,
\qquad
\bigl[ \hat{A}_i(\eta,\vec{x}) , \hat{\Pi}_j(\eta,\vec{x}^{\,\prime}) \bigr]
	=
	\delta_{ij} \, i \delta^{D-1}(\vec{x} \!-\! \vec{x}^{\,\prime}) \, .
\end{equation}
Equations of motion for these field operators are the same as the classical
gauge-fixed equations~(\ref{GaugeFixedEOM}).
Constraints~(\ref{InitialConstraints}) are quantized as conditions on the 
indefinite metric space of states.
Before implementing these, it is advantageous to first 
decompose the field operators into different sectors, and to introduce their
momentum space formulation.


\bigskip
\noindent {\bf Helmholtz decomposition and momentum space.}
It is convenient to break up the spatial components of field operators into 
transverse and longitudinal parts,
\begin{equation}
\hat{A}_i = \hat{A}_i^{\scr T} + \hat{A}_i^{\scr L} \, ,
\qquad \quad
\hat{\Pi}_i = \hat{\Pi}_i^{\scr T} + \hat{\Pi}_i^{\scr L} \, ,
\end{equation}
such  that the individual parts,
\begin{equation}
\hat{A}_i^{\scr T} = \mathbb{P}_{ij}^{\scr T} \hat{A}_j \, ,
\qquad
\hat{A}_i^{\scr L} = \mathbb{P}_{ij}^{\scr L} \hat{A}_j \, ,
\qquad
\hat{\Pi}_i^{\scr T} = \mathbb{P}_{ij}^{\scr T} \hat{\Pi}_j \, ,
\qquad
\hat{\Pi}_i^{\scr L} = \mathbb{P}_{ij}^{\scr L} \hat{\Pi}_j \, ,
\end{equation}
are defined in terms of transverse ad longitudinal projection operators,
\begin{equation}
\mathbb{P}_{ij}^{\scr T} = \delta_{ij} - \frac{\partial_i \partial_j}{\nabla^2} \, ,
\qquad \quad
\mathbb{P}_{ij}^{\scr L} = \frac{\partial_i \partial_j}{\nabla^2} \, ,
\end{equation}
These projectors are idempotent,~$\mathbb{P}_{ij}^{\scr T}\mathbb{P}_{jk}^{\scr T} \!=\! \mathbb{P}_{ik}^{\scr T}$,
$\mathbb{P}_{ij}^{\scr L}\mathbb{P}_{jk}^{\scr L} \!=\! \mathbb{P}_{ik}^{\scr L}$,
and mutually 
orthogonal,~$\mathbb{P}_{ij}^{\scr T}\mathbb{P}_{jk}^{\scr L} \!=\!\mathbb{P}_{ij}^{\scr L}\mathbb{P}_{jk}^{\scr T} \!=\! 0$.
The utility of the Helmholtz decomposition is in decoupling the transverse sector from the
scalar sector (containing longitudinal and temporal components).

The transverse sector field operators have the following momentum space expansion
\begin{subequations}
\begin{align}
&
\hat{A}^{\scr T}_i (\eta,\vec{x})
	= a^{- \frac{D-4}{2} } \! \int\! \frac{d^{D-1} k }{ (2\pi)^{\frac{D-1}{2}} } \,
	e^{i\vec{k} \cdot \vec{x}}
	\sum_{\sigma=1}^{D-2}
	\varepsilon_i(\sigma,\vec{k}) \,
	\hat{\mathcal{A}}_{{\scr T}, \sigma}(\eta,\vec{k}) \, ,
\label{ATfourier}
\\
&
\hat{\Pi}^{\scr T}_i (\eta,\vec{x})
	= a^{\frac{D-4}{2} } \! \int\! \frac{d^{D-1} k }{ (2\pi)^{\frac{D-1}{2}} } \,
	e^{i\vec{k} \cdot \vec{x}}
	\sum_{\sigma=1}^{D-2}
	\varepsilon_i(\sigma,\vec{k}) \,
	\hat{\pi}_{{\scr T}, \sigma}(\eta,\vec{k}) \, ,
\end{align}
\label{TransverseFourier}%
\end{subequations}
where the transverse polarization tensor satisfies
\begin{subequations}
\begin{align}
&
k_i \, \varepsilon_i(\sigma, \vec{k}) = 0 \, ,
&&
\varepsilon_i^*(\sigma,\vec{k}) = \varepsilon_i(\sigma,-\vec{k}) \, ,
\\
&
\varepsilon_i^*(\sigma,\vec{k}) \varepsilon_i (\sigma',\vec{k}) = \delta_{\sigma \sigma'} \, ,
&&
\sum_{\sigma=1}^{D-2} \varepsilon_i^*(\sigma,\vec{k}) \varepsilon_j(\sigma,\vec{k})
	= \delta_{ij} - \frac{ k_i k_j }{ k^2 } \, .
\end{align}
\end{subequations}
Note that powers of the scale factor have been extracted for later convenience
from the Fourier transform in~(\ref{TransverseFourier}).
In momentum space Hermiticity takes a different form,
\begin{equation}
\hat{\mathcal{A}}_{{\scr T}, \sigma}^\dag(\vec{k}) 
	= \hat{\mathcal{A}}_{{\scr T}, \sigma}(-\vec{k}) \, ,
\qquad \qquad
\hat{\pi}_{{\scr T}, \sigma}^\dag(\vec{k}) 
	= \hat{\pi}_{{\scr T}, \sigma}(-\vec{k}) \, ,
\end{equation}
while the non-vanishing canonical commutation relations read
\begin{equation}
\bigl[ \hat{\mathcal{A}}_{{\scr T}, \sigma}(\eta,\vec{k}) , 
	\hat{\pi}_{{\scr T}, \sigma'}(\eta,\vec{k}^{\,\prime}) \bigr]
		= \delta_{\sigma \sigma'} \, i \delta^{D-1}(\vec{k} \!+\! \vec{k}^{\,\prime}) \, .
\end{equation}
The transverse sector equations of motion, decoupled from the scalar sector, are
\begin{equation}
\partial_0 \hat{\mathcal{A}}_{{\scr T}, \sigma}
	= \hat{\pi}_{ {\scr T}, \sigma}
	+ \frac{1}{2} (D\!-\!4) \mathcal{H} \hat{\mathcal{A}}_{{\scr T}, \sigma} \, ,
\qquad
\partial_0 \hat{\pi}_{{\scr T}, \sigma}
	= 
	- k^2 \hat{\mathcal{A}}_{{\scr T}, \sigma}
	- \frac{1}{2} (D\!-\!4) \mathcal{H} \hat{\pi}_{{\scr T}, \sigma}
	\, .
\label{TransverseMomentumEOM}
\end{equation}

The momentum sapce expansions of the scalar sector field operator read
\begin{subequations}
\begin{align}
&
\hat{A}_0 (\eta,\vec{x})
	= a^{- \frac{D-2-2\zeta}{2} } \! \int\! \frac{d^{D-1} k }{ (2\pi)^{\frac{D-1}{2}} } \,
	e^{i\vec{k} \cdot \vec{x}} \,
	\hat{\mathcal{A}}_{0}(\eta,\vec{k}) \, ,
 &
\hat{\mathcal{A}}_{0}^\dag(\vec{k}) 
	= \hat{\mathcal{A}}_{0}(-\vec{k}) \, ,
\label{A0fourier}
\\
&
\hat{\Pi}_0 (\eta,\vec{x})
	= a^{\frac{D-2-2\zeta}{2} } \! \int\! \frac{d^{D-1} k }{ (2\pi)^{\frac{D-1}{2}} } \,
	e^{i\vec{k} \cdot \vec{x}} \,
	\hat{\pi}_{0}(\eta,\vec{k}) \, ,
&
\hat{\pi}_{0}^\dag(\vec{k}) 
	= \hat{\pi}_{0}(-\vec{k}) \, ,
\\
&
\hat{A}^{\scr L}_i (\eta,\vec{x})
	= a^{- \frac{D-2-2\zeta}{2} } \! \int\! \frac{d^{D-1} k }{ (2\pi)^{\frac{D-1}{2}} } \,
	e^{i\vec{k} \cdot \vec{x}}
	\Bigl( - \frac{i k_i}{k} \Bigr)
	\hat{\mathcal{A}}_{\scr L}(\eta,\vec{k}) \, ,
\label{ALfourier}
&
\hat{\mathcal{A}}_{\scr L}^\dag(\vec{k}) 
	= \hat{\mathcal{A}}_{\scr L}(-\vec{k}) \, ,
\\
&
\hat{\Pi}^{\scr L}_i (\eta,\vec{x})
	= a^{\frac{D-2-2\zeta}{2} } \! \int\! \frac{d^{D-1} k }{ (2\pi)^{\frac{D-1}{2}} } \,
	e^{i\vec{k} \cdot \vec{x}}
	\Bigl( - \frac{i k_i}{k} \Bigr)
	\hat{\pi}_{\scr L}(\eta,\vec{k}) \, ,
&
\hat{\pi}_{\scr L}^\dag(\vec{k}) 
	= \hat{\pi}_{\scr L}(-\vec{k}) \, .
\end{align}
\end{subequations}
where different powers of scale factors are extracted from the Fourier transform
in order to obtain simpler equations of motion.
The non-vanishing canonical commutation relations of the scalar sector are
\begin{equation}
\bigl[ \hat{\mathcal{A}}_{0}(\eta,\vec{k}) , 
	\hat{\pi}_{0}(\eta,\vec{k}^{\,\prime}) \bigr]
		= 
\bigl[ \hat{\mathcal{A}}_{\scr \rm L}(\eta,\vec{k}) , 
	\hat{\pi}_{\scr \rm L}(\eta,\vec{k}^{\,\prime}) \bigr]
		= i \delta^{D-1}(\vec{k} \!+\! \vec{k}^{\,\prime}) \, .
\label{scalar momentum commutators}
\end{equation}
while the momentum space equations of motion read
\begin{align}
&
\partial_0 \hat{\mathcal{A}}_0 
	=
	- \xi a^{ 2-2\zeta} \hat{\pi}_0
	+ k \hat{\mathcal{A}}_{\scr \rm L}
	- \frac{1}{2} (D\!-\!2\!-\!2\zeta) \mathcal{H} \hat{\mathcal{A}}_0 \, ,
\label{momentum scalar eom 1}
\\
&
\partial_0 \hat{\pi}_0 =
	k  \hat{\pi}_{\scr \rm L}
	+ \frac{1}{2} (D\!-\!2\!-\!2\zeta) \mathcal{H} \hat{\pi}_0 \, ,
\label{momentum scalar eom 2}
\\
&
\partial_0 \hat{\mathcal{A}}_{\scr \rm L} 
	=
	a^{2-2\zeta} \hat{\pi}_{\scr \rm L}
	- k \hat{\mathcal{A}}_0
	+ \frac{1}{2} (D\!-\!2\!-\!2\zeta) \mathcal{H} \hat{\mathcal{A}}_{\scr \rm L}  \, ,
\label{momentum scalar eom 3}
\\
&
\partial_0 \hat{\pi}_{\scr \rm L} =
	- k \hat{\pi}_0
	- \frac{1}{2} (D\!-\!2\!-\!2\zeta) \mathcal{H} \hat{\pi}_{\scr \rm L} \, .
\label{momentum scalar eom 4}
\end{align}

\bigskip
\noindent {\bf Quantized constraints.}
The primary constraints~(\ref{PrimaryConstraint}) and~(\ref{SecondaryConstraint}) are functions 
of the field operators, and are consequently quantized when those fields are promoted
to field operators. The Fourier space expansion of these two Hermitian constraint operators reads
\begin{subequations}
\begin{align}
\hat{\Psi}_1(\eta,\vec{x}) ={}&
	a^{ \frac{D-2-2\zeta }{2} } \!
	\int\! \frac{d^{D-1}k }{ (2\pi)^{\frac{D-1}{2}} } \,
	e^{ i \vec{k} \cdot \vec{k} }
	\hat{\psi}_1(\eta,\vec{k})
	\, ,
\\
\hat{\Psi}_2(\eta,\vec{x}) ={}&
	a^{ \frac{D-2-2\zeta }{2} } \!
	\int\! \frac{d^{D-1}k }{ (2\pi)^{\frac{D-1}{2}} } \,
	e^{ i \vec{k} \cdot \vec{k} } \, k \,
	\hat{\psi}_2(\eta,\vec{k}) \, ,
\end{align}
\end{subequations}
where,
\begin{equation}
\hat{\psi}_1 = \hat{\pi}_0 \, ,
\qquad \quad
\hat{\psi}_2 = \hat{\pi}_{\scr L}
\, .
\end{equation}
In the quantized theory the constraints are implemented 
as conditions on the space of states. These conditions define the subspace
of physical states, and have to be implemented in the form of a non-Hermitian
linear combination of Hermitian constraint operators~\cite{Glavan:2022pmk},
\begin{equation}
\hat{\mathcal{K}}(\vec{k}) =
	c_1(\eta,k) \hat{\psi}_1(\eta,\vec{k})
	+
	c_2(\eta,k) \hat{\psi}_2(\eta,\vec{k})
	\, ,
\label{Kdef}
\end{equation}
annihilating the physical state,
\begin{equation}
\hat{\mathcal{K}}(\vec{k}) \bigl| \Omega \bigr\rangle = 0 \, .
\qquad \quad
\forall \vec{k} \, .
\label{GBcondition}
\end{equation}
This condition guarantees that any matrix element of the product of Hermitian 
constraints vanishes. The time-independence of the non-Hermitian subsidiary constraint 
operator implies equations of motion for the coefficient functions in~(\ref{Kdef}),
\begin{equation}
\partial_0 c_1 = k c_2 - \frac{1}{2} ( D \!-\! 2 \!-\! 2\zeta ) \mathcal{H} c_1 \, ,
\qquad
\partial_0 c_2 = - k c_1+ \frac{1}{2} (D \!-\! 2 \!-\! 2\zeta) \mathcal{H} c_2 \, .
\label{constraintEOM}
\end{equation}
%

\section{Scalar mode functions and two-point functions}
\label{sec: Scalar mode functions and two-point functions}

The computation of photon mode functions and photon two-point functions 
in subsequent sections greatly benefits from recalling properties of scalar mode 
functions and scalar two-point functions in power-law inflation, that are collected here.

\subsection{Scalar mode functions}
\label{subsec: Scalar mode functions}

The rescaled mode function of a nonminimally coupled scalar field in power-law
inflation satisfies the following ubiquitous equation of motion,
\begin{equation}
\biggl[ \partial_0^2 + k^2 
	+ \Bigl( \frac{1}{4} \!-\! \lambda^2 \Bigr) (1\!-\!\epsilon)^2 \mathcal{H}^2 \biggr] 
	\mathscr{U}_\lambda(\eta,\vec{k}) = 0 \, ,
\label{scalar mode eq}
\end{equation}
where the constant~$\lambda$ parametrized the connection to the nonminnimal coupling.
The general solution to this equation,
\begin{equation}
\mathscr{U}_\lambda(\eta,\vec{k}) = \alpha(\vec{k}) \, U_\lambda(\eta,k)
	+ \beta(\vec{k}) \, U_\lambda^*(\eta,k) \, ,
\label{UgeneralSolution}
\end{equation}
is a linear combination of the fundamental positive-frequency mode function,
\begin{equation}
U_\lambda(\eta,k) = 
	e^{\frac{i\pi}{4} (2\lambda+1)} 
	e^{\frac{ - ik}{ (1-\epsilon) H_0 }}
	\sqrt{ \frac{\pi}{4(1\!-\!\epsilon) \mathcal{H} } } \,
	H^{\scr (1)}_\lambda\biggl( \frac{k}{ (1\!-\!\epsilon) \mathcal{H} } \biggr) \, ,
\label{CTBD}
\end{equation}
and its complex conjugate, where~$H^{\scr (1)}_\lambda$ is the Hankel function of the 
first kind. 
This is the power-law inflation generalization of the Chernikov-Tagirov-Bunch-Davies
mode function in de Sitter space~\cite{Chernikov:1968zm,Bunch:1978yq},
and its normalization is given by the Wronskian,
\begin{equation}
U_\lambda(\eta,k) \partial_0 U_\lambda^*(\eta,k)
	- U_\lambda^*(\eta,k) \partial_0 U_\lambda(\eta,k)
		= i \, .
\label{Wronskian}
\end{equation}
The flat space limit of~(\ref{CTBD}) is obtained by taking~$H_0\!\to\!0$, where first two
subleading orders will be needed in later sections,
\begin{align}
U_\lambda(\eta,k) 
	\ \overset{H_0 \to 0}{\longsim} \ {}&
	\frac{e^{-ik(\eta-\eta_0)}}{ \sqrt{2k} }
	\biggl\{ 1 
		+ \frac{1}{2} \Bigl( \frac{1}{4} \!-\! \lambda^2 \Bigr)
			\Bigl[ \frac{ (1\!-\!\epsilon) H_0 }{ i k} \Bigr]
\nonumber \\
&
	+ \frac{1}{8} \Bigl( \frac{1}{4} \!-\! \lambda^2 \Bigr) 
		\biggl[ \Bigl( \frac{9}{4} \!-\! \lambda^2 \Bigr)
			+ 4 i k(\eta\!-\!\eta_0) \biggr]
		\Bigl[ \frac{ (1\!-\!\epsilon) H_0}{ ik } \Bigr]^2 
	+ \dots
	 \biggr\}
	 \, .
\end{align}

There are two useful recurrence relations for raising and lowering the mode function index,
\begin{align}
\biggl[ \partial_0 + \Bigl( \frac{1}{2} \!+\! \lambda \Bigr) (1\!-\!\epsilon) \mathcal{H} \biggr] U_\lambda 
	= - i k U_{\lambda+1} \, ,
\qquad
\biggl[ \partial_0 + \Bigl( \frac{1}{2} \!-\! \lambda \Bigr) (1\!-\!\epsilon) \mathcal{H} \biggr] U_{\lambda }
	= -ik U_{\lambda-1} \, ,
\label{RecurrenceIdentities}
\end{align}
that follow from recurrence relations for Hankel functions~\cite{Olver:2010,Olver:web},
and allow the Wronskian~(\ref{Wronskian}) to be written in a compact form,
\begin{equation}
{\rm Re} \Bigl[ U_\lambda(\eta,k) U_{\lambda-1}^*(\eta,k)\Bigr]
		= \frac{1}{2k} \, .
\label{AlternateWronskian}
\end{equation}
Further identities that are used in later sections
follow from the mode equation~(\ref{scalar mode eq}) and the 
two recurrence relations~(\ref{RecurrenceIdentities}):
\begin{align}
&
\biggl[ \partial_0^2 + k^2 \! + \Bigl( \frac{1}{4} \!-\! \lambda^2 \Bigr) (1\!-\!\epsilon)^2 \mathcal{H}^2 \biggr] 
	\Bigl( \frac{U_{\lambda-1} }{ \mathcal{H} } \Bigr)
	=
	2(1\!-\!\epsilon) i k U_{\lambda}
	\, ,
\label{ModeIdentity1}
\\
&
\biggl[ \partial_0^2 + k^2 \! + \Bigl( \frac{1}{4} \!-\! \lambda^2 \Bigr) (1\!-\!\epsilon)^2 \mathcal{H}^2 \biggr] 
	\bigl( \mathcal{H} U_{\lambda-1} \bigr)
	\!=
	\!-2 (1\!-\!\epsilon) \mathcal{H}^2 \Bigl[ ik U_\lambda 
		\!+\! 2 (\lambda\!-\!1) (1\!-\!\epsilon) \mathcal{H} U_{\lambda-1}
		\Bigr]
	,
	\,
\label{ModeIdentity2}
\\
&
\biggl[ \partial_0^2 + k^2 \! + \Bigl( \frac{1}{4} \!-\! \lambda^2 \Bigr) (1\!-\!\epsilon)^2 \mathcal{H}^2 \biggr] 
	\frac{\partial U_\lambda}{\partial \lambda}
	= 
	2 \lambda (1\!-\!\epsilon)^2 \mathcal{H}^2 U_\lambda \, .
\label{ModeIdentity3}
\end{align}
The first two follow from simply acting the derivatives on the left-hand side,
and recognizing the equation of motion and the index-raising recurrence relation, 
while the third one follows from a parametric derivative of the equation of motion.

\subsection{Scalar propagators}
\label{subsec: Scalar propagators}

The scalar field two-point function satisfies the following equation of motion in power-law
inflation,
\begin{equation}
\biggl[ \, \dalembertian 
	- \biggl( \Bigl[ \frac{D \!-\! 1 \!-\! \epsilon}{2(1\!-\!\epsilon)} \Bigr]^2 
		\!-\! \lambda^2 \biggr)
	(1\!-\!\epsilon)^2 H^2 \biggr]
	i \bigl[ \tensor*[^{\tt a \!}]{\Delta}{^{\tt \! b }} \bigr](x;x')
	=
	{\tt S}^{\tt ab} \frac{ i\delta^D(x\!-\!x') }{ \sqrt{-g} }
	\, ,
\label{ScalarPropagatorEOM}
\end{equation}
where~${\tt a},{\tt b} \!=\! \pm$ are the so-called Schwinger-Keldysh polarity indices,
and where~${\tt S}^{\scr +-} \!=\! {\tt S}^{\scr +-} \!=\! 0$ 
and~${\tt S}^{\scr ++} \!=\! - {\tt S}^{\scr +-} \!=\! 1$. The polarity indices label 
different types of two-point functions necessary for nonequilibrium quantum field 
theory (e.g.~\cite{Berges:2004yj,NoneqLectures}).
The first of these is the positive-frequency Wightman function, for which the solution can 
be expressed as a sum over modes,
\begin{equation}
i \bigl[ \tensor*[^{\scr \!-\!}]{\Delta}{^{\scr \! +\! }} \bigr]_{\lambda}(x;x')
	= (aa')^{-\frac{D-2}{2}} \! \int\! \frac{d^{D-1}k }{ (2\pi)^{D-1} } \,
		e^{i \vec{k} \cdot (\vec{x} - \vec{x}{\,}' ) } \,
		U_\lambda(\eta,k) U_\lambda^*(\eta',k) \, ,
\label{ScalarSum}
\end{equation}
while its complex conjugate is the negative-frequency Wightman 
function,~$ i \bigl[ \tensor*[^{\scr \!+\!}]{\Delta}{^{\scr \!-\! }} \bigr]_{\lambda}(x;x')
	\!=\! \bigl\{ i \bigl[ \tensor*[^{\scr \!-\!}]{\Delta}{^{\scr \! +\! }} 
	\bigr]_{\lambda}(x;x') \bigr\}^*$.
Likewise, unprimed quantities are related to the unprimed coordinate.
Implicit in this definition are the~$i\delta$ regulators that are introduced by
complexifying the conformal time arguments of the mode functions,
\begin{equation}
\eta \longrightarrow \eta - \frac{i\delta}{2} \, ,
\qquad
\eta' \longrightarrow \eta' + \frac{i\delta}{2} \, ,
\end{equation}
where the substitution is done after the complex conjugation on the second mode function
is performed. These ensure the two-point function is defined as the distributional
limit~$\delta\!\to\!0$ of an analytic function.
The two other two-point functions are the Feynman propagator,
\begin{equation}
i \bigl[ \tensor*[^{\scr \!+\!}]{\Delta}{^{\scr \! +\! }} \bigr]_{\lambda}(x;x')
	=
	\theta(\eta\!-\!\eta')
	i \bigl[ \tensor*[^{\scr \!-\!}]{\Delta}{^{\scr \!+\! }} \bigr]_{\lambda}(x;x')
	+
	\theta(\eta'\!-\!\eta)
	i \bigl[ \tensor*[^{\scr \!+\!}]{\Delta}{^{\scr \!-\! }} \bigr]_{\lambda}(x;x')
	\, ,
\end{equation}
and its complex conjugate,~$ i \bigl[ \tensor*[^{\scr \!-\!}]{\Delta}{^{\scr \!-\! }} \bigr]_{\lambda}(x;x')
	\!=\! \bigl\{ i \bigl[ \tensor*[^{\scr \!+\!}]{\Delta}{^{\scr \! +\! }} 
	\bigr]_{\lambda}(x;x') \bigr\}^*$, sometimes referred to as the Dyson propagator.

For~$\lambda\!<\!(D\!-\!1)/2$ the mode sum in~(\ref{ScalarSum}) is infrared finite,
and it evaluates to~\cite{Janssen:2008px}
\begin{equation}
i \bigl[ \tensor*[^{ \tt a\! }]{\Delta}{^{\! \tt b}} \bigr]_{\lambda}(x;x')
	=
	i \Delta_\lambda(y_{\tt ab},u)
	=
	(aa')^{- \frac{(D-2)\epsilon}{2}} \mathcal{F}_\lambda(y_{\tt ab}) 
	\, ,
\label{DeltaEvaluation}
\end{equation}
where~$\mathcal{F}_\lambda$ is the rescaled propagator function expressed in terms
of the hypergeometric function,
\begin{equation}
\mathcal{F}_\lambda(y)
	=
	\frac{ [(1\!-\!\epsilon) H_0]^{D-2} }{ (4\pi)^{ \frac{D}{2} } }
	\frac{ \Gamma\bigl( \frac{D-1}{2} \!+\! \lambda \bigr) \, \Gamma\bigl( \frac{D-1}{2} \!-\! \lambda \bigr) }
		{ \Gamma\bigl( \frac{D}{2} \bigr) } \,
	{}_2F_1\biggl( \Bigl\{ \frac{D\!-\!1}{2} \!+\! \lambda , \frac{D\!-\!1}{2} \!-\! \lambda \Bigr\} ,
		\Bigl\{ \frac{D}{2} \Bigr\} , 1 \!-\! \frac{y}{2} \biggr) \, , \quad
\label{RescaledPropagatorFunction}
\end{equation}
and where~$y_{\tt ab}$ is the distance function~(\ref{ydef}) with the
appropriate~$i\delta$-prescription
\begin{subequations}
\begin{align}
&
y_{\scr -+} = (1\!-\!\epsilon)^2 \mathcal{H} \mathcal{H}'
	\Bigl[ \| \vec{x} \!-\! \vec{x}^{\,\prime} \|^2 \!-\! 
		\bigl( \eta \!-\! \eta' \!-\! i\delta \bigr)^2 \Bigr]
	\, ,
\qquad
y_{\scr +-} = y_{\scr -+}^* \, ,
\\
&
y_{\scr ++} = (1\!-\!\epsilon)^2 \mathcal{H} \mathcal{H}'
	\Bigl[ \| \vec{x} \!-\! \vec{x}^{\,\prime} \|^2 \!-\! 
		\bigl( |\eta \!-\! \eta'| \!-\! i\delta \bigr)^2 \Bigr]
	\, ,
\qquad
y_{\scr --} = y_{\scr ++}^* \, .
\end{align}
\label{yPrescriptions}
\end{subequations}
Henceforts explicit Schwinger-Keldysh polarity indices are suppressed on the
right-hand sides of expressions, and where relevant are either implied by the 
corresponding polarities on the left-hand side of expressions, or should be clear from the
context.

The rescaled propagator function~(\ref{RescaledPropagatorFunction})
satisfies the hypergeometric equation in disguise,
\begin{equation}
\biggl[ ( 4y\!-\!y^2 ) \frac{\partial^2}{\partial y^2}
	+ D (2\!-\!y) \frac{\partial}{\partial y}
	+ \lambda^2 - \Bigl( \frac{D\!-\!1}{2} \Bigr)^{\!2} \,
	 \biggr] \mathcal{F}_\lambda(y) = 0 \, ,
\label{Feom}
\end{equation}
and has a very useful power series representation around~$y\!=\!0$,
\begin{align}
\mathcal{F}_\lambda(y)
	={}&
	\frac{ \bigl[ (1\!-\!\epsilon) H_0 \bigr]^{D-2} \, 
		\Gamma\bigl( \frac{D-2}{2} \bigr) }{ (4\pi)^{ \frac{D}{2} } }
	\biggl\{
	\Bigl( \frac{y}{4} \Bigr)^{ \! -\frac{D-2}{2}} 
	+ \frac{\Gamma\bigl( \frac{4-D}{2} \bigr) }
		{ \Gamma\bigl( \frac{1}{2} \!+\! \lambda \bigr) \, \Gamma\bigl( \frac{1}{2} \!-\! \lambda \bigr) } \sum_{n=0}^{\infty} 
\label{Fseries}
\\
&
	\times \! \biggl[
	\frac{ \Gamma\bigl( \frac{3}{2} \!+\! \lambda \!+\! n \bigr) \, \Gamma\bigl( \frac{3}{2} \!-\! \lambda \!+\! n \bigr) }
		{ \Gamma\bigl( \frac{6-D}{2} \!+\! n \bigr) \, (n\!+\!1)! }
		\Bigl( \frac{y}{4} \Bigr)^{\! n- \frac{D-4}{2} }
	-
	\frac{ \Gamma\bigl( \frac{D-1}{2} \!+\! \lambda \!+\! n \bigr) \, \Gamma\bigl( \frac{D-1}{2} \!-\! \lambda \!+\! n \bigr) }
		{ \Gamma\bigl( \frac{D}{2} \!+\! n \bigr) \, n! }
		\Bigl( \frac{y}{4} \Bigr)^{\! n }
	\biggr]
	\biggr\} 
	\, .
\nonumber 
\end{align}
Owing to Gauss' relations for hypergeometric functions (9.137 
from~\cite{Gradshteyn:2007}), recurrence relations exist between
contiguous rescaled propagator functions,
\begin{equation}
2 \frac{\partial \mathcal{F}_\lambda}{\partial y} 
	=
	(2\!-\!y) \frac{\partial \mathcal{F}_{\lambda-1}}{\partial y}
	- \Bigl( \frac{D\!-\!3}{2} \!+\! \lambda \Bigr) \mathcal{F}_{\lambda-1} 
	\, ,
\qquad
2 \frac{\partial \mathcal{F}_{\lambda}}{\partial y} 
	=
	(2\!-\!y) \frac{\partial \mathcal{F}_{\lambda+1} }{ \partial y}
	- \Bigl( \frac{D\!-\!3}{2} \!-\! \lambda \Bigr) \mathcal{F}_{\lambda+1}
	\, .
\label{Frecurrence}
\end{equation}
These will be used frequently when constructing and checking the photon two-point 
function. The computation will also call for identities for parametric derivatives of the
rescaled propagator function, obtained by taking parametric derivatives of~(\ref{Feom}),
\begin{equation}
\biggl[ ( 4y\!-\!y^2 ) \frac{\partial^2}{\partial y^2}
	+ D (2\!-\!y) \frac{\partial}{\partial y}
	+ \lambda^2 - \Bigl( \frac{D\!-\!1}{2} \Bigr)^{\!2} \,
	 \biggr] 
	 \frac{\partial \mathcal{F}_\lambda }{\partial \lambda }
	 = 
	 -  2 \lambda \mathcal{F}_\lambda 
	 \, ,
\label{FparamEOM}
\end{equation}
and~(\ref{Frecurrence}),
\begin{subequations}
\begin{align}
&
2 \frac{\partial }{\partial y} \frac{\partial \mathcal{F}_\lambda}{\partial \lambda}
	=
	(2\!-\!y) \frac{\partial }{\partial y}
		\frac{\partial \mathcal{F}_{\lambda-1} }{\partial \lambda}
	- \Bigl( \frac{D\!-\!3}{2} \!+\! \lambda \Bigr) 
		\frac{\partial \mathcal{F}_{\lambda-1} }{\partial \lambda}
	-
	\mathcal{F}_{\lambda-1} 
	\, ,
\\
&
2 \frac{\partial }{\partial y}
	\frac{\partial \mathcal{F}_{\lambda}}{\partial \lambda}
	=
	(2\!-\!y) \frac{\partial  }{ \partial y}
		\frac{\partial \mathcal{F}_{\lambda+1} }{\partial \lambda}
	- \Bigl( \frac{D\!-\!3}{2} \!-\! \lambda \Bigr) 
		\frac{\partial \mathcal{F}_{\lambda+1} }{\partial \lambda} 
	+
	\mathcal{F}_{\lambda+1}
	\, .
\label{FparamRecurrence2}
\end{align}
\label{FparamRecurrence}%
\end{subequations}
Lastly, two special limits will be required,
\begin{align}
&
\Bigl( \lambda \!-\! \frac{D\!-\!3}{2} \Bigr) \mathcal{F}_{\lambda+1}
	\xrightarrow{ \lambda \to \frac{D-3}{2} }
	- 
	\frac{ \bigl[ (1\!-\!\epsilon) H_0 \bigr]^{D-2} }{ (4\pi)^{ \frac{D}{2} } }
	\frac{ \Gamma( D\!-\!1 ) }{ \Gamma\bigl( \frac{D}{2} \bigr) }
	\, ,
\label{SpecialLimit1}
\\
&
\Bigl( \lambda \!-\! \frac{D\!-\!3}{2} \Bigr)^{\!2} 
	\frac{\partial \mathcal{F}_{\lambda+1} }{\partial \lambda}
	\xrightarrow{ \lambda \to \frac{D-3}{2} }
	\frac{ \bigl[ (1\!-\!\epsilon) H_0 \bigr]^{D-2} }{ (4\pi)^{ \frac{D}{2} } }
	\frac{ \Gamma( D\!-\!1 ) }{ \Gamma\bigl( \frac{D}{2} \bigr) }
	\, ,
\label{SpecialLimit2}
\end{align}
that are best obtained from the series representation~(\ref{Fseries}).

While for~$\lambda\!<\!(D\!-\!1)/2$ the CTBD mode function~(\ref{CTBD}) leads to
an infrared finite mode sum for the two-point  functions in~(\ref{ScalarSum}),
this is not generally the case for arbitrary choices of parameter~$\lambda$. 
For~$\lambda\!\ge\!(D\!-\!1)/2$ the mode function in the infrared needs to be modified 
to a different choice of Bogolyubov coefficients in~(\ref{UgeneralSolution}), such that it 
is suppressed in the infrared and leads to an infrared finite mode sum. Under the
relatively mild assumption of this modification being contained to super-Hubble scales,
the construction can effectively be implemented by cutting off the mode sum at 
some infrared~$k_0\!\ll\!H_0$. In this work only the range~$\lambda\!<\! (D\!+\!1)/2$ will 
be relevant. This means that in the range~$(D\!-\!1)/2 \!\le\! \lambda \!<\! (D\!+\!1)/2$
the two-point function evaluates no longer evaluates to~(\ref{DeltaEvaluation}),
but rather to~\cite{Janssen:2008px},
\begin{equation}
i \bigl[ \tensor*[^{\tt a \!}]{\Delta}{^{\! \tt b}} \bigr]_\lambda(x;x')
	=
	i \Delta_\lambda(y,u)
	=
	e^{ - \frac{(D-2)\epsilon}{2(1-\epsilon)} u }
	\Bigl[
	\mathcal{F}_\lambda(y)
	+
	\mathcal{W}_\lambda(u)
	\Bigr]
	\, ,
\label{DeltaW}
\end{equation}
where the additional infrared term is
\begin{equation}
\mathcal{W}_{\lambda}(u)
	=
	\frac{ - \bigl[ (1 \!-\! \epsilon) H_0 \bigr]^{D-2 } \,
			\Gamma(\lambda) \, \Gamma(2\lambda) }
		{ (4\pi)^{\frac{D}{2}} \, \Gamma\bigl( \frac{1}{2} \!+\! \lambda \bigr) \, \Gamma\bigl( \frac{D-1}{2} \bigr) \, \bigl( \frac{D-1}{2} \!-\! \lambda \bigr) } 
	\Bigl[ \frac{ k_0^2 e^{-u} }{ (1\!-\!\epsilon)^2 H_0^2 } 
		\Bigr]^{ \frac{D-1}{2} - \lambda }
		\, .
\end{equation}
%

\section{Dynamics in power-law inflation}
\label{sec: Dynamics in power-law inflation}

In this section the equations of motion for the photon field operators are solved
for two particularly simple choices for the gauge-fixing parameter~$\zeta$.
This is accomplished by making use of the results for the scalar mode functions collected 
in Sec.~\ref{subsec: Scalar mode functions}. Transverse and scalar sectors are considered 
separately.

\subsection{Dynamics of the transverse sector}
\label{subsec: Dynamics of the transverse sector}

The two first order equations of motion for the transverse 
sector~(\ref{TransverseMomentumEOM}) combine into a second order one,
\begin{align}
\biggl[ \partial_0^2 + k^2 + \Bigl( \frac{1}{4} \!-\! \nu^2 \Bigr) \mathcal{H}^2 \biggr] 
	\hat{\mathcal{A}}_{ {\scr T}, \sigma} ={}& 0 \, ,
\\
\hat{\pi}_{{\scr T}, \sigma} ={}&
	\biggl[ \partial_0 
		+ \Bigl( \frac{1}{2} \!-\! \nu \Bigr) (1\!-\!\epsilon) \mathcal{H} 
			\biggr] \hat{\mathcal{A}}_{{\scr T}, \sigma} \, ,
\end{align}
where the index of the equation is
\begin{equation}
\nu = \frac{D\!-\!3 \!-\! \epsilon}{2(1\!-\!\epsilon)} \, .
\label{nu def}
\end{equation}
Making use of the fundamental solution of the scalar mode equation~(\ref{CTBD}),
and the index-lowering recurrence relation~(\ref{RecurrenceIdentities}), the solutions
for the field operators follow
\begin{align}
&
\hat{\mathcal{A}}_{{\scr T}, \sigma} (\eta,\vec{k})
	=
	U_\nu(\eta,k) \, \hat{b}_{\scr T} (\sigma, \vec{k})
	+ U_\nu^*(\eta,k) \, \hat{b}_{\scr T}^\dag (\sigma, -\vec{k}) \, ,
\label{ATsolution}
\\
&
\hat{\pi}_{{\scr T}, \sigma} (\eta,\vec{k})
	=
	- i k U_{\nu-1}(\eta,k) \, \hat{b}_{\scr T} (\sigma, \vec{k})
	+ i k U_{\nu-1}^*(\eta,k) \, \hat{b}_{\scr T}^\dag (\sigma, -\vec{k}) \, .
\label{PiTsolution}
\end{align}
The operator~$\hat{b}_{\scr T}$ and its conjugate are the annihilation and creation
operators satisfying canonical commutation relations,
\begin{equation}
\bigl[ \hat{b}_{\scr T}(\sigma, \vec{k}) ,
	\hat{b}_{\scr T}^\dag (\sigma', \vec{k}^{\,\prime}) \bigr] 
		= \delta_{\sigma\sigma'} \, \delta^{D-1}(\vec{k} \!-\! \vec{k}^{\,\prime}) \, ,
\label{bTcommutator}
\end{equation}
that can be computed with the help of the Wronskian~(\ref{AlternateWronskian}).
Then it is natural to consider the vacuum state of the transverse sector defined
by
\begin{equation}
\hat{b}_{{\scr T},\sigma} (\vec{k}) \bigl| \Omega \bigr\rangle = 0 \, ,
\qquad \quad
\forall \vec{k}, \sigma \, .
\label{bTstate}
\end{equation}
%

\subsection{Dynamics of the scalar sector}
\label{subsec: Dynamics of the scalar sector}

Equations of motion~(\ref{momentum scalar eom 2})
and~(\ref{momentum scalar eom 4}) for canonical momenta
decouple from the other two equations of the scalar sector. 
They can be combined into a second order equation,
\begin{align}
\biggl[ \partial_0^2 + k^2 
	+ \Bigl( \frac{1}{4} \!-\! \nu_\zeta^2 \Bigr) (1\!-\!\epsilon)^2 \mathcal{H}^2 \biggr] \hat{\pi}_0 
	={}& 0 \, ,
\\
\hat{\pi}_{\scr L} ={}&
	\frac{1}{k} \biggl[ \partial_0  
		+ \Bigl( \frac{1}{2} \!-\! \nu_\zeta \Bigr) (1\!-\!\epsilon) \mathcal{H} 
		\biggr] \hat{\pi}_0 
	\, ,
\end{align}
where the index of the equation is
\begin{equation}
\nu_\zeta 
	= \frac{ D\!-\!1\!-\!\epsilon \!-\! 2\zeta }{ 2(1\!-\!\epsilon) }
	=
	\nu
	+
	\frac{ 1 \!-\! \zeta }{1\!-\!\epsilon }
	\, ,
\label{nuZetaDef}
\end{equation}
with~$\nu$ defined in~(\ref{nu def}).
The solutions for canonical momenta immediatly follow
\begin{align}
&
\hat{\pi}_0(\eta,\vec{k}) =
	i k U_{\nu_\zeta}(\eta,k) \, \hat{b}_{\scr P}(\vec{k})
	- i k U_{\nu_\zeta}^*(\eta,k) \, \hat{b}_{\scr P}^\dag(-\vec{k}) 
	\, ,
\label{piL solution}
\\
&
\hat{\pi}_{\scr L}(\eta,\vec{k}) =
	k U_{\nu_\zeta-1}(\eta,k) \, \hat{b}_{\scr P}(\vec{k})
	+ k U_{\nu_\zeta-1}^*(\eta,k) \, \hat{b}_{\scr P}^\dag(-\vec{k}) 
	\, ,
\label{pi0 solution}
\end{align}
where~operators $\hat{b}_{\scr P}(\vec{k})$ and its conjugate account for the initial 
constants of integration, and where the overall normalization was chosen for convenience
to correspond to the one used in~\cite{Glavan:2022pmk}.

The remaining two equations~(\ref{momentum scalar eom 1})
and~(\ref{momentum scalar eom 3})
for vector potential field operators can be combined into an 
inhomogeneous second order equation,
\begin{align}
\biggl[ \partial_0^2 + k^2
	+ \Bigl( \frac{1}{4} \!-\! \nu_\zeta^2 \Bigr) (1\!-\!\epsilon)^2 
		\mathcal{H}^2 \biggr] \hat{\mathcal{A}}_{\scr L} 
	={}&
	a^{2-2\zeta} \Bigl[ 2(1\!-\!\zeta) \mathcal{H} \hat{\pi}_{\scr L}
		- (1\!-\!\xi) k \hat{\pi}_0 \Bigr] \, ,
\\
\hat{\mathcal{A}}_0
	={}&
	- \frac{1}{k} \biggl[ \partial_0 
		+ \Bigl( \frac{1}{2} \!-\! \nu_\zeta \Bigr) (1\!-\!\epsilon) \mathcal{H} 
			\biggr] \hat{\mathcal{A}}_{\scr L} 
	+ \frac{a^{2-2\zeta} }{k} \hat{\pi}_{\scr L} \, .
\end{align}
Using the results from Sec.~\ref{subsec: Scalar mode functions}, the solutions can 
be written as
\begin{align}
\hat{\mathcal{A}}_{\scr L} (\eta,\vec{k})
	={}&
	- i U_{\nu_\zeta }(\eta,k) \, \hat{b}_{\scr H}(\vec{k})
	+ i U_{\nu_\zeta }^*(\eta,k) \, \hat{b}_{\scr H}^\dag(- \vec{k})
	- i v_{\scr L}(\eta,k) \, \hat{b}_{\scr P}(\vec{k})
	+ i v_{\scr L}^*(\eta,k) \, \hat{b}_{\scr P}^\dag(-\vec{k}) 
	\, ,
\label{AL solution}
\\
\hat{\mathcal{A}}_{0} (\eta,\vec{k})
	={}&
	U_{\nu_\zeta-1}(\eta,k) \, \hat{b}_{\scr H}(\vec{k})
	+
	U_{\nu_\zeta-1}^*(\eta,k) \, \hat{b}_{\scr H}^\dag(- \vec{k})
	+ v_{0}(\eta,k) \, \hat{b}_{\scr P}(\vec{k})
	+ v_{0}^*(\eta,k) \, \hat{b}_{\scr P}^\dag(-\vec{k}) \, ,
\label{A0 solution}
\end{align}
where the particular mode functions satisfy,
\begin{align}
\biggl[ \partial_0^2 + k^2
	+ \Bigl( \frac{1}{4} \!-\! \nu_\zeta^2 \Bigr) (1\!-\!\epsilon)^2 
		\mathcal{H}^2 \biggr] v_{\scr L} 
	={}&
	a^{2-2\zeta} \Bigl[ 2(1\!-\!\zeta) i k \mathcal{H}  U_{\nu_\zeta-1}
		+ (1\!-\!\xi) k^2 U_{\nu_\zeta} \Bigr]  
		,
\label{vL eom}
\\
v_0
	={}&
	\frac{i}{k} \biggl[ \partial_0 
		+ \Bigl( \frac{1}{2} \!-\! \nu_\zeta \Bigr) (1\!-\!\epsilon) \mathcal{H} 
			\biggr] v_{\scr L}
	+ a^{2-2\zeta} U_{\nu_\zeta - 1} \, .
\label{v0 eom}
\end{align}
and where~$\hat{b}_{\scr H}(\vec{k})$ and its conjugate account for constants of
integration. Solving these equations is somewhat more involved, and is postponed until 
Sec.~\ref{subsec: Solving for particular mode functions}. However, the normalization 
of particular mode functions can be derived straightforwardly.
The equation of motion~(\ref{vL eom}) for the longitudinal particular mode function, and the 
Wronskian~(\ref{AlternateWronskian}) imply the following relation,
\begin{equation}
\partial_0 \, {\rm Im} \Bigl[ v_{\scr L} (\eta,k) \, \partial_0 U_{\nu_\zeta}^*(\eta,k)
		- \partial_0 v_{\scr L}(\eta,k) \, U_{\nu_\zeta}^*(\eta,k) \Bigr]
	= 
	- (1\!-\!\zeta) a^{2-2\zeta} \mathcal{H} 
	\, .
\end{equation}
Integrating this expression now gives,
\begin{equation}
{\rm Im} \Bigl[ v_{\scr L} (\eta,k) \, \partial_0 U_{\nu_\zeta}^*(\eta,k)
		- \partial_0 v_{\scr L}(\eta,k) \, U_{\nu_\zeta}^*(\eta,k) \Bigr]
	= 
	- \frac{ a^{2-2\zeta} }{ 2 } 
	\, .
\end{equation}
where the constant of integration is fixed according to the conventions
of~\cite{Glavan:2022pmk}. By using Eq.~(\ref{v0 eom}) 
and the recurrence relation for scalar mode function~(\ref{RecurrenceIdentities})
this condition can be rewritten in the form of a Wronskian-like relation,
\begin{equation}
{\rm Re} \Bigl[ v_{\scr L} (\eta,k) U_{\nu_\zeta-1}^*(\eta,k)
		+ v_0(\eta,k) U_{\nu_\zeta}^*(\eta,k) \Bigr]
	= 
	0
	\, .
\label{Wronskian-like}
\end{equation}
These conditions are then used to infer the commutators between the 
constant momentum space operators, where the nonvanishing ones are
\begin{equation}
\bigl[ \hat{b}_{\scr H}(\vec{k}) , \hat{b}_{\scr P}^\dag(\vec{k}^{\,\prime}) \bigr] 
	=
	\bigl[ \hat{b}_{\scr P}(\vec{k}) , \hat{b}_{\scr H}^\dag(\vec{k}^{\,\prime}) \bigr] 
	= - \delta^{D-1}(\vec{k} \!-\! \vec{k}^{\,\prime})
	\, .
\label{bPHcommutator}
\end{equation}
While these operators can be canonicalized to satisfy the canonical relations for
creation/annihilation operators, it is not necessary to do so~\cite{Glavan:2022pmk}.

\subsection{Solving for particular mode functions}
\label{subsec: Solving for particular mode functions}

Equations of motion for the particular mode functions~(\ref{vL eom}) and~(\ref{vL eom})
can be solved for arbitrary values of~$\xi$ and~$\zeta$. However, their solution
is rather unwieldy, and involves~${}_3F_2$ generalized hypergeometric functions
(cf.~Appendix A of~\cite{Domazet:2024dil}). However, for discreet choices fo~$\zeta$
the solutions simplify considerably, and are expressible in terms of the CTBD scalar mode 
functions~(\ref{CTBD}), and a finite number of their derivatives. For this to be the case, 
the time-dependent factor~$a^{2-2\zeta}$ multiplying the source term in~(\ref{vL eom}) 
has to be proportional to an even power (either positive or negative) of the conformal 
Hubble rate,
\begin{equation}
a^{2-2\zeta} = \Bigl( \frac{ \mathcal{H} }{ H_0 } \Bigr)^{\!2n} 
	\, ,
\qquad
n \in \mathbb{Z} \, .
\end{equation}
This puts a condition on the values of the second gauge-fixing parameter~$\zeta$
that lead to practical and simple solutions for the photon mode functions,
\begin{equation}
\zeta = 1 - n(1 \!-\! \epsilon) \, .
\label{zetaN}
\end{equation}
Henceforth, when referring to~$\zeta$ we assume that it takes discreet values
given by the relation above, unles stated otherwise explicitly. 
This imples that the index defined in~(\ref{nuZetaDef})
also takes discreet values,
\begin{equation}
\nu_\zeta = \nu + n \, .
\end{equation}
In such discreet~$\zeta$ gauges the equations of motion for particular mode functions read
\begin{align}
\MoveEqLeft[10]
\biggl[ \partial_0^2 + k^2
	+ \Bigl( \frac{1}{4} \!-\! (\nu \!+\! n)^2 \Bigr) (1\!-\!\epsilon)^2 
		\mathcal{H}^2 \biggr] v_{\scr L} 
	=
	\Bigl( \frac{ \mathcal{H} }{ H_0 } \Bigr)^{\!2n}  
	\Bigl[ 2n(1 \!-\! \epsilon) i k \mathcal{H}  U_{ \nu + n -1}
		+ (1\!-\!\xi) k^2 U_{\nu + n} \Bigr]
		\, ,
\label{EOMnL}
\\
v_0
	={}&
	\frac{i}{k} \biggl[ \partial_0 
		+ \Bigl( \frac{1}{2} \!-\! (\nu \!+\! n) \Bigr) (1\!-\!\epsilon) \mathcal{H} 
			\biggr] v_{\scr L}
	+
	\Bigl( \frac{ \mathcal{H} }{ H_0 } \Bigr)^{\!2n}  
	U_{\nu + n - 1} 
	\, .
\label{EOMn0}
\end{align}
These equations can be solved for any~$n$, which is done in 
Appendix~\ref{sec: Mode functions in discreet zeta gauges}. However,
almost all cases end up being impractical when constructing the two-point function
in Sec.~\ref{sec: Two-point functions}. That is why here only two simple choices are 
considered: the conformal gauge for which~$n\!=\!0$, and
the deceleration gauge for which~$n\!=\!1$. The solutions for 
particular mode functions in these two one-parameter families of gauges are worked 
out in the remainder of this section, in both cases normalized by the requirement 
that in flat space they reduce to
\begin{align}
&
v_{\scr L} \xrightarrow{H_0\to0}
	\frac{1}{4} \Bigl[ - (1\!+\!\xi) + 2 (1\!-\!\xi) ik (\eta \!-\! \eta_0) \Bigr]
	\frac{ e^{-ik(\eta-\eta_0)} }{ \sqrt{2k} }
	\, .
\label{vLflat}
\\
&
v_0 \xrightarrow{H_0\to0}
	\frac{1}{4} \Bigl[ (1\!+\!\xi) + 2 (1\!-\!\xi) ik (\eta \!-\! \eta_0) \Bigr]
	\frac{ e^{-ik(\eta-\eta_0)} }{ \sqrt{2k} }
	\, ,
\label{v0flat}
\end{align}
Henceforth we label the first gauge-fixing parameter differently in each 
gauge:~$\xi\!=\!\alpha$ for the conformal gauge, and~$\xi\!=\!\beta$ for the deceleration 
gauge, in order to distinguish between them more easily, while~$\zeta$ still stands for the
general case.

\bigskip

\noindent {\bf Conformal gauge ($\boldsymbol{n\!=\!0}$).}
The equations of motion for particular mode functions in this case read
\begin{align}
\biggl[ \partial_0^2 + k^2 
	+ \Bigl( \frac{1}{4} \!-\! \nu^2 \Bigr) (1\!-\!\epsilon)^2 \mathcal{H}^2 \biggr] 
		v_{\scr L}
	={}& 
	(1 \!-\! \alpha) k^2 U_{\nu} \, ,
\label{vL conformal eom}
\\
v_0 ={}&
	\frac{i}{k} \biggl[ \partial_0 
		+ \Bigl( \frac{1}{2} \!-\! \nu \Bigr) (1\!-\!\epsilon) \mathcal{H} \biggr] v_{\scr L} 
	+ U_{\nu-1} \, ,
\label{v0 conformal eom}
\end{align}
and are solved by,
\begin{align}
&
v_{\scr L} =
	- \frac{(1\!-\!\alpha) i k}{2(1\!-\!\epsilon)} \biggl[ 
		\frac{1}{\mathcal{H}} U_{\nu-1} - \frac{1}{H_0} U_\nu \biggr]
	- \frac{ 1 \!-\! (1\!-\!\alpha) \nu }{2} U_\nu
	 \, ,
\label{vL n=0 solution}
\\
&
v_0 =
	- \frac{(1\!-\!\alpha) ik}{2 (1\!-\!\epsilon)} \biggl[ 
			\frac{1}{\mathcal{H}} U_\nu - \frac{1}{H_0} U_{\nu-1} \biggr]
	+ \frac{ 1 \!-\! (1\!-\!\alpha) \nu }{2} U_{\nu-1}
	\, .
\label{v0 n=0 solution} 
\end{align}
Checking that these indeed solve~(\ref{vL conformal eom}) 
and~(\ref{v0 conformal eom})
involves using the equation of motion~(\ref{scalar mode eq})
for the scalar mode function and the identity~(\ref{ModeIdentity1}).
It is clear that~$\alpha\!=\!1$ is the simplest choice as it turns 
particular solutions into homogeneous ones, which justifies referring to
this choice as the~{\it simple conformal gauge}. 
This gauge choice is known to be simplest choice in the de Sitter limit~\cite{Woodard:2004ut}.
In a sense this gauge can rightfully be considered the generalization of the
Feynman gauge in flat space to power-law inflation.


\bigskip
\noindent {\bf Deceleration gauge ($\boldsymbol{n \!=\! 1}$).}
The particular mode functions
in this case satisfy,
\begin{align}
\biggl[ \partial_0^2 + k^2
	+ \Bigl( \frac{1}{4} \!-\! (\nu \!+\! 1)^2 \Bigr) (1\!-\!\epsilon)^2 
		\mathcal{H}^2 \biggr] v_{\scr L} 
	={}&
	\Bigl( \frac{ \mathcal{H} }{ H_0 } \Bigr)^{\!2}  
	\biggl[ 2(1 \!-\! \epsilon) i k \mathcal{H}  U_{ \nu }
		+ (1\!-\!\beta) k^2 U_{\nu + 1} \biggr]  
		\, ,
\label{vLdecEOM}
\\
v_0
	={}&
	\frac{i}{k} \biggl[ \partial_0 
		+ \Bigl( \frac{1}{2} \!-\! (\nu \!+\! 1) \Bigr) (1\!-\!\epsilon) \mathcal{H} 
			\biggr] v_{\scr L}
	+
	\Bigl( \frac{ \mathcal{H} }{ H_0 } \Bigr)^{\!2} U_{\nu } \, .
\label{v0decEOM} 
\end{align}
Making use of identities~(\ref{ModeIdentity2}) and~(\ref{ModeIdentity3}) it can be seen 
that solutions are,
\begin{align}
&
v_{\scr L} =
	\frac{ - ik }{2(1\!-\!\epsilon) H_0}
	\biggl\{
	\frac{ \beta }{ \nu\!+\!1 } 
		\biggl[ \frac{ \mathcal{H} }{ H_0 } U_\nu - U_{\nu+1} \biggr]
	+
	\Bigl( 1 \!-\! \frac{\beta}{\beta_s} \Bigr)
		\biggl[ \frac{ik}{\nu (1\!-\!\epsilon) H_0} \frac{\partial U_{\nu+1} }{ \partial \nu }
			+ \frac{ \mathcal{H} }{\nu H_0} U_\nu + U_{\nu+1} \biggr]
	\biggr\}
	\, ,
\label{vLdecSOL}
\\
&
v_0 =
	\frac{ - ik }{ 2 (1\!-\!\epsilon) H_0 }
	\biggl\{
	\frac{ \beta }{ \nu\!+\!1 } 
		\biggl[ \frac{\mathcal{H} }{H_0} U_{\nu+1} - U_\nu \biggr]
	+
	\Bigl( 1 \!-\! \frac{ \beta }{ \beta_s } \Bigr)
		\biggl[ \frac{ik}{ \nu (1\!-\!\epsilon) H_0 } \frac{\partial U_\nu}{\partial \nu}
		+ U_{\nu} \biggr]
	\biggr\}
	\, .
\label{v0decSOL}
\end{align}
where another parameter is introduced,
\begin{equation}
\beta_s 
	= \frac{ \nu \!+\! 1 }{ \nu} 
	= \frac{ D \!-\! 1 \!-\! 3\epsilon }{ D \!-\! 3 \!-\! \epsilon } \, .
\label{SimpleBeta}
\end{equation}
Showing that these solutions indeed satisfy Eqs.~(\ref{vLdecEOM}) and~(\ref{v0decEOM})
requires the use of scalar mode equation of motion~(\ref{scalar mode eq}), 
the related identity~(\ref{ModeIdentity2}),
and the recurrence relations~(\ref{RecurrenceIdentities}).
The special case~$\beta \!=\! \beta_s$ gives the simplest result for the particular
mode functions, and it is thus referred to as the {\it simple deceleration gauge}.
It can be seen as the power-law inflation generalization of the Fried-Yennie gauge
in flat space~\cite{Fried:1958zz,Adkins:1993qm}.

\subsection{Dynamics of constraints}
\label{subsec: Dynamics of constraints}

Equations of motion for the coefficient functions of constraints~(\ref{constraintEOM})
can likewise be combined into a homogeneous second order equation,
\begin{align}
\biggl[ \partial_0^2 + k^2 
	+ \Bigl( \frac{1}{4} \!-\! \nu_\zeta^2 \Bigr) (1\!-\!\epsilon)^2 \mathcal{H}^2 \biggr]
	c_2 ={}& 0 \, ,
\\
c_1 ={}& 
	- \frac{1}{k}
	\biggl[
	\partial_0 + \Bigl( \frac{1}{2} \!-\! \nu_\zeta \Bigr)
		(1\!-\!\epsilon)\mathcal{H} 
	\biggr] c_2 
	\, .
\end{align}
The solutions are chosen to be consistent with the Lorentz 
invariance~\cite{Gupta:1949rh,Bleuler:1950cy}
for~$H_0\!\to\!0$ in both gauges, and the de Sitter invariance~\cite{Glavan:2022dwb}
for~$\epsilon\!\to\!0$ in the deceleration gauge,
\begin{equation}
c_1(\eta,k) = - i U_{\nu_\zeta-1}^*(\eta,k) 
	\, ,
\qquad \quad
c_2(\eta,k) = U_{\nu_\zeta}^*(\eta,k)
	\, .
\end{equation}
While these solutions are valid for arbitrary~$\zeta$, here they are considered for
discreet choices in~(\ref{zetaN}) only, on account of limitations for particular mode functions 
from the preseding subsection.
The non-Hermitian constraint operator~(\ref{Kdef}) simplifies upon this choice,
\begin{equation}
\hat{\mathcal{K}}(\vec{k}) = \hat{b}_{\scr P}(\vec{k}) \, ,
\end{equation}
from where it follows that the scalar sector of the state must be defined by
\begin{equation}
\hat{b}_{\scr P}(\vec{k}) \bigl| \Omega \bigr\rangle = 0 \, ,
\qquad \quad
\hat{b}_{\scr H}(\vec{k}) \bigl| \Omega \bigr\rangle = 0 \, ,
\qquad \quad
\forall \vec{k} \, .
\label{bPHstate}
\end{equation}
%

\section{Two-point functions}
\label{sec: Two-point functions}

The solutions for field operators, and the definitions of the state determined in the
preceding section are sufficient to compute the photon two-point functions in 
the conformal gauge and in the deceleration gauge. These two-point functions have
to satisfy both their respective equations of motion and the respective Ward-Takahashi 
identities, that are recalled first. This is followed by computing the momentum space 
representation of the two-point functions in terms of mode sums. These mode sums are 
then evaluated to obtain the position space two-point functions, that are then written in 
the covariantized form.

\subsection{Generalities}
\label{subsec: Generalities}

The positive-frequency Wightman function for the photon field is defined as 
an expectation value of an off-coincident product of vector field opeartors,
\begin{equation}
i \bigl[ \tensor*[_\mu^{\scr - \!}]{\Delta}{_\nu^{\scr \! +}} \bigr] (x;x')
	= \bigl\langle \Omega \bigr| \hat{A}_\mu(x) \hat{A}_\nu(x') 
		\bigl| \Omega \bigr\rangle \, ,
\label{WightmanDef}
\end{equation}
and its complex conjugate,~$i \bigl[ \tensor*[_\mu^{\scr + \!}]{\Delta}{_\nu^{\scr \! -}} \bigr] (x;x') \!=\! 
	\bigl\{ i \bigl[ \tensor*[_\mu^{\scr - \!}]{\Delta}{_\nu^{\scr \! +}} \bigr] (x;x') \bigr\}^*$, 
is the negative-frequency Wightman 
functio. The Feynman propagator is then defined as an expectation valus of the 
time-ordered product of vector field operators, and is expressible in terms of the
Wightman functions,
\begin{equation}
i \bigl[ \tensor*[_\mu^{\scr + \!}]{\Delta}{_\nu^{\scr \! +}} \bigr] (x;x')
	=
	\theta(\eta \!-\! \eta')
	i \bigl[ \tensor*[_\mu^{\scr - \!}]{\Delta}{_\nu^{\scr \! +}} \bigr] (x;x')
	+
	\theta(\eta' \!-\! \eta)
	i \bigl[ \tensor*[_\mu^{\scr + \!}]{\Delta}{_\nu^{\scr \! -}} \bigr] (x;x')
	\, .
\end{equation}
Its complex conjugate~$i \bigl[ \tensor*[_\mu^{\scr - \!}]{\Delta}{_\nu^{\scr \! -}} \bigr] (x;x') \!=\! 
	\bigl\{ i \bigl[ \tensor*[_\mu^{\scr + \!}]{\Delta}{_\nu^{\scr \! +}} \bigr] (x;x') \bigr\}^*$,
defines the Dyson propagator. Equations of motion~(\ref{GaugeFixedEOM}) for the field 
operators can be written in a more familiar covariant form,
\begin{equation}
\mathcal{D}^{\mu\nu} \hat{A}_\nu = 0 \, ,
\qquad \quad
\mathcal{D}_{\mu\nu} =
	g_{\mu\nu} {\dalembertian} 
	- 
	\nabla_\mu \nabla_\nu 
	+
	\frac{1}{\xi} \bigl( \nabla_\mu \!+\! 2 \zeta n_\mu \bigr)
		\bigl( \nabla_\nu \!-\! 2 \zeta n_\nu  \bigr)
	- 
	R_{\mu\nu}
	\, .
\end{equation}
This equation of motion is inherited by the two-point functions,
\begin{equation}
{\mathcal{D}_\mu}^\rho \,
	i \bigl[ \tensor*[_\rho^{\tt a \!}]{\Delta}{_\nu^{\tt \! b} } \bigr] (x;x')
	=
	{\tt S}^{\tt ab} g_{\mu\nu} \frac{ i \delta^D(x\!-\!x') }{ \sqrt{g} } \, ,
\label{Photon2ptEOM}
\end{equation}
where the source on the right-hand side appears for Feynman and Dyson propagators
on account of step functions in their definition and the canonical commutation relations.

In addition to the equations of motion, the photon two-point functions have to satisfy
the Ward-Takahashi identities. For the two-parameter family of linear gauges~(\ref{gauges})
these were derived in~\cite{Glavan:2022pmk},
\begin{equation}
\bigl( \nabla^\mu \!-\! 2 \zeta n^\mu \bigr)
	i \bigl[ \tensor*[_\mu^{\tt a \!}]{\Delta}{_\nu^{\tt \! b} } \bigr] (x;x')
	=
	- \xi \partial_\nu' \biggl[
	\Bigl( \frac{a'}{a} \Bigr)^{\! \zeta} 
	i \bigl[ \tensor*[^{\tt a\!}]{\Delta}{^{\tt \! b} } \bigr]_{\nu_\zeta } (x;x')
	\biggr]
	\, .
\label{WTidentity}
\end{equation}
The expression in brackets on the right-hand side is the full Faddeev-Popov (FP) ghost
two-point function, comprised of the ratio of scale factors multipliying the scalar
two-point function that satisfies equation of motion~(\ref{ScalarPropagatorEOM}),
\begin{equation}
\biggl[ \, \dalembertian 
	- \biggl( \Bigl[ \frac{D \!-\! 1 \!-\! \epsilon}{2(1\!-\!\epsilon)} \Bigr]^2 
		\!-\! \nu_\zeta^2 \biggr)
	(1\!-\!\epsilon)^2 H^2 \biggr]
	i \bigl[ \tensor*[^{\tt a \!}]{\Delta}{^{\tt \! b }} \bigr]_{\nu_\zeta}(x;x')
	=
	{\tt S}^{\tt ab} \frac{ i\delta^D(x\!-\!x') }{ \sqrt{-g} }
	\, .
\end{equation}
It is worth noting that the FP two-point function is not invariant under simultaneous
complex conjugation and exchange of coordinate arguments.
This is not surprising given that the two FP ghost fields comprising the two-point
function need to satisfy different equations of motion.

\subsection{Mode sums}
\label{subsec: Photon two-point functions as sums over modes}

The first step in computing the Wightman function~(\ref{WightmanDef})
is plugging in the Fourier transforms for the field operators (\ref{ATfourier}),
(\ref{A0fourier}), and (\ref{ALfourier}), and the solutions for the field operators
(\ref{ATsolution}), (\ref{AL solution}) and (\ref{A0 solution}). Then the commutation 
relations~(\ref{bTcommutator}) and~(\ref{bPHcommutator}), and the 
conditions~(\ref{bTstate}) and~(\ref{bPHstate}) defining the state 
produce mode sum expressions for two-point function components ,
\begin{align}
&
i \bigl[ \tensor*[_0^{\scr - \!}]{\Delta}{_0^{\scr \! +}} \bigr] (x;x')
	=
	(aa')^{- \frac{D-4 }{2} - n(1-\epsilon) } \!
	\int\! \frac{d^{D-1} k }{ (2\pi)^{D-1} } \,
	e^{i\vec{k} \cdot (\vec{x} - \vec{x}{\,}') } \,
\nonumber \\
&	\hspace{4.cm}
	\times \!
	\biggl[ - U_{\nu-1+n}(\eta,k) v_{0}^*(\eta' ,k)
		- v_{0}(\eta,k) U_{\nu-1+n}^*(\eta',k) 
	  \biggr] \, ,
\label{ModeSumN00}
\\
&
i \bigl[ \tensor*[_0^{\scr - \!}]{\Delta}{_i^{\scr \! +}} \bigr] (x;x')
	= 
	(aa')^{- \frac{D-4 }{2} - n(1-\epsilon) } \!
	\int\! \frac{d^{D-1} k }{ (2\pi)^{D-1} } \,
	e^{i\vec{k} \cdot (\vec{x} - \vec{x}{\,}') } \,
	\frac{k_i}{k}
\nonumber \\
&	\hspace{4.cm}
\times \!
	\biggl[
	U_{\nu-1+n}(\eta,k) v_{\scr L}^*(\eta',k)
	+ v_{0}(\eta,k) U_{\nu+n}^*(\eta',k) 
	\biggr] \, ,
\label{ModeSumN0i}
\\
&	
i \bigl[ \tensor*[_{i\,}^{\scr - \!}]{\Delta}{_j^{\scr \! +}} \bigr] (x;x')
	= (aa')^{-\frac{D-4}{2}}
	\int\! \frac{d^{D-1} k }{ (2\pi)^{D-1} } \,
		e^{i \vec{k} \cdot (\vec{x} - \vec{x}{\,}' ) } \,
		\Bigl( \delta_{ij} \!-\! \frac{k_i k_j}{k^2} \Bigr)
		U_\nu(\eta,k) \, U_\nu^*(\eta',k)
\\
&	\hspace{0.5cm}
	+
	(aa')^{- \frac{D-4 }{2} - n(1-\epsilon) } \!\!
	\int\! \frac{d^{D-1} k }{ (2\pi)^{D-1} } \,
	e^{i\vec{k} \cdot (\vec{x} - \vec{x}{\,}') } \,
	\frac{k_i k_j}{k^2}
	\biggl[ - U_{\nu+n}(\eta,k) v_{\scr L}^*(\eta' ,k)
		- v_{\scr L}(\eta,k) U_{\nu+n}^*(\eta',k) 
	  \biggr] \, .
\label{ModeSumNij}
\nonumber 
\end{align}
These expressions constitute the momentum space expressions
for the photon two-point functions. In the remainder of the section
these mode sums are evaluated for the two gauges.

\bigskip
\noindent{\bf Conformal gauge.}
Upon plugging in the conformal gauge particular mode 
functions~(\ref{vL n=0 solution}) and~(\ref{v0 n=0 solution}) into the mode 
sums~(\ref{ModeSumN00})--(\ref{ModeSumNij}), and upon using 
recurrence relations for mode functions~(\ref{RecurrenceIdentities}),
the mode sums can be recognized as derivatives acting on the scalar
two-point functions~(\ref{ScalarSum}). 
In the conformal gauge the photon two-point function components read:
\begin{align}
i \bigl[ \tensor*[_0^{\scr - \!}]{\Delta}{_0^{\scr \! +}} \bigr] (x;x')
	={}&
	- (aa')^{-\frac{(D-4)\epsilon}{2}} 
	\frac{ \mathcal{H} \mathcal{H}' }{ H_0^2 } \mathcal{F}_{\nu-1} ( y )
\nonumber \\
&
		-
		\frac{(1\!-\!\alpha) (aa')^{-\frac{(D-4)\epsilon}{2}}  }{2 (1\!-\!\epsilon) H_0^2} 
		\Bigl[ \mathcal{H} \partial_0' + \mathcal{H}' \partial_0
			+ (D\!-\!3)(1\!-\!\epsilon) \mathcal{H} \mathcal{H}' \Bigr]
		\mathcal{F}_{\nu-1} ( y )
	 \, ,
\\
i \bigl[ \tensor*[_0^{\scr - \!}]{\Delta}{_i^{\scr \! +}} \bigr] (x;x')
	= {}&
	\frac{(1\!-\!\alpha) (aa')^{- \frac{(D-4)\epsilon}{2}} }{2 (1\!-\!\epsilon) H_0^2} 
	\partial_i'
	\Bigl[
		\mathcal{H}' \mathcal{F}_\nu( y )
		-
		\mathcal{H} \mathcal{F}_{\nu-1}( y )
	\Bigr] \, ,
\\
i \bigl[ \tensor*[_{i\,}^{\scr - \!}]{\Delta}{_j^{\scr \! +}} \bigr] (x;x')
	={}&
	\delta_{ij} 
	\frac{\mathcal{H} \mathcal{H}' }{ H_0^2 } (aa')^{-\frac{(D-4)\epsilon}{2}} 
	\mathcal{F}_{\nu}( y )
\nonumber \\
&
	+
	\frac{(1\!-\!\alpha) (aa')^{-\frac{(D-4)\epsilon}{2}}  }{2(1\!-\!\epsilon) H_0^2 }
	\frac{\partial_i \partial_j'}{\nabla^2}
	\Bigl[ 
		\mathcal{H} \partial_0' + \mathcal{H}' \partial_0 
		+
		(D\!-\!1) (1\!-\!\epsilon) \mathcal{H} \mathcal{H}'
	\Bigr] \mathcal{F}_{\nu}( y )
	   \, .
\end{align}
The inverse Laplacian in the~$(ij)$ component can be removed with the use of
the following identity (cf.~Appendix C from~\cite{Glavan:2020zne}),
\begin{equation}
\Bigl[ \mathcal{H} \partial_0' + \mathcal{H}' \partial_0 
	+ (D\!-\!1) (1\!-\!\epsilon) \mathcal{H} \mathcal{H}' \Bigr] f(y)
	=
	\frac{1}{2(1\!-\!\epsilon)} \nabla^2 I \bigl[ f(y) \bigr] \, ,
\label{NeatIdentity}
\end{equation}
where~$I[f] \!=\! \int \! dy f(y)$ denotes the primitive function. Furthermore,
the derivatives in other components can be acted explicitly using
\begin{equation}
\mathcal{H} (\partial_0' y) + \mathcal{H}' (\partial_0 y)
	= - 2 (1\!-\!\epsilon) \mathcal{H} \mathcal{H}' (2\!-\!y)
		+ 2 (1\!-\!\epsilon) \bigl( \mathcal{H}^2 \!+\! \mathcal{H}'^2 \bigr) \, ,
\end{equation}
to rewrite the two-point function components as:
\begin{align}
&
i \bigl[ \tensor*[_0^{\scr - \!}]{\Delta}{_0^{\scr \! +}} \bigr] (x;x')
	=
	 (aa')^{-\frac{(D-4)\epsilon}{2}} \biggl\{
	-
	\frac{ \mathcal{H} \mathcal{H}' }{ H_0^2 } 
	+
	\frac{ (1\!-\!\alpha) \mathcal{H} \mathcal{H}' }{ H_0^2 } 
	\biggl[ ( 2\!-\!y )  \frac{\partial }{\partial y}
		- \frac{ D\!-\!3 }{2} \biggr]
\nonumber \\
&	\hspace{6.cm}
	-
	\frac{(1\!-\!\alpha) }{H_0^2} \bigl( \mathcal{H}^2 \!+\! \mathcal{H}'^2 \bigr) 
		\frac{\partial }{\partial y} 
	\biggr\}
	\mathcal{F}_{\nu-1} ( y )
	\, ,
\label{conformal delta 00}
\\
&
i \bigl[ \tensor*[_0^{\scr - \!}]{\Delta}{_i^{\scr \! +}} \bigr] (x;x')
	= 
	\frac{(1\!-\!\alpha) (aa' )^{- \frac{(D-4)\epsilon}{2}} }{2 (1\!-\!\epsilon) H_0^2} 
	\bigl( \partial_i' y \bigr)
	\biggl[
		\mathcal{H}' \frac{\partial\mathcal{F}_\nu( y ) }{\partial y}
		-
		\mathcal{H} \frac{ \partial \mathcal{F}_{\nu-1}( y ) }{ \partial y }
	\biggr] \, ,
\label{conformal delta 0i}
\\
&
i \bigl[ \tensor*[_{i\,}^{\scr - \!}]{\Delta}{_j^{\scr \! +}} \bigr] (x;x')
	= 
	 \delta_{ij} 
	\frac{\mathcal{H} \mathcal{H}' }{ H_0^2 } (aa')^{-\frac{(D-4)\epsilon}{2}} 
	\mathcal{F}_{\nu}( y )
	+
	\frac{(1\!-\!\alpha) (aa')^{-\frac{(D-4)\epsilon}{2}}  }{4(1\!-\!\epsilon)^2 H_0^2 } 
	\partial_i \partial_j' I \bigl[ \mathcal{F}_{\nu}(y) \bigr]
\label{conformal delta ij}
	   \, .
\end{align}
%

\bigskip
\noindent{\bf Deceleration gauge.}
Plugging in the particular mode functions~(\ref{vLdecSOL})
and~(\ref{v0decSOL}) in the deceleration gauge into the mode 
sums~(\ref{ModeSumN00})--(\ref{ModeSumNij}), 
followed by using recurrence relations~(\ref{RecurrenceIdentities}), and
recognizing scalar two-point functions produces:
\begin{align}
i \bigl[ \tensor*[_0^{\scr - \!}]{\Delta}{_0^{\scr \! +}} \bigr] (x;x')
	={}&
	\Bigl( 1 \!-\! \frac{ \beta }{ \beta_s } \Bigr)
	\frac{ (aa')^{ - \frac{(D-4)\epsilon}{2} } }{2\nu(1\!-\!\epsilon)^2 H_0^2 } 
	\nabla^2
	\frac{\partial}{\partial \nu}
	\mathcal{F}_\nu(y) 
\\
&
	-
	\frac{ \beta (aa')^{ - \frac{(D-4)\epsilon}{2} }  }{ 2(\nu\!+\!1) (1\!-\!\epsilon) H_0^2 } 
	\biggl[
		\mathcal{H} \partial_0
		+
		\mathcal{H}' \partial_0' 
		+
		\Bigl( \nu \!+\! \frac{D \!-\! 1}{2} \Bigr) (1\!-\!\epsilon)
			\bigl( \mathcal{H}^2 \!+\! \mathcal{H}'^2 \bigr)
		\biggr]
	\mathcal{F}_\nu(y) 
	\, ,
\nonumber \\
i \bigl[ \tensor*[_0^{\scr - \!}]{\Delta}{_i^{\scr \! +}} \bigr] (x;x')
	={}&
	\frac{ (aa')^{- \frac{ (D-4)\epsilon }{2} } }{ 2 \nu (1\!-\!\epsilon) H_0^2 }
		\partial'_i
		\Bigl[ 
		\mathcal{H} 
		\mathcal{F}_{\nu+1}(y) 
		-
		\mathcal{H}' 
		\mathcal{F}_\nu(y) 
		\Bigr]
\nonumber \\
&
	-
	\Bigl( 1 \!-\! \frac{\beta}{\beta_s} \Bigr)
	\frac{ (aa')^{- \frac{ (D-4)\epsilon }{2} } }{2 \nu (1\!-\!\epsilon)^2 H_0^2} 
	\partial'_i
	\biggl[ \partial_0 - \Bigl( \nu \!-\! \frac{D \!-\! 3}{2} \Bigr) (1\!-\!\epsilon) \mathcal{H} 
	\biggr] 
	\frac{ \partial }{ \partial \nu }
	\mathcal{F}_{\nu+1}(y) \, ,
\\
i \bigl[ \tensor*[_{i\,}^{\scr - \!}]{\Delta}{_j^{\scr \! +}} \bigr] (x;x')
	={}& 
	\delta_{ij}
	\frac{ \mathcal{H} \mathcal{H}' }{ H_0^2 }
	(aa')^{- \frac{(D-4)\epsilon}{2}} 
	\mathcal{F}_\nu(y) 
	-
	\Bigl( 1 \!-\! \frac{\beta}{\beta_s} \Bigr)
		\frac{ (aa')^{- \frac{ (D-4)\epsilon }{2} } }{ 2 \nu (1\!-\!\epsilon)^2 H_0^2 }
		\partial_i \partial_j'
		\frac{\partial}{\partial \nu}
		\mathcal{F}_{\nu+1}(y) 
\nonumber \\
&
	-
	\frac{ (aa')^{- \frac{ (D-4)\epsilon }{2} } }{ 2\nu (1\!-\!\epsilon) H_0^2 }
	\frac{\partial_i \partial_j' }{\nabla^2}
	\Bigl[ 
	\mathcal{H}  \partial_0' 
	+ \mathcal{H}' \partial_0 
	+ ( D \!-\! 1 ) (1\!-\!\epsilon) \mathcal{H} \mathcal{H}' 
	\Bigr] 
	\mathcal{F}_\nu(y) 
	\, .
\end{align}
These components are further rewritten using the identity in~(\ref{NeatIdentity})
for the~$(ij)$ component, and acring the derivatives in the two other components,
\begin{align}
i \bigl[ \tensor*[_0^{\scr - \!}]{\Delta}{_0^{\scr \! +}} \bigr] (x;x')
	={}&
	\frac{ \beta }{ \beta_s }
	\frac{ (aa')^{ - \frac{(D-4)\epsilon}{2} }  }{ 2 \nu H_0^2 } 
	\biggl\{
		\bigl( \mathcal{H}^2 \!+\! \mathcal{H}'^2 \bigr)
		\biggl[
			( 2 \!-\! y ) \frac{\partial}{\partial y}
			\!-\! \nu \!-\! \frac{D \!-\! 1}{2} 
		\biggr]
		\!-\!
		4\mathcal{H} \mathcal{H}' \frac{\partial}{\partial y}
		\biggr\}
	\mathcal{F}_\nu(y) 
\nonumber \\
&	\hspace{-2cm}
	+
	\Bigl( 1 \!-\! \frac{ \beta }{ \beta_s } \Bigr)
	\frac{ (aa')^{ - \frac{(D-4)\epsilon}{2} } }{2\nu H_0^2 } 
	\biggl\{
	4 \bigl( \mathcal{H}^2 \!+\! \mathcal{H}'^2 \bigr) \frac{\partial}{\partial y}
	\!-\!
	2
	\mathcal{H}\mathcal{H}' 
	\biggl[
	2 (2 \!-\! y)
	\frac{\partial}{\partial y}
	\!-\!
	D\!+\!1
	\biggr]
	\biggr\}
	\frac{\partial}{\partial y}
	\frac{\partial}{\partial \nu}
	\mathcal{F}_\nu(y) 
	\, ,
\label{00DecelComputed}
\\
i \bigl[ \tensor*[_0^{\scr - \!}]{\Delta}{_i^{\scr \! +}} \bigr] (x;x')
	={}&
	\frac{ (aa')^{- \frac{ (D-4)\epsilon }{2} } }{ 2 \nu (1\!-\!\epsilon) H_0^2 }
		\bigl( \partial'_i y \bigr)
	\biggl\{
		\mathcal{H} 
		\frac{\partial}{\partial y} \mathcal{F}_{\nu+1}(y) 
		-
		\mathcal{H}' 
		\frac{\partial}{\partial y} \mathcal{F}_\nu(y) 
\nonumber \\
&
	+
	\Bigl( 1 \!-\! \frac{\beta}{\beta_s} \Bigr)
	\biggl[ 
	\mathcal{H} \biggl( (2\!-\!y) \frac{\partial}{\partial y} 
	+ 
	\nu \!-\! \frac{D \!-\! 1}{2}
	\biggr)
	-
	2 \mathcal{H}' \frac{\partial}{\partial y} 
	\biggr] 
	\frac{\partial}{\partial y}
	\frac{ \partial }{ \partial \nu }
	\mathcal{F}_{\nu+1}(y) 
	\biggr\}
	\, ,
\label{0iDecelComputed}
\\
i \bigl[ \tensor*[_{i\,}^{\scr - \!}]{\Delta}{_j^{\scr \! +}} \bigr] (x;x')
	={}&
	\delta_{ij}
	\frac{ \mathcal{H} \mathcal{H}' }{ H_0^2 }
	(aa')^{- \frac{(D-4)\epsilon}{2}} 
	\mathcal{F}_\nu(y)
\nonumber \\
&
	-
	\frac{ (aa')^{- \frac{ (D-4)\epsilon }{2} } }{ 2\nu (1\!-\!\epsilon)^2 H_0^2 }
	\partial_i \partial_j'
	\biggl[
	\frac{ 1 }{2} I\bigl[ \mathcal{F}_\nu(y) \bigr]
	+
	\Bigl( 1 \!-\! \frac{\beta}{\beta_s} \Bigr)
		\frac{\partial}{\partial \nu}
		\mathcal{F}_{\nu+1}(y) 
	\biggr]
	\, .
\label{ijDecelComputed}
\end{align}
%

\subsection{Covariantization}
\label{subsec: Covariantization}

The two-point functions in cosmological spaces can be written in a covariant tensor basis.
This is true regardless of the gauge, as long as the gauge-fixing term respects cosmological 
symmetries.
This basis is constructed from derivatives of the bi-local variables~(\ref{ydef})
and~(\ref{uvDef}),\footnote{There is an additional tensor structure that should be
added in the general case, 
$\bigl[ (\partial_\mu y) (\partial'_\nu u) \!-\! (\partial_\mu u) (\partial'_\nu y) \bigr]$,
that is anti-symmetric under reflection~$x \!\leftrightarrow \! x'$.
While it was necessary to consider this tensor
structure in~\cite{Domazet:2024dil}, here this is redundant,
and if we had added it we would have found that the structure function it multiplies
vanishes.}
\begin{align}
i \bigl[ \tensor*[_\mu^{\scr - \!}]{\Delta}{_\nu^{\scr \! +}} \bigr] (x;x')
	= {}&
	\bigl( \partial_\mu \partial'_\nu y \bigr) \, \mathcal{C}_1 ( y,u )
	+ \bigl( \partial_\mu y \bigr) \bigl( \partial'_\nu y \bigr) \, 
		\mathcal{C}_2 ( y,u )
\nonumber \\
&
	+ \Bigl[ \bigl( \partial_\mu y \bigr) \bigl( \partial'_\nu u \bigr)
		\!+\! \bigl( \partial_\mu u \bigr) \bigl( \partial'_\nu y \bigr) \Bigr] 
			\mathcal{C}_3 ( y,u )
	+ \bigl( \partial_\mu u \bigr) \bigl( \partial'_\nu u \bigr) \, 
		\mathcal{C}_4 ( y,u ) 
		\, .
\label{CovariantForm}
\end{align}
Here and henceforth~$\mathcal{C}_n$ denotes the scalar structure functions of the
covariant representation in either gauge, while~$\mathcal{A}_n$ and~$\mathcal{B}_n$
will denote scalar structure functions in the covariant gauge and in the deceleration gauge,
respectively. Inferring the structure functions for the two gauges is facilitated by 
writing out the components of the covariant representation,
\begin{align}
&
i \bigl[ \tensor*[_{i\,}^{\scr - \!}]{\Delta}{_j^{\scr \! +}} \bigr] (x;x') 
	=
	2 \delta_{ij} (1\!-\!\epsilon)^2 \mathcal{H} \mathcal{H}' \Bigl\{ - \mathcal{C}_1 + I[\mathcal{C}_2] \Bigr\}
	+ \partial_i \partial'_j I^2[\mathcal{C}_2] \, ,
\label{covariant convenient ij}
\\
&
i \bigl[ \tensor*[_{0}^{\scr - \!}]{\Delta}{_i^{\scr \! +}} \bigr] (x;x') 
	=
	(1\!-\!\epsilon) \bigl( \partial'_i y \bigr) \biggl\{
	\mathcal{H}\Bigl[ \mathcal{C}_1 - (2\!-\!y) \mathcal{C}_2 + \mathcal{C}_3 \Bigr]
	+ 2 \mathcal{H}' \mathcal{C}_2
	\biggr\} \, ,
\label{covariant convenient 0i}
\\
&
i \bigl[ \tensor*[_{0}^{\scr - \!}]{\Delta}{_0^{\scr \! +}} \bigr] (x;x') 
	=
	(1\!-\!\epsilon)^2 
	\biggl\{
	2 \bigl( \mathcal{H}^2 \!+\! \mathcal{H}'^2 \bigr) 
	\Bigl[ \mathcal{C}_1 - (2\!-\!y) \mathcal{C}_2 +\mathcal{C}_3 \Bigr]
\nonumber \\
&	\hspace{2.7cm}
	+ \mathcal{H} \mathcal{H}' \Bigl[ 
	- (2\!-\!y) \mathcal{C}_1
	- (4y \!-\! y^2)\mathcal{C}_2 
	+ 8 \mathcal{C}_2
	- 2 (2\!-\!y) \mathcal{C}_3
	+ \mathcal{C}_4 \Bigr]
	\biggr\} \, .
\label{covariant convenient 00}
\end{align}

\bigskip
\noindent {\bf Conformal gauge.}
Comparing expressions~(\ref{covariant convenient ij})--(\ref{covariant convenient 00})
with expressions~(\ref{conformal delta 00})--(\ref{conformal delta ij})
gives the scalar structure functions for the conformal gauge:
\begin{align}
\mathcal{A}_1(y,u) ={}&
	\frac{ e^{-\frac{(D-4)\epsilon}{2(1-\epsilon)} u } }{2(1\!-\!\epsilon)^2 H_0^2 } 
	\times
	\biggl[ - 1 + \frac{(1\!-\!\alpha) }{2} \biggr] \mathcal{F}_\nu(y) 
	\, ,
\label{A1}
\\
\mathcal{A}_2(y,u) ={}&
	\frac{ e^{-\frac{(D-4)\epsilon}{2(1-\epsilon)} u } }{2(1\!-\!\epsilon)^2 H_0^2 } 
	\times
	\frac{(1\!-\!\alpha) }{2} \frac{\partial \mathcal{F}_{\nu}(y) }{\partial y}
	\, ,
\label{A2}
\\
\mathcal{A}_3(y,u) ={}&
	\frac{ e^{-\frac{(D-4)\epsilon}{2(1-\epsilon)} u } }{2(1\!-\!\epsilon)^2 H_0^2 } 
	\times
		\biggl[ 1 - \frac{(1\!-\!\alpha ) }{2} \frac{(D\!-\!4) \epsilon }{2(1\!-\!\epsilon)}  
			\biggr] \mathcal{F}_\nu(y)
	\, ,
\label{A3}
\\
\mathcal{A}_4(y,u) ={}&
	\frac{ e^{-\frac{(D-4)\epsilon}{2(1-\epsilon)} u } }{2(1\!-\!\epsilon)^2 H_0^2 } 
		\times
	\biggl[ 1
		- \frac{(1\!-\!\alpha)}{2} \frac{(D\!-\!4)\epsilon}{2(1\!-\!\epsilon)}
		 \biggr]
		 \Bigl[ (2\!-\!y) \mathcal{F}_\nu(y) - 2 \mathcal{F}_{\nu-1}(y) \Bigr] \, ,
\label{A4}
\end{align}
where the recurrence relations~(\ref{Frecurrence}) 
were used to simplify the last two structure functions.
This is the first main result of this paper. All other two point functions
are inferred from this results by simply changing the~$i\delta$-prescription in
the distance function to the appropriate one, as defined in~(\ref{yPrescriptions}).
In Appendix~\ref{sec: Checks for two-point function} it is shown that this solution
indeed satisfies the appropriate equations of motion and the Ward-Takahashi identity.

\bigskip
\noindent {\bf Deceleration gauge.}
Comparing two-point function 
components~(\ref{00DecelComputed})--(\ref{ijDecelComputed}) with the components of 
the covariantized form~(\ref{covariant convenient ij})--(\ref{covariant convenient 00})
gives the scalar structure functions for the deceleration gauge:
\begin{align}
\mathcal{B}_1(y,u) ={}&
	\frac{ e^{- \frac{(D-4)\epsilon}{2(1-\epsilon)} u }  }{ 2 \nu (1\!-\!\epsilon)^2 H_0^2 } 
	\biggl[ - \Bigl( \nu \!+\! \frac{1}{2} \Bigr) \mathcal{F}_\nu (y) 
		- \Bigl( 1 \!-\! \frac{\beta}{\beta_s} \Bigr)
		\frac{\partial}{\partial y} \frac{\partial }{\partial \nu} \mathcal{F}_{\nu+1}(y)
		\biggr] 
		\, ,
\label{B1}
\\
\mathcal{B}_2(y,u) ={}&
	\frac{ e^{- \frac{(D-4)\epsilon}{2(1-\epsilon)} u }  }{ 2 \nu (1\!-\!\epsilon)^2 H_0^2 } 
	\biggl[ - \frac{1}{2} \frac{\partial}{ \partial y} \mathcal{F}_\nu (y) 
		- \Bigl( 1 \!-\! \frac{\beta}{\beta_s} \Bigr)
		\frac{\partial^2}{\partial y^2} \frac{\partial }{\partial \nu} \mathcal{F}_{\nu+1}(y)
		\biggr] 
		\, ,
\label{B2}
\\
\mathcal{B}_3(y,u) ={}&
	\frac{ e^{- \frac{(D-4)\epsilon}{2(1-\epsilon)} u }  }{ 2 \nu (1\!-\!\epsilon)^2 H_0^2 } 
	\,
	\frac{(D\!-\!4) \epsilon}{ 2(1\!-\!\epsilon) }
	\biggl[ \frac{1}{2} \mathcal{F}_\nu (y)
		+ \Bigl( 1 \!-\! \frac{\beta}{\beta_s} \Bigr)
			\frac{\partial}{\partial y} \frac{\partial }{ \partial \nu } \mathcal{F}_{\nu+1}(y)
			\biggr] 
			\, ,
\label{B3}
\\
\mathcal{B}_4(y,u) ={}&
	\frac{ e^{- \frac{(D-4)\epsilon}{2(1-\epsilon)} u }  }{ 2 \nu (1\!-\!\epsilon)^2 H_0^2 } 
	\,
	\frac{(D\!-\!4) \epsilon}{ 2(1\!-\!\epsilon) }
	\biggl[
	\frac{1}{2} (2\!-\!y) \mathcal{F}_\nu(y) - \mathcal{F}_{\nu+1}(y)
\nonumber \\
&	\hspace{5cm}
		- \Bigl( 1 \!-\! \frac{\beta}{\beta_s} \Bigr)\frac{(D\!-\!4) \epsilon}{ 2(1\!-\!\epsilon) }
			\frac{ \partial }{ \partial \nu } \mathcal{F}_{\nu+1}(y)
	\biggr] \, .
\label{B4}
\end{align}
This is the second main result of this paper, and the other two-point functions are obtained
from here by simply changing the $i\delta$-prescription in distance functions to appropriate
ones. The expressions for the last two structure 
functions were simplified using the equation of motion~(\ref{Feom}) and the recurrence
relations~(\ref{Frecurrence}) for the rescaled propagator function, as well as parametric
derivatives of those expressions~(\ref{FparamEOM}) and~(\ref{FparamRecurrence}).
From the structure functions~(\ref{B1})--(\ref{B4}) it is also clear that the simple 
deceleration gauge~(\ref{SimpleBeta}) indeed is the simplest choice for the gauge-fixing 
parameter in this family of gauges, as all parametric derivatives drop out. Checks that
this solution indeed satisfies appropriate equations of motion and the Ward-Takahashi 
identity are performed in Appendix~\ref{sec: Checks for two-point function}.

\subsection{Alternate forms}

Both gauge choices considered here also admit the following representation
for the two-point function,
\begin{equation}
i \bigl[ \tensor*[_\mu^{\tt a }]{\Delta}{_\nu^{\tt b}} \bigr](x;x')
	= i \bigl[ \tensor*[_\mu^{\tt a }]{\Delta}{_\nu^{\tt b}} \bigr](x;x') 
	\Bigr|_{\xi=\overline{\xi}}
		+ ( \overline{\xi} \!-\!\xi) \partial_\mu \partial_\nu' \Upsilon (y,u)
		\, ,
\label{AlternateForm}
\end{equation}
motivated by the form of the gauge transformation of the vector potential. 
Here~$\overline{\xi}$ is some reference gauge-fixing parameter 
that is chosen arbitrarily. This choice then fixes the bi-scalar in the second term,
that can be found in terms of the structure functions of the two-point functions. 
This is accomplished by first writing out the double derivative acting
on it in terms of the tensor basis from~(\ref{CovariantForm}),
\begin{equation}
\partial_\mu \partial'_\nu \Upsilon
	=
	\bigl( \partial_\mu \partial'_\nu y \bigr) \frac{\partial \Upsilon }{\partial y}
	+
	\bigl( \partial_\mu y \bigr) \bigl( \partial'_\nu y \bigr) 
		\frac{\partial^2 \Upsilon }{\partial y^2}
	+
	\Bigl[
	\bigl( \partial_\mu y \bigr) \bigl( \partial'_\nu u \bigr) 
	\!+\!
	\bigl( \partial_\mu u \bigr) \bigl( \partial'_\nu y \bigr)
	\Bigr]
	\frac{\partial^2 \Upsilon }{\partial u \partial y}
	+
	\bigl( \partial_\mu u \bigr) \bigl( \partial'_\nu u \bigr) \frac{\partial^2 \Upsilon }{\partial u^2}
	\, .
\end{equation}
It is now simplest to infer~$\Upsilon$ by comparing the last term above to fourth 
scalar structure functions in~(\ref{A4}) and~(\ref{B4}). The consistency of this inference
is then checked by comparing the other three scalar structure functions with the respective
terms in the expression above.

\bigskip

\noindent{\bf Conformal gauge.} 
The simplest member of the family of conformal
gauges is~$\alpha\!=\!1$, which may be called the {\it simple conformal gauge}.
Applying the analysis described above produces the following expression,
\begin{equation}
i \bigl[ \tensor*[_\mu^{\tt a}]{\Delta}{_\nu^{\tt b}} \bigr](x;x')
	=
	i \bigl[ \tensor*[_\mu^{\tt a}]{\Delta}{_\nu^{\tt b}} \bigr](x;x') \Bigr|_{\alpha=1}
	-
	\frac{(1\!-\!\alpha)}{2} \partial_\mu \partial'_\nu 
	\biggl\{
	\frac{ e^{- \frac{(D-4)\epsilon}{ 2(1-\epsilon) } u } }{2(1\!-\!\epsilon)^2 H_0^2 }
	\!\times\!
	\frac{2(1\!-\!\epsilon)}{(D\!-\!4)\epsilon}
	\Bigl[ (2\!-\!y) \mathcal{F}_\nu - 2 \mathcal{F}_{\nu-1} \Bigr]
	\biggr\}
	\, .
\label{ConformalAlternateForm1}
\end{equation}
Checking that this form indeed matches the tensor basis 
representation~(\ref{CovariantForm}) with the structure functions in~(\ref{A1})--(\ref{A4})
requires using recurrence relations~(\ref{Frecurrence}).
Despite being correct, this representation is cumbersome because of 
factors~$(D\!-\!4)$ and~$\epsilon$ in the denominator, that make the de Sitter limit
and the flat space limit subtle. 
A more convenient representation can be derived by appealing to 
integrated versions of recurrence relations~(\ref{Frecurrence}), 
\begin{align}
2 \mathcal{F}_{\lambda-1} 
	={}&
	(2\!-\!y) \mathcal{F}_\lambda
	- \Bigl( \frac{D\!-\!3}{2} \!-\! \lambda \Bigr) I \bigl[ \mathcal{F}_\lambda \bigr]
	+
	{\tt const.}
	\, ,
\\
2 \mathcal{F}_{\lambda+1} 
	={}&
	(2\!-\!y) \mathcal{F}_\lambda
	- \Bigl( \frac{D\!-\!3}{2} \!+\! \lambda \Bigr) I \bigl[ \mathcal{F}_\lambda \bigr]
	+
	{\tt const.}
	\, ,
\end{align}
that are determined up to integration constants. From these it follows that
\begin{equation}
(2\!-\!y) \mathcal{F}_\lambda - 2 \mathcal{F}_{\lambda-1}
	=
	\frac{\frac{D-3}{2} - \lambda}{\lambda} 
	\bigl( \mathcal{F}_{\lambda-1} - \mathcal{F}_{\lambda+1} \bigr)
	\, ,
\end{equation}
where the constant of integration was fixed using the power-series representation~(\ref{Fseries}).
This allows to write the representation~(\ref{ConformalAlternateForm1}) 
in a more convenient form,
\begin{equation}
i \bigl[ \tensor*[_\mu^{\tt a}]{\Delta}{_\nu^{\tt b}} \bigr](x;x')
	=
	i \bigl[ \tensor*[_\mu^{\tt a}]{\Delta}{_\nu^{\tt b}} \bigr](x;x') \Bigr|_{\alpha=1}
	+
	\frac{(1\!-\!\alpha)}{2} \partial_\mu \partial'_\nu 
	\biggl[
	\frac{ e^{- \frac{(D-4)\epsilon}{ 2(1-\epsilon) } u } }{2\nu(1\!-\!\epsilon)^2 H_0^2 }
	\bigl( \mathcal{F}_{\nu-1} - \mathcal{F}_{\nu+1} \bigr)
	\biggr]
	\, .
\label{ConformalAlternateForm2}
\end{equation}
It is worth noting that the simple conformal gauge propagator retains a remarkably 
simple form that it has in the de Sitter limit,
\begin{equation}
i \bigl[ \tensor*[_\mu^{\tt a}]{\Delta}{_\nu^{\tt b}} \bigr](x;x') \Bigr|_{\alpha=1}
	=
	aa' \Bigl[
	\bigl( \eta_{\mu\nu} \!+\! \delta_\mu^0 \delta_\nu^0 \bigr) i \Delta_{\nu}(y,u)
	-
	\delta_\mu^0 \delta_\nu^0
	i \Delta_{\nu-1}(y,u)
	\Bigr]
	\, ,
\end{equation}
where the scalar two-point functions are replaced by their power-law inflation 
counterparts defined in~(\ref{DeltaEvaluation}).

\bigskip

\noindent{\bf Deceleration gauge.}
It is natural to adopt the simple deceleration gauge~(\ref{SimpleBeta}) as the
reference gauge for the alternate form of the deceleration gauge two-point functions,
that is given by
\begin{equation}
i \bigl[ \tensor*[_\mu^{\tt a}]{\Delta}{_\nu^{\tt b}} \bigr](x;x')
	=
	i \bigl[ \tensor*[_\mu^{\tt a}]{\Delta}{_\nu^{\tt b}} \bigr](x;x') \Bigr|_{\beta = \beta_s}
	-
	(\beta_s \!-\! \beta) \partial_\mu \partial'_\nu
	\biggl[
	\frac{ e^{-\frac{(D-4)\epsilon }{ 2(1-\epsilon) } u } }{ 2(\nu\!+\!1) (1\!-\!\epsilon)^2 H_0^2 }
		\frac{\partial \mathcal{F}_{\nu+1} }{ \partial \nu }
	\biggr]
	\, .
\label{DecelerationAlternateForm}
\end{equation}
%

\section{Various limits}
\label{sec: Various limits}

In order to establish connection with the existing literature and known results it is useful
to examine several available limits that are considered in this section.

\subsection{De Sitter limit}
\label{subsec: De Sitter limit}

The de Sitter limit is obtained by taking the limit~$\epsilon\!\to\!0$ in which 
the Hubble parameter is constant~$H\!=\!H_0$. The tensor structure functions
retain their form, with the bi-scalars defined in~(\ref{ydef}) and~(\ref{uvDef}) 
taking their de Sitter limits, and the index defined in~(\ref{nu def}) takes a simpler value,
\begin{equation}
\nu \xrightarrow{ \epsilon \to 0 } \nu_0 = \frac{D\!-\!3}{2} \, .
\end{equation}

\medskip
\noindent{\bf Conformal gauge.} 
Taking the limit~$\epsilon\!\to\!0$
of the scalar structure functions~(\ref{A1})--(\ref{A2}) for the conformal gauge
gives,
\begin{align}
\mathcal{A}_1 \xrightarrow{\epsilon \to 0}{}&
	\frac{ 1 }{2 H_0^2 } 
	\biggl[ - 1 + \frac{(1\!-\!\alpha) }{2} \biggr] \mathcal{F}_{\nu_0}(y) 
	\, ,
\\
\mathcal{A}_2 \xrightarrow{\epsilon \to 0}{}&
	\frac{ 1 }{2 H_0^2 } 
	\frac{(1\!-\!\alpha) }{2} \frac{\partial \mathcal{F}_{\nu_0}(y) }{\partial y}
	\, ,
\\
\mathcal{A}_3 \xrightarrow{\epsilon \to 0}{}&
	\frac{ 1 }{2 H_0^2 } \mathcal{F}_{\nu_0}(y)
	\, ,
\\
\mathcal{A}_4 \xrightarrow{\epsilon \to 0}{}&
	\frac{ 1 }{2 H_0^2 } 
		 \Bigl[ (2\!-\!y) \mathcal{F}_{\nu_0}(y) - 2 \mathcal{F}_{\nu_0-1}(y) \Bigr] 
	\, ,
\end{align}
where the last one is recognized to be a constant using the 
power-series representation~(\ref{Fseries}).
For the special case~$\alpha\!=\!1$ it attains its simplest form
and reproduces exactly the simple gauge photon propagator 
in de Sitter~\cite{Woodard:2004ut}.
It is curious that the latter two structure functions do not depend on the
gauge-fixing parameter in the de Sitter limit.

\bigskip

\noindent{\bf Deceleration gauge.} 
In the de Sitter limit this gauge reduces to the general covariant gauge,
and the simple deceleration gauge becomes the simple covariant 
gauge,\footnote{Note that in~$D\!=\!3$ spacetime dimensions the simple covariant
gauge does not exsit since the gauge-fixing parameter~(\ref{BetaS0})
diverges,~$\beta_s^0\!\xrightarrow{D \to 3} \! \infty$, and eliminates the
gauge-fixing functional~(\ref{gauges}).}
\begin{equation}
\beta_s \xrightarrow{\epsilon\to0} \frac{\nu_0 \! +\! 1}{ \nu_0}
	= \frac{D \!-\! 1}{D \!-\! 3} = \beta_s^0 \, .
\label{BetaS0}
\end{equation}
Consequently, the two-point function must reduce to the corresponding 
general covariant gauge two-point function in de Sitter~\cite{Glavan:2022dwb}.
For the first three structure functions~(\ref{B1})--(\ref{B3}) this is 
obviously so,
\begin{align}
\mathcal{B}_1
	\xrightarrow{\epsilon\to0}{}&
	\frac{ 1  }{ 2 \nu_0 H_0^2 } 
	\biggl[ - \Bigl( \nu_0 \!+\! \frac{1}{2} \Bigr) \mathcal{F}_{\nu_0} (y) 
		- \Bigl( 1 \!-\! \frac{\beta}{\beta_s^0} \Bigr)
		\frac{\partial}{\partial y} \frac{\partial }{\partial \nu_0} \mathcal{F}_{\nu_0+1}(y)
		\biggr] 
		\, ,
\\
\mathcal{B}_2
	\xrightarrow{\epsilon\to0}{}&
	\frac{ 1 }{ 2 \nu_0 H_0^2 } 
	\biggl[ - \frac{1}{2} \frac{\partial}{ \partial y} \mathcal{F}_{\nu_0} (y) 
		- \Bigl( 1 \!-\! \frac{\beta}{\beta_s^0} \Bigr)
		\frac{\partial^2}{\partial y^2} \frac{\partial }{\partial \nu_0} \mathcal{F}_{\nu_0+1}(y)
		\biggr] 
		\, ,
\\
\mathcal{B}_3
	\xrightarrow{\epsilon\to0}{}&
	0
	\, ,
\end{align}
while for the fourth structure function the limit should be taken carefully,
in order not to overlook an important de Sitter breaking contribution.
Indeed, by appealing to special limits~(\ref{SpecialLimit1}) and~(\ref{SpecialLimit2})
a nonvanishing limit for the fourth structure function is obtained,
\begin{equation}
\mathcal{B}_4
	\xrightarrow{\epsilon\to0}
	\beta \times \frac{ H^{D-4} }{ (4\pi)^{\frac{D}{2}} }
	\frac{ \Gamma(D\!-\!1) }{ (D\!-\!1) \, \Gamma\bigl( \frac{D}{2} \bigr) }
	\, .
\end{equation}
This matches the result in~\cite{Glavan:2022dwb,Glavan:2022nrd},
that consists of the de Sitter invariant part worked out in~\cite{Allen:1985wd,Tsamis:2006gj,Garidi:2006ey,Youssef:2010dw,Frob:2013qsa},
and the additional nonvanishing de Sitter breaking fourth structure function 
for~$\beta \!\neq\!0$.

\subsection{Flat space limit}
\label{subsec: Flat space limit}

In the limit of flat space~$H_0\!\to\!0$, both gauge-fixing terms reduce to the Lorentz
invariant one, and consequently both propagators must reduce to the same Lorentz
invariant propagator.
The tensor structures from~(\ref{CovariantForm}) in this limit reduce to:
\begin{align}
\bigl( \partial_\mu \partial_\nu' y \bigr)
	\ \overset{H_0 \to 0}{\longsim} \ {}&
	\! - 2 (1\!-\!\epsilon)^2 H_0^2 \eta_{\mu\nu} 
	\, ,
\\
\bigl( \partial_\mu y  \bigr) \bigl( \partial_\nu' y \bigr)
	\ \overset{H_0 \to 0}{\longsim} \ {}&
	\! -4 (1\!-\!\epsilon)^4 H_0^4 \Delta x_\mu \Delta x_\nu
	\, ,
\\
\Bigl[ \bigl( \partial_\mu y \bigr) \bigl( \partial_\nu' u \bigr)
	\!+\! \bigl( \partial_\mu u  \bigr) \bigl( \partial_\nu' y \bigr) \Bigr]
	\ \overset{H_0 \to 0}{\longsim} \ {}&
	2 (1\!-\!\epsilon)^3 H_0^3 \bigl( \Delta x_\mu \delta_\nu^0
		- \delta_\mu^0 \Delta x_\nu \bigr) \, ,
\\
\bigl( \partial_\mu u \bigr) \bigl( \partial_\nu' u \bigr)
	\ \overset{H_0 \to 0}{\longsim} \ {}&
	(1\!-\!\epsilon)^2 H_0^2 \delta_\mu^0 \delta_\nu^0 \, .
\end{align}
The reduced propagator functions reduce to flat space scalar propagators,
\begin{equation}
\mathcal{F}_\lambda(y)
	\xrightarrow{H_0 \to 0}
	\frac{ \Gamma\bigl( \frac{D-2}{2} \bigr) }
		{ 4 \pi^{ \frac{D}{2} } \bigl( \Delta x^2 \bigr)^{ \! \frac{D-2}{2} } }
		\equiv
		i \Delta ( x\!-\!x' )
	\, ,
\label{FlatSpaceFunction}
\end{equation}
derivatives with respect to distance functions become,
\begin{equation}
\frac{\partial}{\partial y} 
	\ \overset{H_0 \to 0}{ \longsim} \
	\frac{1 }{ (1\!-\!\epsilon)^2 H_0^2} \frac{ \partial }{ \partial (\Delta x^2)}
	\, ,
\end{equation}
and parametric derivatives give,
\begin{equation}
\frac{\partial}{\partial\lambda} \mathcal{F}_\lambda(y)
	\ \overset{H_0 \to 0}{ \longsim} \
	\frac{\lambda(1\!-\!\epsilon)^2 H_0^2 \Delta x^2}{ D \!-\! 4 } i \Delta ( x\!-\!x' )
	\, ,
\qquad
\frac{\partial}{\partial y} \frac{\partial}{\partial\lambda} \mathcal{F}_\lambda(y)
	\ \overset{H_0 \to 0}{ \longsim} 
	- \frac{\lambda}{ 2 } i \Delta ( x\!-\!x' )
	\, .
\end{equation}
For the conformal gauge the flat space limits of the structure functions are:
\begin{align}
\mathcal{A}_1(y,u) 
	\ \overset{H_0 \to 0}{ \longsim} \
		{}&
	\frac{ 1 }{2(1\!-\!\epsilon)^2 H_0^2 } 
	\biggl[ - 1 + \frac{(1\!-\!\alpha) }{2} \biggr] 
	i \Delta ( x\!-\!x' )
	\, ,
\\
\mathcal{A}_2(y,u) 
	\ \overset{H_0 \to 0}{ \longsim} \
		{}&
	\frac{ 1 }{2(1\!-\!\epsilon)^4 H_0^4 } 
	\biggl[
	- \frac{(1\!-\!\alpha) (D\!-\!2) }{4 \Delta x^2 } 
	\biggr]
	i \Delta ( x\!-\!x' )
	\, ,
\\
\mathcal{A}_3(y,u)
	\ \overset{H_0 \to 0}{ \longsim} \
		{}&
	\frac{ 1 }{2(1\!-\!\epsilon)^2 H_0^2 } 
		\biggl[ 1 - \frac{(1\!-\!\alpha ) }{2} \frac{(D\!-\!4) \epsilon }{2(1\!-\!\epsilon)}  
			\biggr] i \Delta ( x\!-\!x' )
	\, ,
\\
\mathcal{A}_4(y,u)
	\ \overset{H_0 \to 0}{ \longsim} \
		{}& 
		\frac{ 1 }{2(1\!-\!\epsilon)^2 H_0^2 } \times 0 \, ,
\end{align}
while for the deceleration gauge they reduce to:
\begin{align}
\mathcal{B}_1(y,u)
	\ \overset{H_0 \to 0}{ \longsim} \ {}&
	\frac{ 1 }{ 2 (1\!-\!\epsilon)^2 H_0^2 } 
	\biggl[ -  1 + \frac{(1 \!-\! \beta)}{2} \biggr] 
		i \Delta ( x\!-\!x' )
		\, ,
\\
\mathcal{B}_2(y,u)
	\ \overset{H_0 \to 0}{ \longsim} \ {}&
	\frac{ 1 }{ 2 (1\!-\!\epsilon)^4 H_0^4 } 
	\biggl[ 
		- \frac{(1\!-\!\beta)(D\!-\!2) }{4 \Delta x^2}
		\biggr] 
		i \Delta ( x\!-\!x' )
		\, ,
\\
\mathcal{B}_3(y,u)
	\ \overset{H_0 \to 0}{ \longsim} \ {}&
	\frac{ 1 }{ 2 (1\!-\!\epsilon)^2 H_0^2 } 
	\biggl[
		- \frac{ (1\!-\!\beta)}{2} \frac{(D\!-\!4) \epsilon}{ 2(1\!-\!\epsilon) }
			\biggr] i \Delta ( x\!-\!x' )
			\, ,
\\
\mathcal{B}_4(y,u)
	\ \overset{H_0 \to 0}{ \longsim} \ {}&
	\frac{ 1 }{ 2 (1\!-\!\epsilon)^2 H_0^2 } 
	\times 0
	\, .
\end{align}
It follows that both gauges produce the same flat space limit,
\begin{equation}
i \bigl[ \tensor*[_\mu^{\tt a \!}]{\Delta}{_\nu^{\tt \! b}} \bigr](x;x')
	\xrightarrow{H_0 \to 0}
	\biggl[
	\eta_{\mu\nu}
	-
	\frac{(1\!-\!\xi)}{2}
	\biggl(
	\eta_{\mu\nu}
	-
	(D\!-\!2)
	\frac{ \Delta x_\mu \Delta x_\nu }{ \Delta x^2 }
	\biggr)
	\biggr] 
	i \Delta ( x\!-\!x' )
	\, ,
\label{FlatSpaceLimit}
\end{equation}
where~$\xi\!=\!\alpha$ for the conformal gauge, and~$\xi\!=\!\beta$ for the deceleration
gauge.

\subsection{Four-dimensional limit}
\label{subsec: Four-dimensional limit}

In four spacetime dimensions the index~(\ref{nu def}) reduces to the conformal value,
\begin{equation}
\nu \xrightarrow{D\to4} \frac{1}{2} \, .
\end{equation}
Therefore, some of the rescaled propagator functions reduce to 
the four-dimensional limit of rescaled flat space two-point 
function~(\ref{FlatSpaceFunction}),
\begin{equation}
\mathcal{F}_{\nu}(y) , \mathcal{F}_{\nu-1}(y) 
	\xrightarrow{D\to4}
	\frac{(1\!-\!\epsilon)^2 H_0^2}{ 4 \pi^2 y}
	=
	\frac{ H_0^2}{ \mathcal{H} \mathcal{H}' }
		\frac{ 1}{ 4\pi^2 \Delta x^2 } 
	\, ,
\label{FflatLimit}
\end{equation}
which is easily inferred from the power-series representation~(\ref{Fseries}).

\bigskip

\noindent {\bf Conformal gauge.}
The limits in~(\ref{FflatLimit}) are sufficient to compute the four-dimensional limit
for the conformal gauge. The structure functions~(\ref{A1})--(\ref{A4}) reduce to
\begin{equation}
\mathcal{A}_1 \xrightarrow{D \to 4}
	- 
	\frac{(1\!+\!\alpha) }{16 \pi^2 y}
	\, ,
\qquad
\mathcal{A}_2 \xrightarrow{D \to 4}
	-
	\frac{(1\!-\!\alpha) }{16 \pi^2 y^2}
	\, ,
\qquad
\mathcal{A}_3 \xrightarrow{D \to 4}
	\frac{ 1 }{ 8\pi^2 y  } 
	\, ,
\qquad
\mathcal{A}_4 \xrightarrow{D \to 4}
	- \frac{ 1}{ 8\pi^2 } 
	\, .
\end{equation}
Plugging these into the covariantized representation~(\ref{CovariantForm}),
and acting the derivatives in the basis tensors 
explicitly~\footnote{
Acting explicitly derivatives in the tensor structures in~(\ref{CovariantForm}) 
produces the following expressions:
\begin{align*}
&
\bigl( \partial_\mu \partial'_\nu y \bigr)
	=
	(1\!-\!\epsilon)^2 \mathcal{H} \mathcal{H}' 
	\Bigl[
	-
	2 \eta_{\mu\nu}
	\!+\!
	2 (1\!-\!\epsilon) \bigl(
		\mathcal{H}' \delta_\nu^0 \Delta x_\mu
		\!-\!
		\mathcal{H} \delta_\mu^0 \Delta x_\nu \bigr)
	\!+\!
	\delta_\mu^0 \delta_\nu^0 y
	\Bigr]
	\, ,
\qquad
\bigl( \partial_\mu \partial'_\nu u \bigr)
	= (1\!-\!\epsilon)^2 \mathcal{H} \mathcal{H}' \delta_\mu^0 \delta_\nu^0
	\, ,
\\
&
\bigl( \partial_\mu y \bigr) \bigl( \partial'_\nu y \bigr)
	=
	(1\!-\!\epsilon)^2 \mathcal{H} \mathcal{H}'
	\Bigl[ \delta^0_\mu \delta^0_\nu y^2
		+ 
		2 (1\!-\!\epsilon) y \bigl( \mathcal{H}' \delta^0_\nu \Delta x_\mu
			\!-\! \mathcal{H} \delta^0_\mu \Delta x_\nu \bigr)
		-
		4(1\!-\!\epsilon)^2 \mathcal{H}\mathcal{H}' \Delta x_\mu \Delta x_\nu \Bigr]
		\, ,
\\
&
\Bigl[ \bigl( \partial_\mu y \bigr) \bigl( \partial'_\nu u \bigr)
		+ \bigl( \partial_\mu u \bigr) \bigl( \partial'_\nu y \bigr) \Bigr]
	=
	2 (1\!-\!\epsilon)^2 \mathcal{H} \mathcal{H}'
	\Bigl[ \delta^0_\mu \delta_\nu^0 y
		+ (1\!-\!\epsilon) \bigl( \mathcal{H}' \delta_\nu^0 \Delta x_\mu
		\!-\!
		\mathcal{H} \delta_\mu^0 \Delta x_\nu \bigr)
		\Bigr]
		\, .
\end{align*}
}
reveals that the four-dimensional limit in the conformal gauge,
\begin{equation}
i \bigl[ \tensor*[_\mu^{\tt a \!}]{\Delta}{_\nu^{\tt \! b}} \bigr](x;x')
	\xrightarrow{D \to 4}
	\biggl[
	\eta_{\mu\nu}
	-
	\frac{(1\!-\!\alpha)}{2}
	\biggl(
	\eta_{\mu\nu}
	-
	2 
		\frac{ \Delta x_\mu \Delta x_\nu }{ \Delta x^2 } 
	\biggr)
	\biggr]
	\frac{ 1}{ 4\pi^2 \Delta x^2 } 
	\, ,
\label{FlatSapceLimit}
\end{equation}
corresponds to the flat space limit~(\ref{FlatSpaceLimit}) in four dimensions.
The reason behind this is that the gauge-fixing functional becomes exactly 
conformally invariant in~$D\!=\!4$, so that the quantization can proceed
just as in flat space~\cite{Huguet:2013dia}.

\bigskip

\noindent {\bf Deceleration gauge.}
Apart from the four-dimensional limit of the rescaled propagator functions 
in~(\ref{FflatLimit}), further special identities following 
from~(\ref{SpecialLimit1}) and~(\ref{SpecialLimit2}) are needed,
\begin{equation}
\frac{ (D \!-\! 4) \epsilon }{2(1 \!-\! \epsilon)} \mathcal{F}_{\nu+1}
	\xrightarrow{ D \to 4 }
	-
	\frac{ (1\!-\!\epsilon)^2 H_0^2 }{ 8\pi^2 }
	\, ,
\qquad\quad
\Bigl[ \frac{ (D\!-\!4) \epsilon }{2(1\!-\!\epsilon)} \Bigr]^{\!2}  
	\frac{\partial \mathcal{F}_{\nu+1} }{\partial \nu}
	\xrightarrow{ D \to 4 }
	\frac{ (1\!-\!\epsilon)^2 H_0^2 }{ 8 \pi^2 }
	\, ,
\end{equation}
and one that is derived using the series representation~(\ref{Fseries}),
\begin{equation}
\frac{\partial}{\partial y}
\frac{\partial \mathcal{F}_{\nu+1}  }{\partial \nu }
	\xrightarrow{D \to 4}
	\frac{ (1\!-\!\epsilon)^2 H_0^2  }{ 8 \pi^2 }
	\biggl[
	\frac{ 6 \!-\! y}{(4\!-\!y)^2} \ln\Bigl( \frac{y}{4} \Bigr)
	-
	\frac{2(3 \!-\! y)}{4y \!-\! y^2}
	\biggr]
	\, .
\end{equation}
Using these, and the fact that the simple deceleration gauge reduces 
to~$\beta_s \! \xrightarrow{D\to4}\! 3$, the four structure functions are computed to be:
\begin{align}
&
\mathcal{B}_1 
	\xrightarrow{D\to4}
	\frac{1}{4\pi^2 }
	\biggl\{
		- 
		\frac{ 1}{ y } 
		- 
		\Bigl( 1 \!-\! \frac{\beta}{3} \Bigr)
	\biggl[
	\frac{ 6 \!-\! y }{ 2 (4\!-\!y)^2} \ln\Bigl( \frac{y}{4} \Bigr)
	-
	\frac{3 \!-\! y}{4y \!-\! y^2}
	\biggr]
		\biggr\}
		\, ,
\\
&
\mathcal{B}_2
	\xrightarrow{D\to4}
	\frac{1}{4\pi^2 }
	\biggl\{
		\frac{1}{2y^2}
		- 
		\Bigl( 1 \!-\! \frac{\beta}{3} \Bigr)
		\frac{\partial}{\partial y}
	\biggl[
	\frac{ 6 \!-\! y}{ 2 (4\!-\!y)^2} \ln\Bigl( \frac{y}{4} \Bigr)
	-
	\frac{3 \!-\! y}{4y \!-\! y^2}
	\biggr]
		\biggr\}
		\, ,
\\
&
\mathcal{B}_3
	\xrightarrow{D\to4}
	0
	\, ,
\\
&
\mathcal{B}_4
	\xrightarrow{D\to4}
		\frac{ \beta }{ 24\pi^2 }
		\, .
\end{align}
The alternate form~(\ref{DecelerationAlternateForm}) in~$D\!=\!4$ reads,
\begin{equation}
i \bigl[ \tensor*[_\mu^{\scr - \!}]{\Delta}{_\nu^{\scr \!+}} \bigr](x;x')
	\xrightarrow{D\to4} 
	i \bigl[ \tensor*[_\mu^{\scr - \!}]{\Delta}{_\nu^{\scr \!+}} \bigr](x;x') \Bigr|_{\beta = 3}
	+
	\frac{ (3 \!-\! \beta ) }{ 24 \pi^2 }
	\partial_\mu \partial'_\nu 
	\biggl[
	\frac{ 2(3 \!-\! y) }{4\!-\!y} \ln\Bigl( \frac{y}{4} \Bigr)
	-
	{\rm Li}_2\Bigl( 1 \!-\! \frac{y}{4} \Bigr)
	-
	\frac{u^2}{2}
	\biggr]
	\, ,
\end{equation}
where~${\rm Li}_2$ is the dilogarithm function. In the flat space limit~$H_0\!\to\!0$
this expression correctly reduces to~(\ref{FlatSapceLimit}).

\section{Simple observables}
\label{sec: Simple observables}

Computing simple observables provides another way of checking the 
two-point functions worked out in Sec.~\ref{sec: Two-point functions},
besides examining special limits worked out in the preceding section.
The first of the two simple observables considered here is the tree-level correlator 
of the vector field strength, expressible in terms of derivatives acting on the two-point 
function,
\begin{equation}
\bigl\langle \Omega \bigr| 
	\hat{F}_{\mu\nu}(x) \, \hat{F}_{\rho\sigma}(x') \bigl| \Omega \bigr\rangle
	=
	4 \bigl( \delta^\alpha_{[\mu} \partial_{\nu]} \bigr) \bigl( \delta^\beta_{[\rho} \partial'_{\sigma]} \bigr)
	i \bigl[ \tensor*[_\alpha^{\scr - \!}]{\Delta}{_\beta^{\scr \! +}} \bigr](x;x') 
	\, .
\label{F correlator def}
\end{equation}
It is clear that the dependence on the arbitrary gauge-fixing parameters drops out 
from this correlator when two-point functions written in
forms~(\ref{ConformalAlternateForm2}) and~(\ref{DecelerationAlternateForm}) are used.
Thus, only parts corresponding to the simple conformal gauge and the simple 
deceleration gauge contribute.
The correlator~(\ref{F correlator def}) can be expanded in an appropriate tensor basis,
\begin{align}
\MoveEqLeft[3]
\bigl\langle \Omega \bigr| \hat{F}_{\mu\nu}(x) 
	\hat{F}_{\rho\sigma}(x') \bigl| \Omega \bigr\rangle =
	\bigl( \partial_{\mu} \partial'_{[\rho} y \bigr) 
		\bigl( \partial'_{\sigma]} \partial_\nu y \bigr) \,
		\mathcal{G}_1(y,u)
	+ 
	\bigl( \partial_{[\mu} y \bigr) \bigl( \partial_{\nu]} \partial'_{[\sigma} y \bigr) 
			\bigl( \partial'_{\rho]} y \bigr) \, 
			\mathcal{G}_2(y,u)
\nonumber \\
&
+ \Bigl[ \bigl( \partial_{[\mu} y \bigr) \bigl( \partial_{\nu]} \partial'_{[\sigma} y \bigr) 
			\bigl( \partial'_{\rho]} u \bigr)
		+ \bigl( \partial_{[\mu} u \bigr) \bigl( \partial_{\nu]} \partial'_{[\sigma} y \bigr) 
			\bigl( \partial'_{\rho]} y \bigr) \Bigr] \, \mathcal{G}_3(y,u)
\nonumber \\
&
+ \bigl( \partial_{[\mu} u \bigr) \bigl( \partial_{\nu]} \partial'_{[\sigma} y \bigr)
		\bigl( \partial_{\rho]}' u \bigr) \, \mathcal{G}_4(y,u)
+ \bigl( \partial_{[\mu} y \bigr) \bigl( \partial_{\nu]} u \bigr)
		\bigl( \partial'_{[\rho} y \bigr) \bigl( \partial'_{\sigma]} u \bigr)
		\, \mathcal{G}_5(y,u) \, ,
\label{F basis expansion}
\end{align}
where the structure functions of this expansion
are related to the structure functions of the two-point functions in~(\ref{CovariantForm}),
\begin{subequations}
\begin{align}
&
\mathcal{G}_1 =
	4 \Bigl( \frac{\partial \mathcal{C}_1}{\partial y} 
	- \mathcal{C}_2 \Bigr) \, ,
\qquad \quad
\mathcal{G}_2 =
	\frac{\partial \mathcal{G}_1}{\partial y} \, ,
\qquad \quad
\mathcal{G}_3 =
	\frac{\partial \mathcal{G}_1}{\partial u} \, ,
\\
&
\mathcal{G}_4 =
	4 \Bigl( \frac{\partial^2 \mathcal{C}_1}{\partial u^2} 
	- 2 \frac{\partial \mathcal{C}_3}{\partial u}
	+ \frac{\partial \mathcal{C}_4}{\partial y} \Bigr) \, ,
\qquad \quad
\mathcal{G}_5 
	= - \frac{\partial^2 \mathcal{G}_1}{\partial u^2}
		+ \frac{\partial \mathcal{G}_4}{\partial y} \, .
\end{align}
\end{subequations}
Plugging in the scalar structure functions
in the conformal gauge~(\ref{A1})--(\ref{A4}), or in the
deceleration gauge~(\ref{B1})--(\ref{B4}) produces the same answer
for the two independent structure functions,
\begin{equation}
\mathcal{G}_1 =
	\frac{2 \, e^{- \frac{(D-4)\epsilon}{2(1-\epsilon)} u } }{ (1\!-\!\epsilon)^2 H_0^2 } 
		\biggl[ - \frac{\partial}{\partial y} \mathcal{F}_\nu \biggr]
		\, ,
\qquad
\mathcal{G}_4 =
	\frac{2 \, e^{- \frac{(D-4)\epsilon}{2(1-\epsilon)} u } }{ (1\!-\!\epsilon)^2 H_0^2 } 
	\biggl[ \frac{(D \!-\! 4)\epsilon}{2(1 \!-\! \epsilon)} 
	\biggl( 1-\frac{(D \!-\! 4)\epsilon}{2(1 \!-\! \epsilon)} \biggr)
	\mathcal{F}_\nu \biggr]
	\, ,
\label{Gs}
\end{equation}
and consequently for the remaining three.
These are precisely the structure functions found in~\cite{Domazet:2024dil},
computed using the two-point function in the simple covariant gauge.
Therefore, it follows immediately that the~$D\!\to\!4$ limit of the vector field correlator 
reproduces the flat space result,
\begin{equation}
\bigl\langle \Omega \bigr| \hat{F}_{\mu\nu}(x) \, \hat{F}_{\rho\sigma}(x') \bigl| \Omega \bigr\rangle
	\xrightarrow{D\to4}
	\frac{2}{\pi^2 (\Delta x^2)^2} \biggl[
		\eta_{\mu [\rho} \eta_{\sigma] \nu}
			- 4 \eta_{\alpha[\mu} \eta_{\nu][\sigma} \eta_{\rho] \beta}
				\frac{ \Delta x^\alpha \Delta x^\beta }{ \Delta x^2}
		\biggr] \, ,
\label{FFcoincident}
\end{equation}
as it should since the gauge-invariant photon is conformally coupled to gravity in
four spacetime dimensions. 

The other simple observable is the one-loop energy-momentum tensor,
\begin{equation}
\bigl\langle \Omega \bigr| \hat{T}_{\mu\nu}(x) \bigl| \Omega \bigr\rangle
	= \Bigl( \delta_{(\mu}^\rho \delta_{\nu)}^\sigma
		\!-\! \frac{1}{4} g_{\mu\nu} g^{\rho\sigma} \Bigr)
		g^{\alpha\beta}
		\bigl\langle \Omega \bigr| \hat{F}_{\rho\alpha}(x) 
			\hat{F}_{\sigma\beta}(x) \bigl| \Omega \bigr\rangle \, .
\end{equation}
The energy-momentum tensor above is defined as a variation of the gauge-invariant
Maxwell action~(\ref{CovariantAction}), but it can also be defined as the variation of the 
gauge-fixed action~(\ref{GaugeFixedAction}).  The difference between the two is immaterial 
at the expectation value level, as the difference is guaranteed to vanish 
identically~\cite{Glavan:2022pmk} 
(see also~\cite{Belokogne:2015etf,Belokogne:2016dvd} for the case of the Stueckelberg 
field). Since the structure functions~(\ref{Gs}) match the ones obtained 
in~\cite{Domazet:2024dil}, the energy-momentum tensor also matches,
\begin{equation}
\bigl\langle \Omega \bigr| \hat{T}_{\mu\nu}(x) \bigl| \Omega \bigr\rangle
	=
	0
	\, ,
\end{equation}
consistent with the computation in the simple covariant gauge~\cite{Domazet:2024dil}.
It should be noted that the conformal anomaly~(e.g.~\cite{Brown:1977pq}) 
does not appear when computing just this diagram. The entire one-loop effective action 
should be renormalized for the finite conformal anomaly to appear in the 
energy-momentum tensor.

\section{Discussion}
\label{sec: Discussion}

Proton propagators have been constructed for power-law inflation in two different
one-parameter families of noncovariant gauges~(\ref{gauges}), using canonical 
quantization methods~\cite{Glavan:2022pmk}. This is a considerable 
improvement compared to much more complicated propagator in the simple covariant gauge 
worked out previously in~\cite{Domazet:2024dil}. The two new propagators worked out 
here are far more tractable, as they are expressed in terms of scalar propagators and their 
derivatives, multiplying simple tensor structures. This makes practically feasible to explore
the dynamical symmetry breaking effects found for scalar electrodynamics in de Sitter 
space~\cite{Prokopec:2007ak}, in more realistic inflationary spacetimes
with the non-vanishing principal slow-roll parameter. It also allows to explore the vicinity
of the symmetric phase of the Abelian Higgs model~\cite{Glavan:2023lvw}, for which the 
unitary gauge propagator~\cite{Glavan:2020zne} is not appropriate. The presence of 
arbitrary gauge-fixing parameters provides another useful feature of being able to test for
gauge dependence of computed observables.

The biggest technical simplification compared to the general covariant 
gauge~\cite{Domazet:2024dil} 
comes from dispensing with the need to explicitly evaluate the inverse Laplace operator 
acting on a scalar two-point function. This requirement was used to identify 
discreet choices for the second gauge-fixing parameter~$\zeta$
that lead to simple solutions for mode functions. Two of these choices, dubbed the
conformal gauge~($\zeta\!=\!1$), and the deceleration gauge ($\zeta\!=\!\epsilon$), 
in addition lead to tractable position space photon 
propagators, that are expressed in terms of a finite number of derivatives acting on scalar 
propagators. These propagators are expressed in the covariantized 
form~(\ref{CovariantForm}), with the structure functions given 
in~(\ref{A1})--(\ref{A4}) for the conformal gauge,
and in~(\ref{B1})--(\ref{B4}) for the deceleration gauge. 
Both propagators satisfy both the respective equations of motion, and
the respective Ward-Takahashi identity.
The conformal gauge propagator also represents a new propagator for de Sitter space,
that is a one-parameter generalization of the simple photon gauge propagator~\cite{Woodard:2004ut}, 
while in the deceleration gauge the propagator correctly reduces to the general 
covariant gauge propagator in the de Sitter limit~\cite{Glavan:2022dwb,Glavan:2022nrd}, including the de Sitter breaking term.

\section*{Acknowledgements}
\label{sec: Acknowledgements}

This work was supported by the European Union and the Czech Ministry of Education, 
Youth and Sports (Project: MSCA Fellowship CZ FZU I --- CZ.02.01.01/00/22\textunderscore010/0002906).

\appendix

\section{Mode functions in discreet~$\zeta_n$ gauges}
\label{sec: Mode functions in discreet zeta gauges}

The equation of motion~(\ref{EOMnL}) 
for the particular mode functions satisfy can be solved for arbitrary~$n$
in terms of CTBD scalar mode functions~(\ref{CTBD}) and their derivatives.
These are worked out in this appendix, up to homogeneous solutions
that can be fixed by the flat space limit and the Wronskian normalization.
However, only the solutions for~$n\!=\!0,1$, considered in the
main text, lead to simple expressions when working out the position 
space two-point functions.

\subsection{Particular mode functions for $n\!<\!0$}

For discreet gauges with~$n\!<\!0$ the mode function equation~(\ref{EOMnL})
is solved by the following ansatz
\begin{align}
v_{\scr L} ={}&
	C_n U_{\nu+n}
	+
	\frac{ (1 \!-\! \xi) }{ 2(1 \!-\! 2n ) }
	\Bigl[ \frac{ - i k }{ (1 \!-\! \epsilon) H_0 } \Bigr]
	\Bigl( \frac{\mathcal{H}}{H_0} \Bigr)^{\! 2n-1} 
	U_{\nu+n-1}
\nonumber \\
&
	+
	\sum_{\ell=1}^{1-n}
	D_{n}^{\ell}
	\Bigl[ \frac{ - ik }{ (1 \!-\! \epsilon) H_0 } \Bigr]^{1-\ell}
	\Bigl( \frac{\mathcal{H}}{H_0} \Bigr)^{\! 2n-1+\ell}
	U_{\nu+n-1+\ell}
	\, ,
\label{AnsatzNless}
\end{align}
where~$C_n$ and~$D_{n}^{\ell}$ are coefficients to be determined.
Plugging in this ansatz into the equation of motion,
and applying the following identities
\begin{align}
\MoveEqLeft[3]
\biggl[ \partial_0^2 + k^2
	+ \Bigl( \frac{1}{4} \!-\! (\nu \!+\! n)^2 \Bigr) (1\!-\!\epsilon)^2 
		\mathcal{H}^2 \biggr] 
		\biggl[
		\Bigl( \frac{ \mathcal{H} }{ H_0 } \Bigr)^{\!2n-1} U_{\nu+n-1}
		\biggr]
\nonumber \\
&
	=
	(1 \!-\! \epsilon)^2 H_0^2
	\Bigl( \frac{ \mathcal{H} }{ H_0 } \Bigr)^{\!2n}
	\biggl[
	-
	4n \nu \Bigl( \frac{ \mathcal{H} }{ H_0 } \Bigr) U_{\nu+n-1}
	-
	2 (1 \!-\! 2n ) 
		\Bigl[ \frac{ - i k}{(1\!-\!\epsilon) H_0 } \Bigr] U_{\nu+n} 
	\biggr]
	\, ,
\\
\MoveEqLeft[3]
\biggl[ \partial_0^2 + k^2
	+ \Bigl( \frac{1}{4} \!-\! (\nu \!+\! n)^2 \Bigr) (1\!-\!\epsilon)^2 
		\mathcal{H}^2 \biggr]
	\biggl[
	\Bigl( \frac{\mathcal{H}}{H_0} \Bigr)^{\! 2n-1+\ell}
	U_{\nu+n-1+\ell}
	\biggr]
	=
	(1 \!-\! \epsilon)^2 H_0^2
	\Bigl( \frac{\mathcal{H}}{H_0} \Bigr)^{\! 2n+\ell }
\nonumber \\
&
	\times\!
	\biggl[
	4
	(n \!-\!1 \!+\! \ell ) ( 2n \!-\! 1 \!+\! \ell \!+\! \nu )
	\Bigl( \frac{\mathcal{H}}{H_0} \Bigr)
	U_{\nu+n-1+\ell }
	+
	2 ( 2n \!-\! 1 \!+\! \ell) 
	\Bigl[ \frac{ - ik }{ (1 \!-\! \epsilon) H_0 } \Bigr] 
	U_{\nu+n-2+\ell}
	\biggr]
	\, ,
\end{align}
yields a recurrence equation for coefficients of the sum,
\begin{equation}
D_n^{1}
	=
	- \frac{1}{2} \biggl[ 1 \!-\! \frac{ (1 \!-\! \xi) \nu }{ 1 \!-\! 2n } \biggr]
	\, ,
\qquad \quad
D_n^{\ell+1}
	=
	\frac{ 2( 1 \!-\! n \!-\! \ell) ( 1 \!-\! 2n \!-\! \ell \!-\! \nu ) }{ - 2n \!-\! \ell }
	D_n^{\ell}
	\, ,
\end{equation}
that is solved by
\begin{equation}
D_n^\ell
	=
	- 2^{\ell-2} \,
	\frac{ \Gamma( 1 \!-\! n ) }{ \Gamma( 2 \!-\! n \!-\! \ell ) }
	\frac{ \Gamma( 1 \!-\! 2n \!-\! \nu ) }{ \Gamma( 2 \!-\! 2n \!-\! \nu \!-\! \ell ) }
	\frac{\Gamma(1 \!-\! 2n \!-\! \ell)}{\Gamma( - 2n )}
	\biggl[ 1 \!-\! \frac{ (1 \!-\! \xi) \nu }{ 1 \!-\! 2n } \biggr]
	\, .
\end{equation}
The other particular mode function is then obtained by acting the time
derivative in~(\ref{EOMn0}) and using recurrence relations~(\ref{RecurrenceIdentities}).
Coefficient~$C_n$ remains undetermined by the equation of motion
as it multiplies the homogeneous solution,
and is rather fixed by requiring the flat space limits~(\ref{vLflat}) and~(\ref{v0flat}),
and the normalization condition~(\ref{Wronskian-like}).

\subsection{Particular mode functions for $n\!>\!1$}

For discreet gauges with~$n\!>\!1$ the solution to~(\ref{EOMnL}) is captured by a 
more involved ansatz,
\begin{align}
&
v_{\scr L} 
	=
	E_n U_{\nu+n}
	+
	\Bigl[ \frac{-ik}{(1\!-\!\epsilon) H_0 } \Bigr]
	\Bigl( \frac{\mathcal{H} }{ H_0} \Bigr)^{\! 2n-1}
	\biggl[
	\frac{1}{2\nu} 
		U_{\nu+n-1}
	+
	\sum_{\ell=1}^{n-1} F_n^\ell
		\Bigl[ \frac{-ik}{(1\!-\!\epsilon) H_0 } \Bigr]^{\ell}
	\Bigl( \frac{\mathcal{H} }{ H_0} \Bigr)^{\! -\ell}
		U_{\nu+n+1-\ell}
	\biggr]
\nonumber \\
&
	+
	\Bigl[ \frac{-ik}{(1\!-\!\epsilon) H_0 } \Bigr]^{n+1}
	\Bigl( \frac{\mathcal{H} }{ H_0} \Bigr)^{\! n-1}
	\biggl[
	G_n
	\frac{\partial U_{\nu+1} }{ \partial\nu }
	+
	\sum_{\ell=1}^{n-1} 
	\Bigl[ \frac{-ik}{(1\!-\!\epsilon) H_0 } \Bigr]^{\ell}
	\Bigl( \frac{\mathcal{H} }{ H_0} \Bigr)^{\!-\ell}
	\biggl(
	J_n^\ell
	\frac{\partial U_{\nu-1+\ell} }{ \partial \nu }
	+
	K_n^\ell
	U_{\nu-1+\ell}
	\biggr)
	\biggr]
	\, ,
\end{align}
with~$E_n$, $F_n^\ell$, $G_n$, $J_n^\ell$, and~$K_n^\ell$ the coefficients to
be determined. Plugging this ansatz into the equation of motion~(\ref{EOMnL}),
and using the following identities:
\begin{align}
&
\biggl[ \partial_0^2 + k^2
	+ \Bigl( \frac{1}{4} \!-\! (\nu \!+\! n)^2 \Bigr) (1\!-\!\epsilon)^2 
		\mathcal{H}^2 \biggr] 
		\biggl[
		\Bigl( \frac{ \mathcal{H} }{ H_0 } \Bigr)^{\!2n-1} U_{\nu+n-1}
		\biggr]
\nonumber \\
&	\hspace{1cm}
	=
	(1 \!-\! \epsilon)^2 H_0^2 \Bigl( \frac{ \mathcal{H} }{ H_0 } \Bigr)^{\!2n} 
	\biggl[
	- 4n\nu \Bigl( \frac{\mathcal{H} }{H_0} \Bigr)
		U_{\nu+n-1}
	+
	2 (2n \!-\! 1) 
		\Bigl[ \frac{ - i k}{(1\!-\!\epsilon) H_0} \Bigr]
		U_{\nu+n} 
	\biggr]
	\, ,
\\
&
\biggl[ \partial_0^2 + k^2 
	+ \Bigl( \frac{1}{4} \!-\! (\nu\!+\!n)^2 \Bigr) (1\!-\!\epsilon)^2 \mathcal{H}^2 \biggr]
	\biggl[ \Bigl( \frac{ \mathcal{H} }{H_0} \Bigr)^{\!2n-1-\ell} U_{\nu+n+1-\ell} \biggr]
	=
	(1\!-\!\epsilon)^2 H_0^2
	\Bigl( \frac{\mathcal{H}}{H_0} \Bigr)^{\! 2n-\ell}
\nonumber \\
&	\hspace{1.5cm}
	\times\!
	\biggl[
	4 (n\!-\! \ell) (\nu \!+\! 2n \!-\! \ell )
	\Bigl( \frac{\mathcal{H}}{H_0} \Bigr) U_{\nu+n+1-\ell}
	+
	2 (2n\!-\!1\!-\! \ell) \Bigl[ \frac{ - ik}{(1\!-\!\epsilon) H_0} \Bigr] U_{\nu+n-\ell}
	\biggr]
	\, ,
\\
&
\biggl[ \partial_0^2 + k^2
	+ \Bigl( \frac{1}{4} \!-\! (\nu \!+\! n)^2 \Bigr) (1\!-\!\epsilon)^2 \mathcal{H}^2 \biggr] 
	\biggl[ \Bigl( \frac{\mathcal{H}}{H_0} \Bigr)^{\! n-1} 
		\frac{\partial U_{\nu+1} }{\partial \nu} \biggr]
\nonumber \\
&	\hspace{1cm}
	=
	(1\!-\!\epsilon)^2 H_0^2
	\Bigl( \frac{  \mathcal{H}}{H_0} \Bigr)^{\!n}
	\biggl[
	2 (\nu \!+\! n)
		\Bigl( \frac{\mathcal{H}}{H_0} \Bigr) U_{\nu+1}
	+
		2 (n\!-\!1) 
		\Bigl[ \frac{ - ik }{ (1\!-\!\epsilon) H_0} \Bigr]
		\frac{\partial U_{\nu} }{\partial \nu}
	\biggr]
	\, ,
\\
&
\biggl[ \partial_0^2 + k^2
	+ \Bigl( \frac{1}{4} \!-\! (\nu \!+\! n)^2 \Bigr) (1\!-\!\epsilon)^2 
		\mathcal{H}^2 \biggr]
		\biggl[
		\Bigl( \frac{\mathcal{H}}{H_0} \Bigr)^{\!n-1-\ell} 
		\frac{\partial U_{\nu-1+\ell} }{\partial \nu}
		\biggr]
	=
	(1 \!-\! \epsilon)^2 H_0^2
		\Bigl( \frac{\mathcal{H}}{H_0} \Bigr)^{\!n-\ell} 
\\
&	\hspace{0.3cm}
	\times\!
	\biggl[
	2 ( \nu \!-\! n \!+\! 2\ell)
	\Bigl( \frac{\mathcal{H}}{H_0} \Bigr)
		U_{\nu-1+\ell}
	- 
	4 (n \!-\! \ell) ( \nu \!+\! \ell ) 
	\Bigl( \frac{\mathcal{H}}{H_0} \Bigr)
		\frac{\partial U_{\nu-1+\ell} }{\partial \nu}
	+
	2 (n \!-\! 1 \!-\! \ell) 
		\Bigl[
			\frac{ - i k }{ (1 \!-\! \epsilon) H_0 }
			\Bigr]
			\frac{\partial U_{\nu+\ell} }{\partial \nu}
		\biggr]
		,
\nonumber 
\\
&
\biggl[ \partial_0^2 + k^2
	+ \Bigl( \frac{1}{4} \!-\! (\nu \!+\! n)^2 \Bigr) (1\!-\!\epsilon)^2 
		\mathcal{H}^2 \biggr]
		\biggl[
		\Bigl( \frac{\mathcal{H}}{H_0} \Bigr)^{\!n-1-\ell} 
		U_{\nu-1+\ell}
		\biggr]
\\
&	\hspace{1cm}
	=
	(1 \!-\! \epsilon)^2 H_0^2
		\Bigl( \frac{\mathcal{H}}{H_0} \Bigr)^{\!n-\ell} 
	\biggl[
	- 
	4 (n \!-\! \ell) ( \nu \!+\! \ell ) 
	\Bigl( \frac{\mathcal{H}}{H_0} \Bigr)
		U_{\nu-1+\ell}
	+
	2 (n \!-\! 1 \!-\! \ell) 
		\Bigl[
			\frac{ - i k }{ (1 \!-\! \epsilon) H_0 }
			\Bigr]
			U_{\nu+\ell}
		\biggr]
		\, .
\nonumber 
\end{align}
produces recurrence relations between different coefficients,
\begin{subequations}
\begin{align}
&
	F_n^1
	=
	- \frac{ 2n \!-\! 1 \!+\! (1\!-\!\xi ) \nu }{ 4 \nu (n\!-\! 1) (\nu \!+\! 2n \!-\! 1 ) } 
	\, ,
\qquad
	F_n^{\ell+1} 
	=
	\frac{ - (2n\!-\!1\!-\! \ell) F_n^\ell  }{ 2 (n \!-\! 1\!-\! \ell) (\nu \!+\! 2n \!-\! 1 \!-\! \ell)  } \, ,
\\
&
	G_n 
	=
	- \frac{ n F_n^{n-1} }{ \nu\!+\!n }
	\, ,
\qquad
	J_n^1
	=
	\frac{ G_n }{ 2 ( \nu \!+\! 1 )  }
	\, ,
\qquad
	J_n^{\ell+1} 
	=
	\frac{ J_n^\ell }{ 2 ( \nu \!+\! 1 \!+\! \ell )  } 
	\, ,
\\
&
	K_n^1
	=
	\frac{ ( \nu \!-\! n \!+\! 2) J_n^1 }{ 2 (n \!-\! 1) ( \nu \!+\! 1 ) }
	\, ,
\qquad
	K_n^{\ell+1}
	=
	\frac{(n \!-\! 1 \!-\! \ell) K_n^\ell + ( \nu \!-\! n \!+\! 2 \!+\! 2\ell) J_n^{\ell+1} }
		{ 2( \nu \!+\! 1 \!+\! \ell )(n \!-\! 1 \!-\! \ell) }
	\, .
\end{align}
\end{subequations}
These are solved by:
\begin{align}
F_n^\ell ={}&
	\frac{ \Gamma(n \!-\! \ell) \, \Gamma(2n) \, \Gamma(\nu \!+\! 2n \!-\! \ell) }
		{ (-2)^{\ell} \, \Gamma(n) \,  \Gamma(2n \!-\! \ell) \, \Gamma(\nu \!+\! 2n) }
	\!\times\!
	\biggl[
	\frac{ 2n \!-\! 1 \!+\! (1\!-\!\xi ) \nu }{ 2 (2n \!-\! 1) \nu } 
	\biggr]
	\, ,
\\
G_n ={}&
	\frac{ - \Gamma(2n) \, \Gamma(\nu \!+\! n) }
		{ (-2)^{n-1} \, \Gamma^2(n) \, \Gamma(\nu \!+\! 2n ) }
	\!\times\!
	\biggl[
	\frac{ 2n \!-\! 1 \!+\! (1\!-\!\xi ) \nu }{ 2 (2n \!-\! 1) \nu } 
	\biggr]
	\, ,
\\
J_n^{\ell} 
	={}&
	\frac{ - \Gamma(\nu \!+\! 1) }{ 2^\ell \, \Gamma(\nu \!+\! 1 \!+\! \ell) }
	\!\times\!
	\frac{ \Gamma(2n) \, \Gamma(\nu \!+\! n) }
		{ (-2)^{n-1} \, \Gamma^2(n) \, \Gamma(\nu \!+\! 2n ) }
	\!\times\!
	\biggl[
	\frac{ 2n \!-\! 1 \!+\! (1\!-\!\xi ) \nu }{ 2 (2n \!-\! 1) \nu } 
	\biggr]
	\, ,
\\
K_n^\ell ={}&
	\frac{ \Gamma(\nu \!+\! 1) }{ 2^\ell \, \Gamma(\nu \!+\! 1 \!+\! \ell) }
	\!\times\!
	\frac{ \Gamma(2n) \, \Gamma(\nu \!+\! n) }
		{ (-2)^n \, \Gamma^2(n) \Gamma(\nu \!+\! 2n) }
	\times\!
	\biggl[
	\frac{ \nu \!-\! n \!+\! 2 }{ (n \!-\! 1) (\nu \!+\! 1) }
	+
	\psi(\nu \!+\! 2) 
\nonumber \\
&	\hspace{2cm}
	- \psi(\nu \!+\! 1 \!+\! \ell) + \psi(2 \!-\! n) - \psi(1 \!+\! \ell \!-\! n)
	\biggr]
	\!\times\!
	\biggl[
	\frac{2n \!-\! 1 \!+\! (1 \!-\! \xi)\nu}{ 2(2n \!-\! 1) \nu }
	\biggr]
	\, ,
\end{align}
where~$\psi(z)$ is the digamma function. Coefficient~$E_n$ remains undetermined
as it multiplies a homogeneous solution that is determined only by the flat space 
limit~(\ref{vLflat}) and~(\ref{v0flat}), and the normalization condition~(\ref{Wronskian-like}).

\section{Checks for position space two-point functions}
\label{sec: Checks for two-point function}

The position space solutions for two-point functions given in 
Sec.~(\ref{subsec: Covariantization}) have to satisfy both the respective equation of 
motion~(\ref{Photon2ptEOM}) and the respective  Ward-Takahashi 
identity~(\ref{WTidentity}). Checking that this is indeed true entails plugging in the covariant 
representation~(\ref{CovariantForm}) for two-point functions into the said equations,
and showing they are satisfies for the structure functions given in 
Sec.~\ref{subsec: Covariantization},~(\ref{A1})--(\ref{A4}) for
the conformal gauge and~(\ref{B1})--(\ref{B4}) for the deceleration gauge.
This is facilitated by the following expressions for derivatives of the bi-local variables,
\begin{subequations}
\begin{align}
&
\nabla_\mu \bigl( \partial_\nu y \bigr) = 
	g_{\mu\nu} (1\!-\!\epsilon) H^2 \bigl( 2 \!-\! y \!-\! 2 \epsilon \, e^{-v} \bigr)
	-
	\frac{\epsilon}{ 1\!-\!\epsilon }
	\Bigl[ \bigl( \partial_\mu y \bigr) \bigl( \partial_\nu u \bigr) 
		\!+\! \bigl( \partial_\mu u \bigr) \bigl( \partial_\nu y \bigr) \Bigr]
		\, ,
\\
&
\nabla_\mu \bigl( \partial_\nu u \bigr) = 
	- g_{\mu\nu} (1\!-\!\epsilon) H^2
	- \frac{1\!+\!\epsilon}{1\!-\!\epsilon}
		\bigl( \partial_\mu u \bigr) \bigl( \partial_\nu u \bigr) 
		\, ,
\qquad \quad
\nabla_\mu \bigl( \partial_\nu' u \bigr) = 0
\, ,
\end{align}
\label{derivatives}%
\end{subequations}
and by the contraction identities for basis tensors given in 
Table~\ref{TensorContractions},
%
\begin{table}[h!]
\renewcommand{\arraystretch}{1.3}
\centering
\begin{tabular}{l r}
\hline
$\quad
	g^{\mu\nu} \bigl( \partial_{\mu} y \bigr) \bigl( \partial_{\nu} y \bigr)$
	&
	$(1\!-\!\epsilon)^2 H^2 \bigl( 4y \!-\! y^2 \bigr) \quad$
\\
\hline
$\quad
	g'^{\rho\sigma} \bigl( \partial_{\rho}' y \bigr) \bigl( \partial_{\sigma}' y \bigr)$
	&
	$(1\!-\!\epsilon)^2 H'^2 \bigl( 4y \!-\! y^2 \bigr) \quad$
\\
\hline
$\quad
	g^{\mu\nu} \bigl( \partial_{\mu} y \bigr) \bigl( \partial_{\nu} u \bigr) $
	&
	$(1\!-\!\epsilon)^2 H^2 \bigl( 2 \!-\! y \!-\! 2 e^{-v} \bigr) \quad$
\\
\hline
$\quad
	g'^{\rho\sigma} \bigl( \partial_{\rho}' y \bigr) \bigl( \partial_{\sigma}' u \bigr) $
	&
	$(1\!-\!\epsilon)^2 H'^2 \bigl( 2 \!-\! y \!-\! 2 e^{v} \bigr) \quad$
\\
\hline
$\quad
	g^{\mu\nu} \bigl( \partial_{\mu} u \bigr) \bigl( \partial_{\nu} u \bigr) $
	&
	$- (1\!-\!\epsilon)^2 H^2 \quad$
\\
\hline
$\quad
	g'^{\rho\sigma} \bigl( \partial_{\rho}' u \bigr) \bigl( \partial_{\sigma}' u \bigr) $
	&
	$ - (1\!-\!\epsilon)^2 H'^2 \quad$
\\
\hline
$\quad
	g^{\mu\nu} \bigl( \partial_{\mu} y \bigr) \bigl( \partial_{\nu} \partial_{\!\rho}' y \bigr) $
	&
	$ (1\!-\!\epsilon)^2 H^2 (2\!-\!y) \bigl( \partial_{\rho}' y \bigr) \quad$
\\
\hline
$\quad
	g'^{\rho\sigma} \bigl( \partial_{\mu} \partial_{\rho}' y \bigr) \bigl( \partial_{\sigma}' y \bigr)$
	&
	$ (1\!-\!\epsilon)^2 H'^2 (2\!-\!y) \bigl( \partial_{\mu} y \bigr) \quad$
\\
\hline
$\quad
	g^{\mu\nu} \bigl( \partial_{\mu} u \bigr) \bigl( \partial_{\nu} \partial_{\rho}' y \bigr) $
	&
	$- (1\!-\! \epsilon)^2 H^2 
		\bigl[ \bigl(\partial'_{\rho}y \bigr) + 2 e^{-v} \bigl( \partial'_{ \rho} u \bigr) \bigr]
		\quad$
\\
\hline
$\quad
	g'^{\rho\sigma} \bigl( \partial_{\mu} \partial_{\rho}' y \bigr) \bigl( \partial'_{\sigma} u \bigr) $
	&
	$ - (1\!-\! \epsilon)^2 H'^2 
		\bigl[ \bigl(\partial_{\mu}y \bigr) + 2 e^v \bigl( \partial_{ \mu} u \bigr) \bigr]
		\quad$
\\
\hline
$\quad
	g^{\mu\nu} \bigl( \partial_{ \mu} \partial_{\rho}' y \bigr) 
	\bigl( \partial_{\nu} \partial_{\sigma}' y \bigr) $
	&
	$\qquad (1\!-\!\epsilon)^2 H^2 \bigl[ 4 (1\!-\!\epsilon)^2 H'^2 g'_{\rho\sigma}
			- \bigl( \partial'_\rho y \bigr) \bigl( \partial'_\sigma y \bigr) \bigr] \quad$
\\
\hline
$\quad
	g'^{\rho\sigma} \bigl( \partial_{\mu} \partial_{\rho}' y \bigr) 
	\bigl( \partial_{\nu} \partial_{\sigma}' y \bigr) \qquad$
	&
	$\qquad (1\!-\!\epsilon)^2 H'^2 \bigl[ 4 (1\!-\!\epsilon)^2 H^2 g_{\mu\nu}
			- \bigl( \partial_\mu y \bigr) \bigl( \partial_\nu y \bigr) \bigr] \quad$
\\
\hline
\end{tabular}
\caption{
Contractions of tensor structures (table adopted from~\cite{Glavan:2020zne}).}
\label{TensorContractions}
\end{table}
%
and by the equations for the scalar two-point functions given in
Sec.~\ref{subsec: Scalar propagators}.
These checks are further facilitated by computer algebra programs such as 
Cadabra~\cite{Peeters:2007wn,Peeters:2006kp,Peeters:2018dyg} and Wolfram
Mathematica.

\subsection{Ward-Takahashi identities}
\label{subsec: Ward-Takahashi identities}

The left-hand side of the Ward-Takahashi identity~(\ref{WTidentity})
can be expanded in the basis of two vectors,
\begin{equation}
\bigl( \nabla^\mu \!-\! 2 \zeta n^\mu \bigr)
	i \bigl[ \tensor*[_\mu^{\tt a\! }]{\Delta}{_\nu^{\tt\!  b}} \bigr](x;x')
	=
	(1\!-\!\epsilon)^2 H^2 
	\Bigl[
	\bigl( \partial'_\nu y \bigr) \mathcal{Z}_1(y,u,v)
	+
	\bigl( \partial'_\nu u \bigr) \mathcal{Z}_2(y,u,v)
	\Bigr]
	\, ,
\end{equation}
where the two structure functions are expressed in terms of the photon structure 
functions,
\begin{align}
\mathcal{Z}_1 ={}&
	\biggl[
	\bigl( 4y \!-\!  y^2 \bigr) \frac{\partial }{\partial y}
		+ ( 2 \!-\! y )
		\biggl( \frac{\partial }{\partial u} 
			\!+\! \frac{ D \!+\! 1 \!-\! 3\epsilon \!-\! 2 \zeta }{1 \!-\! \epsilon} 
		\biggr)
	\biggr]
	\mathcal{C}_2
	+
	\biggl[
	(2 \!-\! y) \frac{ \partial }{ \partial y }
	- \frac{ \partial }{ \partial u }
\nonumber \\
&
	-
	\frac{ D \!-\! 2\epsilon \!-\! 2\zeta }{ 1 \!-\! \epsilon }
	\biggr]
	\bigl( \mathcal{C}_1 + \mathcal{C}_3 \bigr)
	- 2 e^{-v}
	\biggl[
		\biggl( \frac{\partial }{\partial u} 
			\!+\! \frac{ (D \!-\! 2) \epsilon \!-\! 2\zeta }{ 1 \!-\! \epsilon } 
		\biggr)
	\mathcal{C}_2
	+
	\frac{\partial \mathcal{C}_3 }{\partial y}	
	\biggr]
	\, ,
\label{WT Z1}
\\
\mathcal{Z}_2 ={}&
	\biggl[
	\bigl( 4y \!-\! y^2 \bigr) \frac{\partial }{\partial y}
	+
	( 2 \!-\! y )
	\biggl( \frac{\partial }{\partial u}
	\!+\! \frac{ D \!-\! 2 \epsilon \!-\! 2\zeta }{ 1\!-\! \epsilon } 
	\biggr)
	\biggr]
	\mathcal{C}_3
	+
	\biggl[
	\bigl( 2 \!-\! y \bigr) \frac{\partial }{\partial y}
	- \frac{\partial }{\partial u}
	- \frac{D \!-\! 1\!-\! \epsilon \!-\! 2\zeta}{1-\epsilon}
	\biggr]
	\mathcal{C}_4
\nonumber \\
&
	- 
	2 e^{-v}
	\biggl[
	\biggl(
	\frac{\partial }{\partial u}
	\!+\! \frac{ (D \!-\! 2) \epsilon \!-\! 2 \zeta }{ 1 \!-\! \epsilon } 
	\biggr)
	\bigl( \mathcal{C}_1 + \mathcal{C}_3 \bigr)
	+
	\mathcal{C}_3
	+
	\frac{\partial \mathcal{C}_4 }{\partial y}
	\biggr]
	\, .
\label{WT Z2}
\end{align}

\bigskip
\noindent {\bf Conformal gauge.}
Plugging in the structure functions~(\ref{A1})--(\ref{A4}) into
expressions~(\ref{WT Z1}) and~(\ref{WT Z2}), and using
equation of motion for the rescaled propagator function~(\ref{Feom})
and recurrence relations~(\ref{Frecurrence}) produces,
\begin{align}
&
(1\!-\!\epsilon)^2 H^2 \mathcal{Z}_1
	=
	- \alpha
	\times
	e^{\frac{-v}{1-\epsilon}}
	\times
	e^{- \frac{(D-2)\epsilon}{2(1-\epsilon)} u }
	\times
	\frac{\partial}{\partial y} \mathcal{F}_\nu(y)
	\, ,
\\
&
(1\!-\!\epsilon)^2 H^2 \mathcal{Z}_2
	=
	-
	\alpha 
	\times
	e^{ \frac{-v}{1-\epsilon} }
	\times
	e^{- \frac{(D-2)\epsilon}{2(1-\epsilon)} u }
	\times
	\Bigl( \frac{D \!-\! 1}{2} \!-\! \nu \Bigr)
	\mathcal{F}_\nu(y)
	\, ,
\end{align}
which is precisely the right-hand side of the Ward-Takahashi 
identity~(\ref{WTidentity}) for the conformal gauge~$\zeta\!=\!1$.

\bigskip
\noindent{\bf Deceleration gauge.}
Plugging in the structure functions~(\ref{B1})--(\ref{B4}) into the 
expressions~(\ref{WT Z1}) and~(\ref{WT Z2}), and then using
the second recurrence relation~(\ref{Frecurrence}), and the 
equation~(\ref{Feom}) for the rescaled propagator function and its parametric 
derivative~(\ref{FparamEOM})
produces,
\begin{align}
&
(1\!-\!\epsilon)^2 H^2 \mathcal{Z}_1
	=
	- \beta
	\times
	e^{\frac{- \epsilon v}{1-\epsilon}}
	\times
	e^{- \frac{(D-2)\epsilon}{2(1-\epsilon)} u } 
	\times
	\frac{\partial}{\partial y} \mathcal{F}_{\nu+1}(y) 
	\, ,
\\
&
(1\!-\!\epsilon)^2 H^2 \mathcal{Z}_2
	=
	- \beta
	\times
	e^{ \frac{- \epsilon v }{1-\epsilon} } 
	\times
	e^{ - \frac{(D-2)\epsilon}{2(1-\epsilon)} u }
	\times
	\Bigl( \frac{D \!-\! 3}{2} \!-\! \nu \Bigr)
	\mathcal{F}_{\nu+1}(y)
	\, ,
\end{align}
which matches the right-hand-side of the Ward-Takahashi identity~(\ref{WTidentity}) 
for the deceleration gauge~$\zeta \!=\! \epsilon$.

\subsection{Equations of motion}
\label{subsec: Equations of motion}

Equations of motion~(\ref{Photon2ptEOM}) for the two-point functions have to be 
satisfied both off-coincidence, and also correctly reproduce the local source
term on the right-hand side for the Feynman propagator.

\subsubsection{Equations of motion off coincidence}
\label{subsubsec: Equations of motion off coincidence}

When checking equations of motion~(\ref{Photon2ptEOM}) 
are satisfied off-coincidence it is more convenient
to first rewrite them with the help of the Ward-Takahashi identity~(\ref{WTidentity}),
\begin{align}
\MoveEqLeft[5]
	\biggl[
	\delta_\mu^\rho {\dalembertian}
		- 2\zeta n^\rho \nabla_\mu
		- 2\zeta \bigl( \nabla_\mu  n^\rho \bigr)
		-{ R_\mu}^\rho
	\biggr]
	i \bigl[ \tensor*[_\rho^{\tt a \!}]{\Delta}{_\nu^{\tt \! b}} \bigr](x;x')
\nonumber \\
&
	=
	{\tt S}^{\tt ab}
	\frac{ i \delta^D(x \!-\! x') }{\sqrt{-g}}
	+
	\Bigl[ (1\!-\!\xi)\partial_\mu \!+\! 2 \zeta n_\mu \Bigr]
	\partial_\nu' \biggl[
	\Bigl( \frac{a'}{a} \Bigr)^{\! \zeta} 
	i \bigl[ \tensor*[^{\tt a\!}]{\Delta}{^{\tt \! b} } \bigr]_{\nu_\zeta } (x;x')
	\biggr]
	\, .
\label{ResidualEOM}
\end{align}
The left-hand side of this equation can be expanded in the appropriate tensor basis,
\begin{align}
\MoveEqLeft[6.]
	\biggl[
	\delta_\mu^\rho {\dalembertian} 
		- 2\zeta n^\rho \nabla_\mu
		- 2\zeta \bigl( \nabla_\mu  n^\rho \bigr)
		-{ R_\mu}^\rho
	\biggr]
	i \bigl[ \tensor*[_\rho^{\tt a\! }]{\Delta}{_\nu^{\tt\!  b}} \bigr](x;x')
	=
	(1\!-\!\epsilon)^2 H^2 \biggl\{
	\bigl( \partial_\mu \partial'_\nu y \bigr) \mathcal{E}_1(y,u,v)
\nonumber \\
&
	+
	\bigl( \partial_\mu y \bigl) \bigr( \partial'_\nu y \bigr) \mathcal{E}_2(y,u,v)
	+ \Bigl[ \bigl( \partial_\mu y \bigr) \bigl( \partial'_\nu u \bigr) 
		\!+\! \bigl( \partial_\mu u \bigr) \bigl( \partial'_\nu y \bigr) \Bigr] \,
		 \mathcal{E}_3(y,u,v)
\nonumber \\
&
	+
	\Bigl[ \bigl( \partial_\mu y \bigr) \bigl( \partial'_\nu u \bigr) 
		\!-\! \bigl( \partial_\mu u \bigr) \bigl( \partial'_\nu y \bigr) \Bigr] \, 
		\overline{\mathcal{E}}_3(y,u,v)
	+
	\bigl( \partial_\mu u \bigr) \bigl( \partial'_\nu u \bigr) \, \mathcal{E}_4(y,u,v) 
	\biggr\}
	\, .
\end{align}
This is the same tensor basis used to expand the two-point 
function~(\ref{CovariantForm}), 
but supplemented by an additional tensor structure odd under 
reflecting coordinates. The structure functions are expressed in terms of the 
photon structure functions by acing the derivatives onto the covariant 
form~(\ref{CovariantForm}),
\begin{align}
\mathcal{E}_1
	={}&
	\biggl[
	( 4y \!-\! y^2 ) \frac{\partial^2 }{\partial y^2}
	+ 
	2 ( 2 \!-\! y )
		\biggl( \frac{\partial }{ \partial u} 
		\!+\! \frac{ D \!-\! 4 \epsilon }{ 2 ( 1 \!-\! \epsilon ) }  \biggr) 
			\frac{\partial}{\partial y} 
	-
	\biggl(\! \frac{\partial }{ \partial u}
		\!+\! \frac{ D \!-\! 1 \!-\! 3 \epsilon }{ 1 \!-\! \epsilon }\! \biggr)
		\frac{\partial }{ \partial u}
	-
	\frac{ D \!-\! 2\epsilon \!-\! 2\zeta }{ 1 \!-\! \epsilon } 
	\biggr] \mathcal{C}_1
\nonumber \\
&
	+
	\frac{2(1 \!-\! \zeta)}{ 1 \!-\! \epsilon }
	\Bigl[
	( 2 \!-\! y ) \mathcal{C}_2
	\!-\!
	\mathcal{C}_3
	\Bigr]
	-
	4 e^{-v}
	\biggl[
	\biggl( \frac{\partial }{ \partial u} 
		\!+\! \frac{ ( D \!-\! 4 ) \epsilon }{ 2 ( 1 \!-\! \epsilon ) }  \biggr) 
			\frac{\partial \mathcal{C}_1 }{\partial y} 
	-
	\frac{ \zeta \!-\! \epsilon }{ 1 \!-\! \epsilon } 
	\mathcal{C}_2
	\biggr]
	,
\label{E1}
\\
\mathcal{E}_2
	={}&
	\biggl[
	( 4y \!-\! y^2 ) \frac{\partial^2}{\partial y^2}
	+
	2 ( 2 \!-\! y )
		\biggl( \frac{\partial}{\partial u}
			\!+\! 
			\frac{D \!+\! 4 \!-\! 6\epsilon \!-\! 2\zeta}{2(1 \!-\! \epsilon)} 
			\biggr) 
			\frac{\partial}{\partial y}
	- 
	\biggl( \frac{\partial}{\partial u} \!+\!
		\frac{ D \!+\! 1 \!-\! 5\epsilon }{1 \!-\! \epsilon } \biggr)
		\frac{\partial}{\partial u}
\nonumber \\
&
	- 
	\frac{2(D \!-\! 3\epsilon \!-\! \zeta) }{ 1 \!-\! \epsilon } 
	\biggr]
	\mathcal{C}_2
	-
	\frac{2(1 \!-\! \zeta)}{1\!-\!\epsilon} \frac{\partial }{\partial y}
		\bigl( \mathcal{C}_1 \!+\! \mathcal{C}_3 \bigr)
	-
	4e^{-v}
	\biggl( \frac{\partial}{\partial u}
			\!+\! \frac{ (D \!-\! 2) \epsilon \!-\! 2 \zeta }{2(1 \!-\! \epsilon)} \biggr) 
			\frac{\partial \mathcal{C}_2 }{\partial y}
	,
\label{E2}
\\
\mathcal{E}_3
	={}&
	- 
	\biggl[
	\frac{ \epsilon }{ 1 \!-\! \epsilon } ( 2 \!-\! y )
		\frac{\partial }{\partial y}
	+
	\frac{ 1 \!-\! \epsilon \!-\! \zeta}{1 \!-\! \epsilon} \frac{\partial }{\partial u}
	- 
	\frac{(D \!-\! 2\epsilon \!-\! 2\zeta)\epsilon}{(1 \!-\! \epsilon)^2} 
	\biggr]
	\mathcal{C}_1
	-
	\biggl[
	\frac{ \epsilon }{ 1 \!-\! \epsilon }
	( 4y \!-\! y^2 ) \frac{\partial }{\partial y}
\nonumber \\
&
	-
	( 2 \!-\! y )
		\biggl( 
		\frac{ 1 \!-\! \epsilon \!-\! \zeta}{1 \!-\! \epsilon} \frac{\partial }{\partial u}
		\!-\! 
		\frac{(D \!+\! 1 \!-\! 3\epsilon \!-\! 2 \zeta ) \epsilon }{(1 \!-\! \epsilon)^2}
		 \biggr)
	\biggr]
	\mathcal{C}_2
	+
	\biggl[
	2 ( 2 \!-\! y )
		\biggl(
		\frac{\partial}{\partial u}
		\!+\!
		\frac{ D\!+\!1 \!-\! 5 \epsilon \!-\! \zeta }{2(1\!-\!\epsilon)} 
		\biggr) \frac{\partial}{\partial y}
\nonumber \\
&
+
	( 4y \!-\! y^2 ) \frac{\partial^2}{\partial y^2}
	-
	\biggl(
	\frac{\partial}{\partial u} 
		\!+\! \frac{ D\!-\!4\epsilon \!-\! \zeta }{1\!-\!\epsilon}
		\biggr)
		\frac{\partial}{\partial u}
	-
	\frac{ D\!-\!1 \!-\! \zeta \!-\!(2D\!+\!1 \!-\! 3\zeta) \epsilon \!+\!4\epsilon^2 }
		{ (1\!-\!\epsilon)^2 }
	\biggr]
	\mathcal{C}_3
\nonumber \\
&
	-
	\frac{1 \!-\! \zeta}{ 1\!-\!\epsilon } \frac{\partial \mathcal{C}_4}{\partial y}
	-
	2e^{-v}
	\biggl[
	\frac{ \epsilon - \zeta }{ 1 \!-\! \epsilon } \frac{\partial\mathcal{C}_1}{\partial y}
	+
	\biggl( 
	\frac{1 \!-\! \epsilon \!-\! \zeta}{1 \!-\! \epsilon} \frac{\partial}{\partial u}
	- 
	\frac{(1 \!+\! \epsilon)(\epsilon \!-\! \zeta) \!+\!(D \!-\! 4)\epsilon^2 }{(1 \!-\! \epsilon)^2} 
	\biggr)
	\mathcal{C}_2
\nonumber \\
&
	+
	2 \biggl(
		\frac{\partial}{\partial u}
		- 
		\frac{1 \!+\! \zeta \!-\! (D\!-\!3)\epsilon }{2(1\!-\!\epsilon)}
		\biggr) \frac{\partial \mathcal{C}_3}{\partial y}
	\biggr]
	\, ,
\label{E3}
\\
\overline{\mathcal{E}}_3
	={}&
	\biggl[
	\frac{ \epsilon }{ (1 \!-\! \epsilon) } ( 2 \!-\! y )
		\frac{\partial }{\partial y}
	+
	\frac{1 \!-\! \epsilon \!-\! \zeta}{1 \!-\! \epsilon} \frac{\partial}{\partial u}
	- \frac{(D \!-\! 2\epsilon \!-\! 2\zeta )\epsilon}{(1 \!-\! \epsilon)^2} 
	\biggr]
	\mathcal{C}_1
	+
	\biggl[
	\frac{\epsilon}{1\!-\!\epsilon} \bigl( 4y \!-\! y^2 \bigr) \frac{\partial}{\partial y}
\nonumber \\
&
	-
	( 2 \!-\! y )
		\biggl(
		\frac{1 \!-\! \epsilon \!-\! \zeta}{1 \!-\! \epsilon} \frac{\partial}{\partial u}
		-
		\frac{(D \!+\! 1 \!-\! 3\epsilon \!-\! 2 \zeta ) \epsilon }{(1\!-\!\epsilon)^2}
		\biggr)
	\biggr]
	\mathcal{C}_2
	+
	\biggl[
	\frac{ 1\!+\!\epsilon \!-\! \zeta }{ 1\!-\!\epsilon } ( 2 \!-\! y ) \frac{ \partial }{ \partial y}
	+
	\frac{ 1 \!-\! \epsilon \!-\! \zeta}{1 \!-\! \epsilon} \frac{\partial }{\partial u}
\nonumber \\
&
	-
	\frac{ ( D \!-\! 2 \!-\! 2\epsilon ) \epsilon }{(1\!-\!\epsilon)^2}
	-
	\frac{ ( 1 \!-\! \zeta ) ( 1 \!+\! \epsilon ) }{ ( 1 \!-\! \epsilon )^2 }
	\biggr]
	\mathcal{C}_3
	-
	\frac{1 \!-\! \zeta }{ 1\!-\!\epsilon } \frac{\partial \mathcal{C}_4}{\partial y}
	-
	2e^{-v}
	\biggl[
	\frac{ \epsilon \!-\! \zeta  }{ 1 \!-\! \epsilon } \frac{\partial\mathcal{C}_1}{\partial y}
	+
	\frac{1 \!+\! \epsilon \!-\! \zeta}{1 \!-\! \epsilon} \frac{\partial \mathcal{C}_3}{\partial y}
\nonumber \\
&
	+
	\biggl(
	\frac{1\!-\!\epsilon\!+\!\zeta}{1 \!-\! \epsilon}
	\frac{\partial }{\partial u}
		\!+\!
		\frac{(D \!-\! 3 \!+\!\epsilon) \epsilon }{(1\!-\!\epsilon)^2}
			-
	\frac{\zeta (1 \!+\! \epsilon) }{(1 \!-\! \epsilon)^2}
		\biggr) \mathcal{C}_2
	\biggr]
	\, ,
\label{E3b}
\\
\mathcal{E}_4 ={}&
	\biggl[
	-
	\frac{ 2 \epsilon }{ 1 \!-\! \epsilon } \bigl( 4y \!-\! y^2 \bigr) \frac{ \partial }{\partial y}
	+
	2 ( 2 \!-\! y ) 
	\biggl( 
	\frac{1 \!-\! \epsilon \!-\!  \zeta}{1 \!-\! \epsilon} \frac{\partial }{\partial u} 
	-
	\frac{ (D\!-\!2\epsilon \!-\! 2 \zeta) \epsilon }{ (1\!-\!\epsilon)^2 } 
	\biggr)
	\biggr]
	\mathcal{C}_3
	+
	\biggl[
	( 4y \!-\! y^2 ) \frac{\partial^2}{\partial y^2}
\nonumber\\
&
	+
	2 ( 2 \!-\! y )
		\biggl( 
			\frac{ \partial }{ \partial u}
			+
			\frac{ D \!-\! 2 \!-\! 4\epsilon }{ 2(1\!-\!\epsilon) }  
			\biggr) \frac{ \partial }{ \partial y}
	- \biggl( \frac{ \partial }{ \partial u }
			+ \frac{ D\!-\!1\!-\!3\epsilon \!-\! 2\zeta }{ 1\!-\!\epsilon }
			\biggr)
			\frac{ \partial }{ \partial u}
	+
	\frac{2(D\!-\!1\!-\!\epsilon \!-\! 2 \zeta)\epsilon }{(1\!-\!\epsilon)^2}
	\biggr] \mathcal{C}_4
\nonumber \\
&
	-
	4e^{-v}
	\biggl[
	\biggl( \frac{\partial }{\partial u}
			\!-\! \frac{ 2 \!-\! (D\!-\!4)\epsilon }{ 2(1\!-\!\epsilon) } \biggr)
		\Bigl( \mathcal{C}_3 \!+\! \frac{ \partial \mathcal{C}_4 }{ \partial y} \Bigr)
	-
	\biggl(
	\frac{\zeta}{1 \!-\! \epsilon}
		\frac{\partial }{\partial u}
	+
	\frac{ (1\!+\!\epsilon) [ (D\!-\!2) \epsilon \!-\!  2\zeta ] }{ 2 (1\!-\!\epsilon)^2 }
		\biggr)
		\bigl( \mathcal{C}_1 \!+\! \mathcal{C}_3 \bigr)
	\biggr]
	\, .
\label{E4}
\end{align}

\bigskip
\noindent{\bf Conformal gauge.}
Recurrence relations~(\ref{Frecurrence}) and the equation~(\ref{Feom}) for the
rescaled propagator function are sufficient to infer that 
plugging~(\ref{A1})--(\ref{A4}) into the structure functions~(\ref{E1})--(\ref{E1})
simplifies them to:
\begin{align}
&
(1\!-\!\epsilon)^2 H^2 \mathcal{E}_1 
	=
	e^{ \frac{-v}{1-\epsilon} } e^{ - \frac{(D-2)\epsilon}{2(1-\epsilon)} u }
	\times
	( 1 \!-\! \alpha ) \frac{\partial \mathcal{F}_\nu}{\partial y}
	\, ,
\\
&
(1\!-\!\epsilon)^2 H^2 \mathcal{E}_2 
	=
	e^{ \frac{-v}{1-\epsilon} } e^{ - \frac{(D-2)\epsilon}{2(1-\epsilon)} u }
	\times
	( 1 \!-\! \alpha ) \frac{\partial^2 \mathcal{F}_\nu}{\partial y^2}
	\, ,
\\
&
(1\!-\!\epsilon)^2 H^2 \mathcal{E}_3 
	=
	e^{ \frac{-v}{1-\epsilon} } e^{ - \frac{(D-2)\epsilon}{2(1-\epsilon)} u }
	\times
	\biggl[
	\Bigl( \frac{D \!-\! 1}{2} \!-\! \nu \Bigr) (1\!-\!\alpha)
	+
	\frac{\alpha}{1\!-\!\epsilon}
	\biggr]
	\frac{\partial \mathcal{F}_\nu}{\partial y}
	\, ,
\\
&
(1\!-\!\epsilon)^2 H^2 \overline{\mathcal{E}}_3 
	=
	e^{ \frac{-v}{1-\epsilon} } e^{ - \frac{(D-2)\epsilon}{2(1-\epsilon)} u }
	\times
	\frac{ (- \alpha) }{ 1 \!-\! \epsilon } \frac{\partial \mathcal{F}_\nu}{\partial y}
	\, ,
\\
&
(1\!-\!\epsilon)^2 H^2 \mathcal{E}_4 
	=
	e^{ \frac{-v}{1-\epsilon} } 
	e^{- \frac{ (D-2) \epsilon }{ 2(1-\epsilon)} u }
	\Bigl( \frac{D \!-\! 1}{2} \!-\! \nu \Bigr)
	\biggl[ \Bigl( \frac{D \!-\! 1}{2} \!-\! \nu \Bigr) (1\!-\!\alpha) 
		\!+\! \frac{ 2 \alpha }{ 1 \!-\! \epsilon } \biggr]
	\mathcal{F}_\nu
	\, .
\end{align}
These precisely account for the right-hand side of~(\ref{ResidualEOM}) off-coincidence.

\bigskip
\noindent{\bf Deceleration gauge.}
Plugging in the structure functions~(\ref{B1})--(\ref{B4}) into~(\ref{E1})--(\ref{E4})
and using relations~(\ref{Feom})--(\ref{FparamRecurrence}) produces:
\begin{align}
&
(1\!-\!\epsilon)^2 H^2 \mathcal{E}_1 
	=
	e^{ \frac{-\epsilon v}{1-\epsilon} } e^{ - \frac{(D-2)\epsilon}{2(1-\epsilon)} u }
	\times
	( 1 \!-\! \beta ) \frac{\partial \mathcal{F}_{\nu+1} }{\partial y}
	\, ,
\\
&
(1\!-\!\epsilon)^2 H^2 \mathcal{E}_2 
	=
	e^{ \frac{-\epsilon v}{1-\epsilon} } e^{ - \frac{(D-2)\epsilon}{2(1-\epsilon)} u }
	\times
	( 1 \!-\! \beta ) \frac{\partial^2 \mathcal{F}_{\nu+1} }{\partial y^2}
	\, ,
\\
&
(1\!-\!\epsilon)^2 H^2 \mathcal{E}_3 
	=
	e^{ \frac{-\epsilon v}{1-\epsilon} } e^{ - \frac{(D-2)\epsilon}{2(1-\epsilon)} u }
	\times
	\biggl[
	\frac{ \epsilon \beta }{ 1 \!-\! \epsilon } 
	+
	\Bigl( \frac{D \!-\! 3}{2} \!-\! \nu \Bigr) (1 \!-\! \beta)
	\biggr]
	\frac{\partial \mathcal{F}_{\nu+1} }{\partial y}
	\, ,
\\
&
(1\!-\!\epsilon)^2 H^2 \overline{\mathcal{E}}_3 
	=
	e^{ \frac{-\epsilon v}{1-\epsilon} } e^{ - \frac{(D-2)\epsilon}{2(1-\epsilon)} u }
	\times
	\frac{ ( - \epsilon \beta ) }{ 1 \!-\! \epsilon } \frac{\partial \mathcal{F}_{\nu+1} }{\partial y}
	\, ,
\\
&
(1\!-\!\epsilon)^2 H^2 \mathcal{E}_4 
	=
	e^{ \frac{-\epsilon v}{1-\epsilon} } e^{ - \frac{(D-2)\epsilon}{2(1-\epsilon)} u }
	\times
	\Bigl( \frac{D \!-\! 3}{2} \!-\! \nu \Bigr)
	\biggr[
	\Bigl( \frac{D \!-\! 3}{2} \!-\! \nu \Bigr) (1 \!-\! \beta )
	+
	\frac{ 2 \epsilon \beta }{ 1 \!-\! \epsilon }
	\biggr]
	\mathcal{F}_{\nu+1}
	\, .
\end{align}
These are precisely what appears on the right-hand side of~(\ref{ResidualEOM}) 
for~$\zeta\!=\!\epsilon$ and~$\nu_\zeta \!=\! \nu \!+\! 1$,
bearing in mind that the scalar two-point function appearing there is 
given by~(\ref{DeltaW}).

\subsubsection{Equations of motion at coincidence}
\label{subsubsec: Equations of motion at coincidence}

When checking that local sources in equations of motion~(\ref{Photon2ptEOM}) 
are correctly reproduced, it is convenient to first apply a simple identity,
\begin{align}
\MoveEqLeft[3]
\bigl( \partial_\mu y \bigr) \bigl( \partial_\nu' y \bigr) f(y,u)
	=
	\partial_\mu \partial'_\nu I^2 \bigl[ f(y,u) \bigr]
	-
	\bigl( \partial_\mu \partial'_\nu y \bigr) I \bigl[ f(y,u) \bigr]
\\
&
	-
	\Bigl[ \bigl( \partial_\mu y \bigr) \bigl( \partial'_\nu u \bigr) 
		\!+\! \bigl( \partial_\mu u \bigr) \bigl( \partial'_\nu y \bigr) \Bigr]
		\frac{\partial}{\partial u} I\bigl[f(y,u) \bigr]
	-
	\bigl( \partial_\mu u \bigr) \bigl( \partial'_\nu u \bigr)
	\frac{\partial^2}{\partial u^2} I^2\bigl[ f(y,u) \bigr]
	\, ,
\nonumber 
\end{align}
to the covariant representation of the propagator~(\ref{CovariantForm}),
so that upon it reads
\begin{align}
i \bigl[ \tensor*[_\mu^{\scr + \!}]{\Delta}{_\nu^{\scr \! +}} \bigr] (x;x')
	={}&
	\partial_\mu \partial'_\nu I^2 \bigl[ \mathcal{C}_2(y,u) \bigr]
	+
	\bigl( \partial_\mu \partial'_\nu y \bigr) 
		\Bigl( \mathcal{C}_1 ( y,u )
		-
		I \bigl[ \mathcal{C}_2(y,u) \bigr]
		\Bigr)
\nonumber \\
&
	+ \Bigl[ \bigl( \partial_\mu y \bigr) \bigl( \partial'_\nu u \bigr)
		\!+\! \bigl( \partial_\mu u \bigr) \bigl( \partial'_\nu y \bigr) \Bigr] 
	\Bigl(
	\mathcal{C}_3 ( y,u )
	-
	\frac{\partial}{\partial u} I\bigl[\mathcal{C}_2(y,u) \bigr]
	\Bigr)
\nonumber \\
&
	+ \bigl( \partial_\mu u \bigr) \bigl( \partial'_\nu u \bigr) \, 
	\Bigl(
	\mathcal{C}_4 ( y,u ) 
	-
	\frac{\partial^2}{\partial u^2} I^2\bigl[ \mathcal{C}_2(y,u) \bigr]
	\Bigr)
	\, .
\end{align}
It is a simple matter to work out the leading order contributions close to coincidence
for the combinations of structure functions appearing in the representation above.
For the conformal gauge these are
\begin{align}
&
I^2[\mathcal{A}_2]
	\ \overset{x' \to x}{\longsim} \
	\frac{1\!-\!\alpha }{2} I [ \mathscr{C} ]
	\, ,
\qquad
\mathcal{A}_1 - I[\mathcal{A}_2]
	\ \overset{x' \to x}{\longsim} \
	- \mathscr{C} 
	\, ,
\nonumber \\
&
\mathcal{A}_3 - \frac{\partial}{\partial u} I [\mathcal{A}_2]
	\ \overset{x' \to x}{\longsim} \
	\mathscr{C}
	\, ,
\qquad
\mathcal{A}_4 - \frac{\partial^2}{\partial u^2} I^2 [\mathcal{A}_2]
	\ \overset{x' \to x}{\longsim} \
	\Bigl( \frac{D\!-\!3}{2} \!-\! \nu \Bigr) I[\mathscr{C}]
	\, ,
\end{align}
while for the deceleration gauge they read
\begin{align}
&
I^2[\mathcal{B}_2]
	\ \overset{x' \to x}{\longsim} \
	\frac{1 \!-\! \beta}{2}
	I [ \mathscr{C} ]
	\, ,
\qquad
\mathcal{B}_1 - I[\mathcal{B}_2]
	\ \overset{x' \to x}{\longsim} \
	- \mathscr{C}
	\, ,
\nonumber \\
&
\mathcal{B}_3 - \frac{\partial}{\partial u} I [\mathcal{B}_2]
	= 0
	\, ,
\qquad
\mathcal{B}_4 - \frac{\partial^2}{\partial u^2} I^2 [\mathcal{B}_2]
	\ \overset{x' \to x}{\longsim} \
	\Bigl( \nu \!-\! \frac{D\!-\!3}{2} \Bigr) I[\mathscr{C}]
	\, ,
\end{align}
where,
\begin{equation}
\mathscr{C}(y,u) 
	= \frac{ i \Delta_{\scr 1/2}(y,u) }{ 2 (1\!-\!\epsilon)^2 H H' }
	\, .
\end{equation}
Then the identities generating the temporal delta function,
\begin{equation}
\partial_0 \partial_0 y_{\scr ++}
	=
	\partial_0 \partial_0 y
	+
	4 (1\!-\!\epsilon)^2 \mathcal{H}^2 \! \times \! i\delta \!\times\! \delta(\eta\!-\!\eta')
	\, ,
\quad \
\partial_0 \partial_0' y_{\scr ++}
	=
	\partial_0 \partial_0' y
	-
	4 (1\!-\!\epsilon)^2 \mathcal{H}^2 \! \times \!i\delta \!\times\! \delta(\eta\!-\!\eta')
	\, ,
\qquad
\end{equation}
and the identity generating the~$D$-dimensional delta function,
\begin{equation}
- 4(1\!-\!\epsilon)^2 H^2 \!\times\!  i\delta \!\times\! \delta(\eta\!-\!\eta') 
	\times \frac{\partial}{\partial y}
	i \Delta_{\scr 1/2}\bigl(y_{\scr ++}, u \bigr)
	=
	\frac{i \delta^{D}(x\!-\!x') }{ \sqrt{-g} }
	\, ,
\end{equation}
are used infer the following identities for generating local terms,
\begin{align}
&
{\mathcal{D}_\mu}^\rho \Bigl\{
	\partial_\rho \partial'_\nu  I \bigl[ \mathscr{C} (y_{\scr ++} , u) \bigr]
	\Bigr\}
	\ \overset{x' \to x}{\longsim} \
	\frac{2}{\xi} \bigl( a^2 \delta_\mu^0 \delta_\nu^0 \bigr)
	\frac{ i \delta^{D}(x\!-\!x') }{ \sqrt{-g} }
\\
&
{\mathcal{D}_\mu}^\rho \Bigl\{
	\bigl( \partial_\rho \partial'_\nu y \bigr) \mathscr{C} (y_{\scr ++} , u)
	\Bigr\}
	\ \overset{x' \to x}{\longsim} \
	-
	\biggl[
	g_{\mu\nu}
	+
	\Bigl(
	1 \!-\! \frac{1}{\xi}
	\Bigr)
	\bigl( a^2 \delta_\mu^0 \delta_\nu^0 \bigr)
	\biggr]
	\frac{ i \delta^{D}(x\!-\!x') }{ \sqrt{-g} }
	\, ,
\\
&
{\mathcal{D}_\mu}^\rho
	\Bigl\{
	\Bigl[ \bigl( \partial_\rho y \bigr) \bigl( \partial'_\nu u \bigr) 
		\!+\!
		\bigl( \partial_\rho u \bigr) \bigl( \partial'_\nu y \bigr) 
		\Bigr] \mathscr{C} (y_{\scr ++} , u)
	\Bigr\}
	\ \overset{x' \to x}{\longsim} \ 
	0 \, ,
\\
&
{\mathcal{D}_\mu}^\rho
	\Bigl\{
	\bigl( \partial_\rho u \bigr) \bigl( \partial'_\nu u \bigr) 
		I \bigl[ \mathscr{C} (y_{\scr ++} , u) \bigr]
	\Bigr\}
	\ \overset{x' \to x}{\longsim} \ 
	0 \, .
\end{align}
When applied to the both gauges these correctly reproduce the local source for the 
Feynman propagator in the equation of motion.


\end{document}